%% file: Narratives_experiment.tex
\documentclass[12pt,a4paper]{article}
\usepackage[utf8]{inputenc}
\usepackage{amsmath,amssymb,amsthm}
\usepackage{xcolor}
\usepackage{tikz}
\usepackage{subcaption}
\usepackage{enumerate}
\usepackage[shortlabels]{enumitem}
\usetikzlibrary{decorations.pathmorphing,decorations.pathreplacing,angles,quotes,patterns}
\usepackage{fontawesome5}
\usepackage[top=1.25in, left=1.25in, right=1.25in, bottom=1.25in]{geometry}
\usepackage[onehalfspacing]{setspace}
\usepackage{hyperref}
\usepackage{footmisc}
\usepackage{booktabs}
\usepackage{pgfplots}
\usepackage{xcolor}
\pgfplotsset{compat=1.18}
\usepgfplotslibrary{fillbetween}
\usepgfplotslibrary{groupplots}

\usepackage{siunitx}
\usepackage{threeparttable}
\usepackage{pdflscape}
\usepackage{subcaption}

\usepackage{tabularx,booktabs,colortbl,xcolor,makecell}
\usepackage[utf8]{inputenc}
\usepackage[T1]{fontenc}
\usepackage{lmodern}
\usepackage{graphicx}
\usepackage{float}
\usepackage{array}
\usepackage{colortbl}
\usepackage{mdframed}
\usepackage{setspace}
\usepackage{longtable}
\usepackage{caption}
\usepackage{sansmath}
\usepackage{helvet}

\input{results_artifacts/Plot_colors}

\newcommand{\blue}[1]{\textcolor{instructionsblue}{#1}}
\newcommand{\redc}[1]{\textcolor{instructionsred}{#1}}
\newcommand{\blk}[1]{\textcolor{instructionsblack}{#1}}

\newcolumntype{Y}{>{\raggedright\arraybackslash}X}
\newcolumntype{C}[1]{>{\centering\arraybackslash}p{#1}}
\newcommand{\ci}[1]{\multicolumn{1}{C{2.05cm}}{\scriptsize #1}}

\newmdenv[
  backgroundcolor=cellgray,
  linecolor=instructionsblue,
  linewidth=0.6pt,
  roundcorner=4pt,
  innerleftmargin=10pt,
  innerrightmargin=10pt,
  innertopmargin=6pt,
  innerbottommargin=6pt
]{infobox}

\newcommand{\E}{\mathbb{E}}

\newcommand{\Report}{\mathcal{R}}
\newcommand{\N}{\ensuremath{\mathbb{N}}}

\DeclareMathOperator*{\argmax}{arg\,max}

\DeclareMathOperator{\lab}{lab}

\newcounter{hypo}
\newtheorem{proposition}[]{Proposition}

\newtheorem{hypothesis}[hypo]{Hypothesis}
\theoremstyle{definition}
\theoremstyle{definition}\newtheorem{assumption}[]{Assumption}
{\theoremstyle{definition}}
{\theoremstyle{definition}}

\newtheorem{result}[]{Result}

\usepackage[round]{natbib}
\allowdisplaybreaks

\usepackage[normalem]{ulem}
\usepackage{cancel}

\title{\vspace{-1.5cm}Strategic communication of narratives: An experiment}
\author{Gerrit Bauch\thanks{Center for Mathematical Economics, Bielefeld University, PO Box 10 01 31, 33501 Bielefeld, Germany. Email: gerrit.bauch@uni-bielefeld.de.}, \hskip5pt Arthur Dolgopolov\thanks{Department of Economics and Finance,
University of Rome Tor Vergata, Via Columbia 2 00133 Rome, Italy. Email: Dolgopolov@Economia.uniroma2.it.}, \hskip5pt and Manuel Foerster\thanks{Center for Mathematical Economics, Bielefeld University, PO Box 10 01 31, 33501 Bielefeld, Germany. Email: manuel.foerster@uni-bielefeld.de.
\\\faIcon{creative-commons}\faIcon{creative-commons-by}\faIcon{creative-commons-nc}\faIcon{creative-commons-sa}\quad The authors thank Yves Breitmoser, Valeria Burdea, Herbert Dawid, Anna Hager, and Stefan Trautmann for fruitful discussions. %
We also thank various members of the Center for Mathematical Economics for their assistance in pre-testing the experiment. Gerrit Bauch gratefully acknowledges financial support by the German Research Foundation (DFG) [RTG 2865/1 – 492988838]. %
Manuel Foerster gratefully acknowledges financial support from the German Research Foundation (DFG) under grant FO 1272/2-1.}}

\date{\today}

\begin{document}
\maketitle

\begin{abstract}
\noindent%
We investigate the strategic communication of narratives under model uncertainty. The sender has private information about the true data-generating process of publicly observable data. The receiver is uncertain about how to interpret the data, but aware of the sender's incentives to strategically provide interpretations (\emph{``narratives''}). We theoretically show that the size of the conflict of interest between the sender and the receiver is a crucial determinant of equilibrium communication. In particular, the stronger the sender's bias, (i) the more senders exaggerate their information and (ii) the more receivers correct the sender's action recommendations. In a laboratory experiment, we find evidence in line with both predictions, suggesting that people in complex and uncertain environments take a narrator's strategic incentives into account. Additional analyses reveal that narrative likelihood does not drive receiver behavior and that narratives are, on average, slightly persuasive only when bias is low.
\end{abstract}

{\footnotesize{\it Keywords:} Narratives, model uncertainty, 
strategic communication, economic experiments

{\it JEL: C72, C91, D81, D82, D83} }

\section{Introduction}

Persuading others often involves providing narratives about past data. In the presence of model uncertainty, the interpretation of such data is particularly demanding. A strategic narrator may exploit this by providing narratives favorable to her cause. Whether such strategies are successful will largely depend on the recipient's strategic sophistication. The economic literature on the communication of narratives has thus far emphasized the role of the narrative's fit to the data of persuasiveness \citep{schwartzstein2021using}. From a game-theoretic point of view, however, recipients understand the narrator's strategic incentives and correct for her bias.

To what extent people in complex and uncertain environments take a narrator's strategic incentives into account is an open question. To investigate this issue, we introduce a formal model of strategic communication of narratives under model uncertainty based on \cite{bauch2024strategiccommunicationnarratives}, and conduct an experiment to test it. 

In our model, two agents observe public data. Each of the finitely many data points may either depend on the unknown state of nature or be independent of it and thus noise. The biased \emph{sender} (she) learns the exact underlying process that has generated the data (the \emph{true model}), while the \emph{receiver} (he) faces model uncertainty in the sense that he does not know how to interpret the data. %
The sender then submits a cheap-talk report to the receiver, providing a \emph{narrative} (or \emph{model}) for how to interpret the data. In the context of the experiment, we can interpret the sender's narrative also as an \emph{action recommendation}. %
Upon observing the sender's narrative, the receiver takes an action. We assume that the receiver resolves uncertainty \`a la \cite{gilboa1989maxmin}. %
The receiver is thus ambiguity averse and maximizes his worst-case expected utility among all models he cannot rule out.

Our model generates a central comparative static result: A larger conflict of interest between the agents leads to coarser communication, in the sense that the sender employs fewer distinct narratives in equilibrium. The size of the conflict of interest between the sender and the receiver thus is a crucial determinant of equilibrium communication.
This result is driven by two key forces: First, the biased sender seeks to persuade the receiver into taking an action that is favorable to her. 
Second, the receiver takes the sender's strategic incentives into account and corrects for the bias. Specifically, the model predicts that as the sender's bias increases, her action recommendations will deviate more from the receiver's optimal action under the true model. In turn,  the receiver's actions will also deviate more from the sender's recommendations. %
While the literature thus far has focused on the sender's ability to persuade the receiver, we focus on the receiver's strategic sophistication and ask to what extent people in complex environments take a narrator’s strategic incentives into account.

To test these predictions, we conduct a laboratory experiment. An informed `sender' is matched with an uninformed `receiver'. Both subjects observe three draws from one of two urns, either an urn with \emph{unknown} composition or one with \emph{known} 50-50 composition. The receiver's task is to guess the composition of the unknown urn. Before she does so, the sender learns from which of the urns each ball has been drawn. She can then communicate a narrative to the receiver without direct payoff consequences. %
To test the predictions, the experiment varies the size of the conflict of interest between the sender and the receiver. While there is no conflict in the \emph{no bias} treatment, we consider a biased sender in the \emph{low bias} and the \emph{high bias} treatments. 

We find evidence for both predictions. In line with the first prediction, sender subjects in the \emph{low bias} and the \emph{high bias} treatments recommend higher actions relative to the receiver's optimal action than sender subjects in the \emph{no bias} treatment. Moreover, receiver subjects in the \emph{high bias} treatment take lower actions relative to the sender's recommendations than receiver subjects in the \emph{low} and the \emph{no bias} treatments. Our results show not only that narrators are strategic, but that people take a narrator's strategic incentives into account, even in complex and uncertain environments. Moreover, they reveal subtle differences in the responses of sender and receiver subjects to conflicting interests. While receivers under-react to a low bias relative to senders, they over-react to a high bias.

Additional analyses show that\textemdash in line with the non-monotone dynamic suggested by the theoretical model, but different from other cheap-talk experiments \citep[see][]{groseclose2021humans}\textemdash 
the sender's narrative is slightly persuasive only under low bias. Receivers compensate for most of the distortion when the sender's bias is too high. %
As a consequence, it is the sender subjects who bear most of the welfare losses from a large conflict of interest in our experiment.

We then compare our approach with an alternative model where subjects evaluate the sender's narrative by its likelihood of generating the observed data \`a la \cite{schwartzstein2021using}. First, we find that this alternative model does not yield our main hypothesis on receiver behavior from a theory perspective. Second, it also does not explain receiver behavior in our experiment: narrative likelihood is not positively correlated with receiver trust and predicted actions are often close to the boundaries of the action space, while subjects rarely choose such actions.

{\bfseries Related literature. }%
On the theoretical side, our work builds on the companion paper \cite{bauch2024strategiccommunicationnarratives}, which models the communication of narratives  as a cheap-talk game under model uncertainty. Narratives are viewed as likelihood functions, which establish a probabilistic link between observable data and the parameter of interest. %
This relates our work to \cite{schwartzstein2021using}, in which a biased narrator provides narratives to a receiver, aiming at making a data set plausible in her favor \citep[see also][]{aina2023tailored,jain2023informing,eliaz2021strategic}. The receiver adopts the narrative if it fits the observable data better than some default model, and thus ignores the sender's strategic incentives. As a main result of theirs, the persuader finds it easier to manipulate the receiver's beliefs if the default model fits the data poorly.

In contemporary and closely related work, \cite{barron2024narrative} conduct a laboratory experiment to test the model of \cite{schwartzstein2021using}. Specifically, they consider an investor-advisor setup and study how sender subjects can shape the way receiver subjects interpret public data by proposing a narrative. They find that narratives are persuasive and that the likelihood of a narrative is a key determinant of persuasiveness. Moreover, individuals use this fact to tailor narratives to their own benefit. %
Our work differs from \cite{barron2024narrative} in three important aspects. First, they consider a probabilistic investor-advisor setup, while we consider an urn setup in which receivers face model uncertainty. Second, in their setup the conflict of interest between the sender and the receiver is either zero or maximal and, except for one treatment, hidden from the receiver, while we focus on comparative statics with respect to a known conflict of interest. Third, their analysis focuses on the persuasiveness of narratives, while we are mainly interested in understanding the extent to which people take a narrator’s strategic incentives into account. %
Despite these differences, our findings on narrative likelihood and narrative persuasiveness complement their findings, suggesting that the set-up and information on conflicting preferences have a crucial influence on receiver behavior.

Our contribution is on the empirical side, where we are the first to study the extent to which people in complex and uncertain environments take a narrator's strategic incentives into account. Following the seminal work on cheap talk by \cite{crawford1982strategic}, a large body of experimental literature has studied cheap talk communication \citep{blume1998experimental,blume2002learning,wang2010pinocchio}. \cite{wang2010pinocchio} find that subjects reveal more information than predicted in equilibrium (``over-communication'') and that information transmission, measured by the receiver payoff, decreases in the sender's bias. \cite{foerster2021casting} study communication about externalities in a psychological game, finding that subjects downplay social impact to cast doubt on the effectiveness of action to excuse passivity; see \cite{benabou2020narratives} for a closely related theoretical model. Different from this literature, communication in our experiment concerns the interpretation of public data rather than the payoff-relevant parameter.

The remainder of the article is organized as follows. Section \ref{Section:theory} introduces the theoretical model and derives the equilibria and hypotheses. Section \ref{Section:Procedures} contains the experimental procedures. In Section \ref{Section:Results}, we present the experimental results. 

\section{Theory}\label{Section:theory}

\subsection{Model}\label{Subsection: model and notation}
Two agents, a \emph{sender} ($S$ or she) and a \emph{receiver} ($R$ or he), both observe a history of past outcomes (or a public signal) $h=(h_1,h_2,\ldots,h_K)\in H=\{0,1\}^K$ about the state of nature $\theta\in\Theta=\{0,1,\ldots, L\}$, with $K,L\in \N$. %
The common prior over the state is a distribution with strictly positive probability mass function $f_0$ on $\Theta$.
A \emph{model} or \emph{narrative} $m\in M$ parametrizes a likelihood function $\{\pi_m(\cdot|\theta)\}_{\theta\in\Theta}$, where $\pi_m(h|\theta)$ denotes the likelihood of history $h$ given state $\theta$ under model $m$. %

The state $\theta$ is drawn according to $f_0$ while there is model uncertainty about the \emph{true model} $m^T$. %
The history $h$ is generated according to the true model $m^T$ and the state $\theta$. %
Specifically, we assume that $m^T\in M = 2^{\{1, \dots, K\}}$ and $h_1,\dots,h_K$ are independent conditional on $\theta$, with
\begin{align}
\Pr(h_k=1\ |\ \theta)=\left\{\begin{array}{cl} \frac{\theta}{L}, & \mbox{if }k\in m^T\\ \frac{1}{2}, & \mbox{else} \end{array}\right. \text{for all } k=1,2,\ldots, K. \label{modified_beta-binomial}
\end{align}
We can interpret \eqref{modified_beta-binomial} as a draw from one of two urns, either an urn with \emph{unknown} composition (if $k\in m^T$) or one with \emph{known} 50-50 composition (if $k\notin m^T$). The likelihood function corresponding to model $m$ then is
\begin{align*}
    \pi_{m}(h \mid \theta) = \left(\frac{1}{2} \right)^{K-\# m} \cdot \left(\frac{\theta}{L}\right)^{\sum_{k \in m} h_k}   
    \cdot \left(1-\frac{\theta}{L}\right)^{\# m-\sum_{k \in m} h_k}.
\end{align*}

While the sender also learns the true model $m^T$ and updates her prior to the posterior $f_{m^T,h}$ using Bayes' rule, the receiver faces full model uncertainty about the true model $m^T\in M$, i.e.,\ he initially is willing to entertain any model.\footnote{We refer the reader to our companion paper \cite{bauch2024strategiccommunicationnarratives} on how to formally model the belief forming process in this case.}

After the sender has observed the history $h$ and the true model $m^T$, she submits a cheap-talk report (or message) $r\in \Report := M$, suggesting a model to the receiver. %
Upon observing the report $r$, the receiver can rule out all models under which $r$ is not sent, resulting in a remaining \emph{minimal feasible set} $\tilde M$. $\tilde{M}$ contains all models the receiver cannot exclude from his considerations, i.e., the indicative meaning of the submitted report is ``$m^T\in\tilde M$''. %
Finally, the receiver takes an action $a\in A=\{0,1,\ldots, L\}$.

\paragraph{Payoffs.}

The sender's payoff is given by
\begin{align*}
    u_S(a,\theta,b)= \max\{ 1300 - 300 \cdot \lvert \theta + b - a \rvert, 0\},
\end{align*}
which depends on the receiver's action $a\in A$, the state $\theta\in \Theta$, and the sender's bias $b\in\{0,1,\ldots, L\}$, a commonly known constant that measures the conflict of interest between sender and receiver. Note that $u_S$ quasi-concave in $a$ and capped from below by $0$. %
Let $U_{m,h}(a, b) := \E[u_S(a, \theta, b) \mid m,h]$ denote the sender's expected utility after an update on the information revealed through the datum $(m,h)$.
In all considered cases, we get a unique maximizer by choosing the action $a(m,h,b)$ in  $\argmax_{a \in A}U_{m,h}(a,b)$ closest to $5$ and refer to it as the sender's \emph{bliss point} under $m$.
The receiver is unbiased, i.e., his payoff function is $u_R(a,\theta) =u_S(a,\theta,0)$. %
Interpreting \eqref{modified_beta-binomial} as a draw from one out of two urns, the receiver's objective is thus to guess the composition of the unknown urn.

Despite obtaining information through the sender's message, the receiver still faces uncertainty in the form of the minimal feasible set $\tilde{M}$. We assume that the receiver resolves the remaining uncertainty \`a la \cite{gilboa1989maxmin} and chooses her action according to her worst-case expected utility given $\tilde M$,\footnote{This is a special instance of an \emph{ambiguity rule} introduced in \cite{bauch2024strategiccommunicationnarratives}.}
\begin{align}\label{max_min}
U_{\tilde{M}, h}(a) = \min_{m \in \tilde M} \E[u_R(a,\theta) \mid m,h].
\end{align}

\paragraph{Equilibrium concept.}
In a strategic interaction, we expect both agents to optimally respond to the other agents' strategies. %
Start by fixing a history $h$. A (pure) strategy for the sender is then a function $\sigma \colon M \to \Report$ %
that assigns a report to each model. Ignoring off-equilibrium messages, a (pure) strategy for the receiver is a function $\rho \colon \sigma(M) \to A$ that assigns an action to each report received.\footnote{Note that since the history $h$ is known to both agents ex ante, i.e., before they make their move, one can treat every $h$ as indexing a different game.} %
Upon observing a report $r \in \sigma(M)$, and given the sender strategy $\sigma$, the receiver can discard all models under which $r$ is not sent. The minimal feasible set is thus $\tilde{M} = \sigma^{-1}(r)$. %
An \emph{equilibrium} $(\sigma,\rho)$ (under max-min expected utility) entails mutual best replies, i.e., $\sigma(m) \in \argmax_{r \in \Report} U_{m,h}(\rho(r), b)$ and $\rho(r) = \argmax_{a \in A} U_{\sigma^{-1}(r),h}(a)$.

\paragraph{Parameters used in the experiment.} 
In the experiment, we set $K=3$, $L=10$, $f_0=\mathcal{U}(\Theta)$. We vary the bias across three treatments to test our hypotheses, namely, \emph{no bias} ($b=0$), \emph{low bias} ($b=1$), and \emph{high bias} ($b=3$).

\subsection{Equilibrium analysis}\label{Subsection:Analysis}

In this section, we first characterize the equilibria of our model. Second, we derive comparative statics with respect to the conflict of interest. We refer to an equilibrium which induces $n$ distinct actions as an \emph{$n$-step equilibrium}. In the following, we restrict attention to equilibria in which messages and induced actions are in a one-to-one correspondence.\footnote{Indeed, considering such \emph{reduced} equilibria $(\sigma,\rho)$, for which $\sigma(m) = \sigma(m')$ whenever $\rho(\sigma(m)) = \rho(\sigma(m'))$, is without loss, cf.\ \cite{bauch2024strategiccommunicationnarratives}.} Thus, messages can be identified with action recommendations.

To better understand equilibria, we totally order the set of models $M$ with respect to the receiver's bliss points $a(m,h,0)$ they induce. An equilibrium turns out to be a consecutive partition of $M$ into up to $N(h,b)$ intervals under this order, together with the receiver's actions under model uncertainty.

\begin{proposition}[Equilibrium characterization]\label{Proposition: N step equilibrium}
For any $h$ and $b$, there is a natural number $N=N(h,b)$ such that there exists an $n$-step equilibrium if and only if $1 \leq n \leq N$. Any $n$-step equilibrium $(\sigma,\rho)$ is characterized by a consecutive partition $\{M_1,M_2,\ldots,M_n\}$ of $M$ and distinct $r_1,r_2,\ldots,r_n\in \Report$ such that, for all $i=1,2,\ldots,n$,
\begin{enumerate}[(i)]
  \item $r_i=\sigma(m)$ if $m\in M_i$,
  \item\label{Theorem: N step equilibrium:ii} $\rho$ satisfies $\rho(r_i)=\argmax\limits_{a} U_{M_i,h}(a)$, and
  \item $r_i \in \argmax_{r} U_{m^T,h}(\rho(r),b)$ for all $m^T\in M_i$.
\end{enumerate}
\end{proposition}

All proofs in this section follow immediately from the corresponding results in \cite{bauch2024strategiccommunicationnarratives} and are omitted. 

We now turn to the comparative statics with respect to the conflict of interest. Information revelation gets coarser the higher the conflict of interest.

\begin{proposition}[Comparative statics]\label{pro:comp_stat}
The maximum step size $N(h,b)$ is non-increasing in $b$.
\end{proposition}

\subsection{Hypotheses}\label{Subsection:Hypotheses}

We next derive our hypotheses. We henceforth focus on $n$-step equilibria $(\sigma,\rho)$, identified by their partition of $M$ (see Proposition \ref{Proposition: N step equilibrium}), with the maximum number of steps $n=N(h,b)$ for any pair of history and bias $(h,b)$. These are focal points and offer the largest scope for informative communication. Let $\Gamma(h,b)$ denote the set of these equilibria.

The theory so far has remained silent about which concrete messages we would expect the sender to send, as these were abstract objects which have meaning only in equilibrium. In order to derive hypotheses on how the receiver's response to the sender's narrative changes with the conflict of interest, we have to narrow down the sender's choice. Consider an $n$-step equilibrium characterized by the consecutive partition $\{M_1,M_2,\ldots,M_n\}$. If $m^T \in M_k$, we would expect the sender to present a narrative from $M_k$, i.e., report some $r_k\in M_k$. Specifically, we assume that the sender chooses the narrative which would, from an ex-ante perspective, maximize her worst-case expected utility conditional on $m^T \in M_k$ if the receiver, instead of following equilibrium behavior, took her message at face value.

\begin{assumption}[Sender narrative]\label{ass1}
If $\sigma$ satisfies $\tilde{M} = \sigma^{-1}(r)$, then
\begin{equation*}
    r \in \argmax_{m\in \tilde M} \min_{m^T\in \tilde M}U_{m^T,h}(a(m,h,0),b).
\end{equation*}
\end{assumption}

Note first that Assumption \ref{ass1} has no impact on the theoretical results in Section \ref{Subsection:Analysis}, as it only changes the labels, but does not alter equilibrium outcome. Second, the sender's rule to determine her narrative is consistent with how the receiver resolves uncertainty. Third, the $\argmax$ set is indeed a singleton for all instances considered in the experiment, so that the sent report is unique.
\medskip

Our first and main hypothesis concerns the receiver's behavior. We are interested in how the receiver's response to the sender's narrative changes with the conflict of interest. In the experiment, we have shown the sender's ``action recommendation'' $a(\sigma(m),h,0)$, corresponding to her narrative $\sigma(m)$ under model $m$, to both sender and receiver. For any given $m, h,b$, let
\begin{align}\label{Delta^R_m}
\Delta^R_m(h,b):=\frac{1}{\lvert \Gamma(h,b)\rvert}\sum_{(\sigma,\rho)\in \Gamma(h,b)} a(\sigma(m),h,0)-\rho(\sigma(m))
\end{align}
denote the difference between the action recommendation and the receiver's equilibrium action depending on the true model $m$, averaged across equilibria. It captures how much and in which direction the equilibrium action differs from the recommendation on average. Consider the average of \eqref{Delta^R_m} over models and over models and histories,
\begin{align*}
\Delta^R(h,b):=\frac{1}{|M|}\sum_{m\in M}\Delta^R_m(h,b) \text{ and }\Delta^R(b):=\frac{1}{|H|}\sum_{h\in H}\Delta^R(h,b),
\end{align*}
respectively.

\begin{proposition}\label{pro:H2}
Suppose that $K=3$, $L=10$, $b\in \{0,1,3\}$, and $f_0=\mathcal U(\Theta)$. 
The downward adjustment $\Delta^R(h,b)$ is strictly increasing in $b$ for all $h\in H$. 
Consequently, the same is true of $\Delta^R(b)$.
\end{proposition}

Note that in the setting of the experiment, equilibria are uninformative if the bias is two or larger. Consequently, the downward adjustment is necessarily identical across these biases. The proofs of this and the subsequent results are relegated to Appendix \ref{app_proofs}. %
Proposition \ref{pro:H2} gives rise to our first hypothesis:

\begin{hypothesis}[Strategic Receiver]
\label{hyp:receiver}
The difference between the receiver’s optimal action induced by the sender's message at face value and the observed receiver action is increasing in the bias.
\end{hypothesis}

Our second hypothesis concerns the sender's behavior. We are interested in how the sender's narrative changes with the conflict of interest. For any given $m, h,b$, let
\begin{align}\label{Delta^S_m}
\Delta^S_m(h,b):=\frac{1}{\lvert \Gamma(h,b)\rvert}\sum_{(\sigma,\rho)\in \Gamma(h,b)} a(\sigma(m),h,0)-a(m,h,0)
\end{align}
denote the difference between the action recommendation and the receiver's optimal action under the true model depending on the true model $m$, averaged across equilibria. Consider the average of \eqref{Delta^S_m} over models and over models and histories,
\begin{align*}
\Delta^S(h,b):=\frac{1}{|M|}\sum_{m\in M}\Delta^S_m(h,b) \text{ and }\Delta^S(b):=\frac{1}{|H|}\sum_{h\in H}\Delta^S(h,b),
\end{align*}
respectively.

\begin{proposition}\label{pro:H1}
Suppose that $K=3$, $L=10$, $b\in \{0,1,3\}$, and $f_0=\mathcal U(\Theta)$. 
The upward shift $\Delta^S(h,b)$ is strictly increasing in $b$ for all $h\in H$. 
Consequently, the same is true of $\Delta^S(b)$.
\end{proposition}

Proposition \ref{pro:H1} yields our second hypothesis:
\begin{hypothesis}[Strategic Sender]
\label{hyp:sender}
The difference between the receiver's optimal action induced by the sender’s message at face value and the receiver’s optimal action under the true data-generating process is increasing in the bias.
\end{hypothesis}

The two main hypotheses assert that senders distort narratives upward and receivers adjust actions downward. Our third hypothesis asks whether, on net, senders are persuasive, i.e., if they succeed to systematically move receiver actions away from their optimal action under the true model. Formally, this is captured by the difference $\rho(\sigma(m))-a(m,h,0)$ of the equilibrium action and the receiver's bliss point. Averaging over models and histories, we consider $\Delta^S(b) - \Delta^R(b)$. 

\begin{proposition}\label{pro:H3}
Suppose that $K=3$, $L=10$, $b\in \{0,1,3\}$, and $f_0=\mathcal U(\Theta)$. The difference $\Delta^S(b) - \Delta^R(b)$ is increasing in $b$ from $b=0$ to $b=1$ and decreasing from $b=1$ to $b=3$.
\end{proposition}

Intuitively, the sender in our model is not able to systematically move the receiver away from the action induced by the true model under either fully informative ($b=0$) or uninformative ($b\geq 2$) communication. With partially informative communication ($b=1$), however, she is able to do so. We thus obtain:\footnote{Observe that Hypothesis \ref{hyp:persuasiveness} differs slightly from the one we had pre-registered because we have opted for a simpler, non-behavioral theoretical model. The pre-registered hypothesis obtains once we introduce behavioral receivers who take the sender's message at face value.}

\begin{hypothesis}[Persuasiveness]
\label{hyp:persuasiveness}
The difference between the observed receiver action and the receiver’s optimal action under the true data-generating process is first increasing and then decreasing in the bias.
\end{hypothesis}

\section{Experimental Procedures}\label{Section:Procedures}

Experimental sessions were conducted at the WISO Experimental Lab at the University of Hamburg. A total of 238 participants took part in the experiment, earning €14.85 on average, including a €6 show-up fee. Participants were randomly assigned the roles of senders and receivers, which they retained throughout the session. They were re-matched as `perfect strangers', playing one round with each partner. Each session was assigned to one of three treatments, \emph{no bias}, \emph{low bias}, and \emph{high bias}, corresponding to bias levels $b \in \{0,1,3\}$. The experiment was conducted in German and implemented in oTree \citep{chen2016otree}.\footnote{We pre-registered the experiment at \href{https://aspredicted.org/qu6gd7.pdf}{https://aspredicted.org/qu6gd7.pdf}.} %
An overview of the experiment's treatments, payoffs, and subjects is summarized in Table \ref{tab:treatments}.

\begin{table}[ht]
\centering
\caption{Data overview}
\label{tab:treatments}
\resizebox{\textwidth}{!}{
\begin{tabular}{r|rr|rrrr|rrr}
\toprule
 & \multicolumn{2}{c|}{Failed quiz} & \multicolumn{4}{c|}{Mean payoff} & \multicolumn{3}{c}{Total} \\
Bias & Once & Twice & SVO & Urn Interaction & Quiz & Total & Participants & Matches & Sessions \\
\midrule
0 & 29 & 13 & 1.49 & 7.68 & 0.73 & 15.91 & 48 & 290 & 2 \\
1 & 62 & 23 & 1.54 & 7.35 & 0.74 & 15.65 & 90 & 421 & 6 \\
3 & 73 & 30 & 1.53 & 5.37 & 0.7 & 13.61 & 100 & 538 & 5 \\
\bottomrule
\end{tabular}
}
\end{table}

Each participant completed two parts of the experiment, referred to as the \emph{Slider Task} and the \emph{Urn Interaction}. Before each task, participants were provided with the instructions for that part and could access them during the task. The Slider Task implements the Social Value Orientation (SVO) elicitation method for social preferences \citep{murphy2011measuring} with 6 sliders.\footnote{We use Max R.\ P.\ Grossmann’s sliders without anchoring and with feedback in the version from June 8, 2024, see \href{https://gitlab.com/gr0ssmann/otree_slider/-/tree/5b649865a7b31529512576d0ddf446a27adcbaaa/}{https://gitlab.com/gr0ssmann/otree\textunderscore%
slider/-/tree/5b649865a7b31529512576d0ddf446a27adcbaaa/}.} Payment was based on two randomly selected sliders per participant: one as an active participant and one as a passive participant. 

The Urn Interaction implements the theoretical model with $K=3$, $L=10$, and $f_0=\mathcal U(\Theta)$. Participants were presented with two urns, A and B. Urn~A always contained exactly five red and five black balls, and this composition was known to both players. Urn~B contained an unknown number of red balls $\theta$, representing the state of nature, which was drawn uniformly and independently from $\Theta=\{0,1,\ldots,10\}$ in each round.

\begin{figure}[ht]
    \centering
    \includegraphics[width=0.47\linewidth]{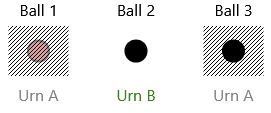}

    \vspace{0.3em}
    \small\emph{(a) What the sender observes}
    
    \vspace{1em}

    \includegraphics[width=0.5\linewidth]{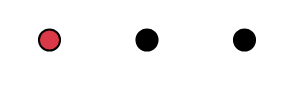}
    
    \vspace{0.3em}
    \small\emph{(b) What the receiver observes}

      \vspace{0.3em}
    
    \caption{Information available to the sender and the receiver prior to communication (translated).}
    \label{fig:balls_experiment}
\end{figure}

In each round, three balls were drawn with replacement. Each draw came from either Urn~A or Urn~B, according to an unknown sequence. These ambiguous draws were generated 
following the method of \cite{stecher2011generating} on the finite set $M$. The sender observed both the color of each ball and the urn from which it was drawn, while the receiver observed only the colors, not the urn identities, see Figure \ref{fig:balls_experiment}.

After observing the draws, the sender sends a message to the receiver. In the message, she reports, for each draw, from which urn the ball (allegedly) originated. This screen is shown below in Figure \ref{fig:sender_screen}. The receiver then submits an action in $\{0,1,\ldots,10\}$, estimating the number of red balls in Urn~B. The sender's payoff was determined by the distance of the receiver's action from the sender's `favorite number' $\theta + b$. The payoff scale was:\ 1300 points if the receiver's action exactly matched $\theta + b$, 1000 points if it was one ball away, 700 points if two balls away, 400 points if three balls away, 100 points if four balls away, and 0 points if five or more balls away. The receiver's payoff was determined by the distance between the receiver's action and the true state $\theta$, using the same payoff scale. A 100 points corresponded to €1.

\begin{figure}[ht]
    \centering
    \includegraphics[width=0.75\textwidth]{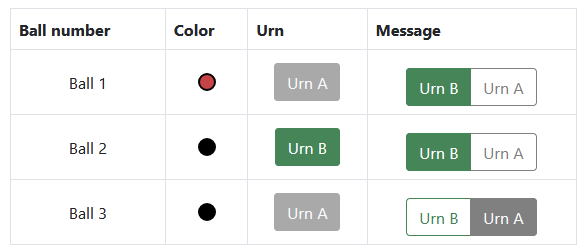}
    \caption{Sender's message screen (translated).}
    \label{fig:sender_screen}
\end{figure}

Before the main phase of the Urn Interaction, all participants completed an unincentivized practice round. After the practice round, the urns from which the balls had actually been drawn and the true number of red balls in Urn~B were revealed. In the incentivized part, participants received no feedback between rounds. Each participant also answered a set of comprehension questions before the Urn Interaction, earning a €1 bonus if all answers were correct within two attempts. Their answers were corrected for them after the second attempt.\footnote{The results do not qualitatively change if participants who failed the quiz twice are dropped. Specifically, all significant rows in Tables 3 and 6 remain significant at least at the 5\% level.} One round was randomly selected to determine the final payoff for each participant.

During the Urn Interaction, participants were provided with a calculation aid on the screen. This aid allowed them to select any assignment of the balls to the two urns and then displayed the payoff-maximizing action for the receiver, assuming the balls were drawn according to the selected assignment (Figure \ref{fig:calculation_aid}). Receivers could also see the action that would maximize their payoff based on the sender’s message, conditional on the message being truthful. The purpose of this aid was to reduce computational complexity and arithmetic errors. Participants were free to ignore it.

\begin{figure}[ht]
    \centering
    \begin{infobox}
    Suppose that the balls were drawn from the urns you selected in the right-hand column above this message. Then estimating \redc{$X = 5$} red balls in Urn~B maximizes your expected payoff.
    \end{infobox}
    \caption{Calculation aid (translated).}
    \label{fig:calculation_aid}
\end{figure}

The complete instructions (in German) can be found in Appendix \ref{app:instructions}.

\section{Experimental Results}\label{Section:Results}

Before testing the two hypotheses directly, we first provide descriptive evidence on truthful narratives and receiver trust. Figure~\ref{fig:truthfullness} shows the share of rounds in which senders report the true model and the share of rounds in which receivers choose the action implied by the sender’s message, fully trusting the sender’s recommendation. Both measures decline monotonically with the level of the bias, providing initial
evidence that subjects respond to the sender's strategic incentives.

\input{results_artifacts/Figure_4_Truthful_outcomes_by_bias}

We now turn to the main results. 
The results are organized around the two hypotheses. Section \ref{subsec:receiver_results} discusses receivers and tests
whether their response to the sender's narrative (and thus action recommendation) changes with the sender's bias (Hypothesis~\ref{hyp:receiver}). Section \ref{subsec:sender_results} studies senders and tests whether their 
narrative changes with the bias (Hypothesis~\ref{hyp:sender}). We then ask whether these two effects translate into overall persuasion and welfare effects, and whether the results can be explained by alternative behavioral channels.

\subsection{Receiver behavior}
\label{subsec:receiver_results}
We first examine whether receiver subjects increasingly choose actions below those implied by the sender's narrative as the sender's bias increases, consistent with accounting for the sender’s strategic incentives. %
To this end, we denote by $\mathcal{D}^{\lab}_{m}(h,b)$ the collection of observed pairs of the model $r$ sent by sender subjects and action $\tilde{a}$ taken by receiver subjects for treatment bias $b$, history $h$ and true underlying model $m$ within the experiment. 
Define
\begin{equation*}
    n_m(h,b)=\bigl|\mathcal D_m^{\lab}(h,b)\bigr|,\qquad n(h,b)=\sum_{m\in M}n_m(h,b), \qquad n(b)=\sum_{h\in H}n(h,b).
\end{equation*}

We then define the empirical version of the receivers' downward adjustment $\Delta_m^R(h,b)$, $\Delta^R(h,b)$, and $\Delta^R(b)$ (see \eqref{Delta^R_m}) as
\begin{align}
	\tilde{\Delta}_m^R(h,b)
	&=
	\frac{1}{n_m(h,b)}
	\sum_{(r,\tilde a)\in\mathcal D_m^{\lab}(h,b)}
	a(r,h,0)-\tilde a,\notag
	\\
	\tilde{\Delta}^R(h,b)
	&=
	\frac{1}{n(h,b)}
	\sum_{m\in M}
	\sum_{(r,\tilde a)\in\mathcal D_m^{\lab}(h,b)}
	a(r,h,0)-\tilde a,\notag
	\\
	\tilde{\Delta}^R(b)
	&=
	\frac{1}{n(b)}
	\sum_{h\in H}\sum_{m\in M}
	\sum_{(r,\tilde a)\in\mathcal D_m^{\lab}(h,b)}
	a(r,h,0)-\tilde a.	\label{Delta^R_b empirical}
\end{align}
Comparing the empirical version to the theoretical one, we thus measure the (equilibrium) action $\rho(\sigma(m))$ by the actual action $\tilde{a}$ of the receiver and action recommendation $\sigma(m)$ sent under Assumption \ref{ass1} as the message $r$ actually sent by the sender. %
Note that we also make sure that every experimental observation receives equal weight, since
\begin{align*}
	\tilde{\Delta}^R(h,b)
	&=
	\sum_{\substack{m\in M\\n_m(h,b)>0}}
	\frac{n_m(h,b)}{n(h,b)}
	\tilde{\Delta}_m^R(h,b),
	\quad
	\tilde{\Delta}^R(b)=
	\sum_{\substack{h\in H\\n(h,b)>0}}
	\frac{n(h,b)}{n(b)}
	\tilde{\Delta}^R(h,b).
\end{align*}
Unlike the theoretical aggregates, which weight models and histories
uniformly, the empirical aggregates therefore weight them according to their frequencies in the experimental sample. None of the results qualitatively change if we use uniform weights for empirical analysis.

\input{results_artifacts/Figure_5_Empirical_distribution_of_Delta_R}

To account for heterogeneity of subjects, we sometimes instead aggregate observations at subject level and treat each subject’s mean difference $\tilde{\Delta}_m^R(h,b)$ as a single observation. Figure~\ref{fig:stacked_KDE_rec} shows the distribution of $\tilde{\Delta}^R$ by treatments $b$, both using subject-level means and using individual rounds pooled across subjects. The differences between treatment means are stronger when considering individual rounds as observations, but they are also present for the subject-level means. A substantial mass at zero corresponds to receivers who ignore strategic considerations and fully trust the sender’s message. As the bias increases, the distributions shift to the right, most noticeably for the highest bias level. 

\input{results_artifacts/Figure_6_Mean_Delta_R_subject_level}

\input{results_artifacts/Table_2_Delta_R_difference_between_the_senders_recommendation_and_the_receivers_action}

\input{results_artifacts/Table_3_Pairwise_tests_for_Hypothesis_receiver}

Figure~\ref{fig:hyp2_chart} and Table~\ref{tab:receivers} report the average of subject-level means of $\tilde{\Delta}^R$ by treatment, together with 95\% confidence intervals. The mean in the \emph{no bias} treatment is close to zero and its confidence interval includes zero. The means for the other two treatments are higher, consistent with Hypothesis~\ref{hyp:receiver}. Pairwise tests in Table~\ref{tab:receivers-tests} show that the increase from \emph{no bias} to \emph{high bias}, and from \emph{low bias} to \emph{high bias}, is statistically significant under both parametric t-tests and Wilcoxon--Mann--Whitney $U$ tests. The increase from no bias to low bias is positive
but not statistically significant. This yields our first main
experimental result.

\begin{result}
Receivers adjust the sender's recommendation downward when bias is large. The \emph{high bias} treatment differs significantly from both the \emph{no bias} and the \emph{low bias} treatment, while the difference between \emph{no bias} and \emph{low bias} is not significant.
\end{result}

Figure \ref{fig:trust_R} provides complementary evidence using a simpler, binary measure of receiver's trust in the narrative. Across receiver subjects, it shows the distribution of the share of rounds in which the receiver exactly follows the sender's recommendation. The levels predicted by theory are indicated by dashed vertical lines. As expected, this measure also declines with bias. In the \emph{low bias} treatment, some receivers follow the sender in every
round, whereas in the \emph{high bias} treatment, no receiver does so, and roughly one quarter of receivers never follow the recommendation exactly.

\input{results_artifacts/Figure_7_Percent_of_trusting_decisions_within_receiver_subjects}

We now estimate the effect of a bias on $\tilde{\Delta}^R(h,b)$ separately by history $h$. We run a linear regression with factors for histories and biases, including their interaction, and cluster standard errors at the subject level. The results are reported in Table~\ref{tab:by_history_receivers}. For each history, the first column reports the baseline level when bias is zero, while the remaining columns report the change in the fitted value when bias increases from $0$ to $1$ and from $1$ to $3$. Standard errors and significance stars for these contrasts are computed using the clustered variance-covariance matrix.

\input{results_artifacts/Table_4_Effect_of_increasing_bias_on_Delta_R_by_history}

Whenever the estimated effect of the bias is statistically detectable, it increases $\tilde{\Delta}^R(h,b)$. The ``Bias $0$'' column also shows how the subjects avoid extreme actions and tend to adjust their actions toward the middle in the absence of a bias. A more detailed graphical comparison of $\tilde{\Delta}^R(h,b)$ across treatments for different histories and models can be found in Figure \ref{fig:hyp2_whiskers_combined} in Appendix \ref{app:additional_figures}.

\subsection{Sender behavior}
\label{subsec:sender_results}

We next 
examine whether sender subjects increasingly choose narratives that implicitly recommend actions above those implied by the true model as the sender's bias increases. %
Using the data sets $\mathcal{D}^{\lab}_m(h,b)$ and observation counts defined in the receiver analysis, we define the empirical counterparts of $\Delta_m^S(h,b)$, $\Delta^S(h,b)$, and $\Delta^S(b)$ (see \eqref{Delta^S_m}) as the average difference of the recommended action by the sender minus the receiver-optimal action given the true underlying model $m$ for each treatment bias $b$ and history $h$:
\begin{align}
    \tilde{\Delta}_m^S(h,b)
    &=
    \frac{1}{n_m(h,b)}
    \sum_{(r,\tilde a)\in\mathcal{D}_m^{\lab}(h,b)}
    a(r,h,0)-a(m,h,0),\notag
    \\
    \tilde{\Delta}^S(h,b)
    &=
    \frac{1}{n(h,b)}
    \sum_{m\in M}
    \sum_{(r,\tilde a)\in\mathcal{D}_m^{\lab}(h,b)}
    a(r,h,0)-a(m,h,0),\notag
    \\
    \tilde{\Delta}^S(b)
    &=
    \frac{1}{n(b)}
    \sum_{h\in H}\sum_{m\in M}
    \sum_{(r,\tilde a)\in\mathcal{D}_m^{\lab}(h,b)}
    a(r,h,0)-a(m,h,0).
    \label{Delta^S_b empirical}
\end{align}
As before, we analyze both, single observations, as well as the aggregates at a subject level with the appropriate identification.

\input{results_artifacts/Figure_8_Empirical_distribution_of_Delta_S}

Figure~\ref{fig:stacked_KDE_sen} plots the distribution of $\tilde{\Delta}^S$ by treatment. Similarly to many cheap-talk experiments, a substantial mass at zero indicates truthful sender behavior. As the bias increases, the distributions shift visibly to the right, therefore indicating systematic upward distortion in senders’ recommendations.

Following the receiver analysis, we again use subject-level means as the unit of observation. Figure~\ref{fig:hyp1_means_chart} and Table~\ref{tab:senders} report the average $\tilde{\Delta}^S$ by treatment, together with 95\% bootstrap confidence intervals. The average in the \emph{no bias} treatment is close to zero and is significantly lower than in the positive-bias treatments, consistent with Hypothesis~\ref{hyp:sender}. Pairwise tests in Table~\ref{tab:senders-tests} confirm this pattern using both Welch $t$-tests and Wilcoxon--Mann--Whitney $U$ tests. This pattern differs from receivers' behavior, for whom the behavioral response appears only once the bias is sufficiently large. Senders already adjust their behavior under low bias, which suggests that senders are more sensitive to whether bias is present at all than to its magnitude. This yields our second main result.

\input{results_artifacts/Figure_9_Mean_Delta_S_subject_level}

\input{results_artifacts/Table_5_Delta_S_difference_between_the_senders_recommendation_and_receivers_optimal_action_for_the_true_model}

\input{results_artifacts/Table_6_Pairwise_tests_for_Hypothesis_sender}

\begin{result}
Senders distort their narratives upward when a bias is present. Both the \emph{low bias} and the \emph{high bias} treatment differ significantly from the \emph{no bias} treatment, while the difference between \emph{low bias} and \emph{high bias} is not significant.
\end{result}

\input{results_artifacts/Figure_10_Percent_of_truthful_decisions_within_sender_subjects}

As expected from the existing experimental literature on cheap talk \citep{cai2006overcommunication,sanchez2007experimental}, there is significant over-communication once the conflict of interest is large despite incentives not to reveal all information. We measure over- or under-communication by the correlation between the action induced by the true model, $a(m^T,h,0)$, and the action induced by the reported model, $a(r,h,0)$. %
Figure~\ref{fig:report_informativeness_history} shows this statistic separately by history.
In \emph{no bias}, communication is fully revealing in equilibrium and the correlation is predicted to be $1$ for every history. Even though $\tilde \Delta^S(h,b)$ is around zero, communication is noisy and observed correlation is less than $1$. In \emph{low bias} the observed correlations are also below the theoretical benchmarks. 
Finally, in \emph{high bias}, equilibrium is pooling, so correlation is predicted to be zero, but the observed correlation is positive, with a confidence interval bounded away from zero for two histories out of four. 
Thus, informativeness decreases with bias as expected, but senders reveal less than theory predicts in \emph{no bias} and \emph{low bias}, and more in \emph{high bias}. This tendency to still reveal information (or the true model) when conflict of interest is high is part of the reason why $\tilde{\Delta}^S$ responds less strongly to bias than predicted by theory.

\input{results_artifacts/Figure_Informativeness_of_the_senders_report_by_history}

\cite{sanchez2007experimental}'s results suggest that subjects can be divided into two types: %
strategic subjects and subjects who are averse to lying. The same effect is visible in Figure \ref{fig:truth_S}, which shows that almost a fifth of senders always reported truthfully in the \emph{low bias} treatment, and still about 5\% of senders did so in the \emph{high bias} treatment, despite incentives to misreport. The levels predicted by theory are indicated by dashed vertical lines.

Together, Figures~\ref{fig:trust_R} and \ref{fig:truth_S} suggest an asymmetry between the two sides of the interaction. Some senders appear persistently truthful even when bias creates incentives to distort the message. By contrast, receiver behavior shows much less evidence of a corresponding fully trusting type: in the \emph{high bias} treatment, none of the receiver subjects fully trust the sender in every round. Thus, the mixed-type heterogeneity that is visible among senders is not mirrored as strongly among receivers.

\input{results_artifacts/Table_7_Effect_of_increasing_bias_on_Delta_S_by_history}

We also estimate the effect of a bias on $\tilde{\Delta}^S(h,b)$ separately by history. The results are reported in Table~\ref{tab:by_history_senders}. %
Higher bias generally increases $\tilde{\Delta}^S(h,b)$, as expected. Partly because power is lower when individual histories are considered, the effect of gradual bias increases appears non-uniform across histories. The overall effect of a bias increase from $0$ to $3$ is strongest and statistically significant for the two histories with most black balls. For these histories, the theoretical $\Delta^S(h,b)$ also predicts the highest shift. The ``Bias $0$'' column shows that $\Delta^S$ is not significantly different from zero in the \emph{no bias} treatment. Indeed, while receivers (Table~\ref{tab:by_history_receivers}) tend to avoid extreme actions even in \emph{no bias} treatment, there is no obvious reason for senders to systematically distort their narratives in absence of a bias. A more detailed graphical comparison of $\Delta^S$ across treatments for different histories and models can be found in Figure \ref{fig:hyp1_whiskers_combined} in Appendix \ref{app:additional_figures}.

\subsection{Persuasiveness}

The two main results describe shifts in opposite directions in the presence of bias:\ senders distort narratives upward, while receivers adjust actions downward. We therefore ask whether, on net and on average, senders systematically move receiver actions away from the action that would be optimal under the true model. To this end, we define the empirical version of the difference between the receiver's action and the action induced by the true model, averaged across histories, models and equilibria, $\Delta^S(b)-\Delta^R(b)$, as
$$\tilde{\Delta}^S(b)- \tilde{\Delta}^R(b),$$
where $\tilde{\Delta}^S(b)$ and $\tilde{\Delta}^R(b)$ are given by  \eqref{Delta^R_b empirical} and \eqref{Delta^S_b empirical}, respectively. This value is a measure of the total persuasiveness of the sender: it measures how far, on average, the sender is able to move the receiver away from the action induced by the true model under a given bias $b$. For the classical cheap-talk setting, \cite{groseclose2021humans} suggests that ``the final policy that receivers choose, in expectation, should equal the policy that they would have chosen had the senders sent no signal''. In other words, senders should not be able to systematically persuade the receivers and shift the expected policy in their favor.\footnote{In classical cheap talk, the average action without communication and the average action under the true model coincide. In narrative persuasion they are distinct and we use the latter.} However, he also points out that evidence from the experiment by \cite{cai2006overcommunication} suggests that in practice senders are capable of doing so.

In our setting, the theoretical value $\Delta^S-\Delta^R$ is non-monotone in the bias, which is our Hypothesis \ref{hyp:persuasiveness}. We find evidence of this in the experiment as well: although none of the treatment levels of $\tilde{\Delta}^S-\tilde{\Delta}^R$ differ significantly from zero at 95\% confidence, pairwise comparisons show that the value indeed increases when the bias increases to 1 and then decreases for bias 3 (Figure \ref{fig:persuasiveness_subject_level}). This is confirmed by one-sided tests and close to 95\% confidence for two-sided tests in Table \ref{tab:persuasiveness_order_tests}. A disaggregated value of $\tilde{\Delta}^S-\tilde{\Delta}^R$ by history also closely follows the theoretical prediction (Figure \ref{fig:hyp3_whiskers_h} in Appendix \ref{app:B1}).
 
Thus, although senders distort their messages, receivers largely
offset this distortion, except for the \emph{low bias} treatment as expected from theory.

\begin{result}
Senders are persuasive only when bias is low.
\end{result}

\input{results_artifacts/Figure_11_Mean_observed_persuasiveness_receiver_subject_level}

\input{results_artifacts/Table_8_Pairwise_tests_of_increase_then_decrease_persuasiveness}

\subsection{Comparison with MLEU}
\label{sec:comparison_w_MLEU}

As an alternative model to max-min expected utility, subjects could instead evaluate the sender's narrative by its likelihood of generating the observed data \citep{schwartzstein2021using}; indeed, there is empirical evidence that the likelihood of a narrative may be a key determinant of persuasiveness \citep{barron2024narrative}. Formally, we can replace max-min expected utility \eqref{max_min} by \emph{maximum likelihood expected utility (MLEU)},
\begin{equation*}
U_{\tilde{M}, h}(a)= \E[u_R(a, \theta) \mid \tilde{m},h],
\end{equation*}
where $\tilde{m}\in\argmax_{m\in \tilde M} \Pr(h|m,f_0)=\sum_{\theta\in \Theta} \pi_m(h|\theta)f_0(\theta)$. Under MLEU, the receiver maximizes his expected utility w.r.t.\ a narrative from the minimal feasible set $\tilde M$ that is most likely to explain the observed data $h$.\footnote{In case of ties, any strict ordering on the set of models can serve as a tiebreaker. The subsequent conclusions are robust to changes in the ordering of models.} The model otherwise is as described in Section \ref{Section:theory}.\footnote{The subsequent conclusions are robust to additionally replacing Assumption \ref{ass1} by one based on maximum likelihood.} 

Theoretically, we find that the MLEU prediction for \(\Delta^R\) does not follow Hypothesis \ref{hyp:receiver}. Then $\Delta^R(h,b)$ even strictly decreases in $b\in\{0,1,3\}$ if $h$ is such that $h^\Sigma=2$  (see also Figure \ref{fig:MLEU_R_whiskers} in Appendix \ref{MEU_vs_MLEU}). In this case, the receiver's adjustment is upward instead of downward when bias is low, $b=1$, as the model $m$ inducing the highest receiver action is such that $k\in m$ if and only if $h_k=1$, and thus also has the highest likelihood. 

Moreover, the MLEU model also does not explain receiver behavior well in our experiment. First, narrative likelihood is not positively associated with receiver trust. The correlations between likelihood and receiver trust reported in Table \ref{tab:likelihood_receiver_correlations} point mostly in the opposite direction: narratives with higher likelihood are not more likely to be followed exactly, nor are they associated with smaller receiver adjustments (lower $|\tilde{\Delta}^R|$). This confirms that likelihood is a poor predictor of the perceived credibility of the message.

\input{results_artifacts/Table_9_Correlations_between_message_likelihood_and_receiver_trust_measures}

Second, MLEU predicts receiver actions that are too extreme. This effect is clearest in the \emph{high bias} treatment. When the receiver discards the sender's message as uninformative, i.e., in the pooling equilibrium, the maximum-likelihood model often implies actions close to the boundaries of the action space, but subjects rarely choose such extreme actions (see also Figure \ref{fig:MLEU_policy_R} in Appendix \ref{MEU_vs_MLEU}).

Finally, we use the distance between the observed $\tilde{\Delta}^R$ and the theoretical prediction $\Delta^R$ as a metric to compare the two models. %
Across 1,249 observations from 119 receiver subjects, MEU provides a better fit under this metric. At the subject level, the mean absolute error is 1.55 under MEU and 2.24 under MLEU. MLEU is closer to the observed $\tilde{\Delta}^R(h,b)$ than MEU in only about 22.6\% of matched observations. By treatment, MEU and MLEU perform almost identically at \emph{low bias} (1.23 vs. 1.30), and MEU fits better for \emph{high bias} (2.11 vs. 3.68). A detailed comparison of the two models for $\Delta^R$ and $\Delta^S$ is shown in Figures \ref{fig:MLEU_R_whiskers_combined} and \ref{fig:MLEU_S_h_whiskers} in Appendix \ref{MEU_vs_MLEU}.

\subsection{Welfare}

We next consider the welfare consequences of strategic narrative communication. While this is not part of the core analysis, it clarifies how the sender's distortion and the receiver's adjustment affect realized payoffs.

Our first observation is that senders bear most of the welfare loss from high bias. Figure \ref{fig:welfare_triptych} shows the average payoffs for different bias levels and two counterfactuals. The red circles show the welfare based on the actual payoffs received by the subjects in the experiment. The \emph{low bias} treatment does not differ significantly from \emph{no bias} in terms of welfare. However, the \emph{high bias} treatment shows that stronger incentives to distort the narrative create communication problems large enough to reduce payoffs, especially for the senders. Because increasing the bias mechanically changes the attainable payoffs, comparisons of absolute payoffs across treatments do not isolate the losses caused by conflict of interest. For reference, green triangles show the average payoffs that would result if receivers knew the true model (but not the state) and chose the optimal action according to expected utility given the true model. We therefore compare the payoffs to two benchmarks.

First, blue squares show the average payoffs that would result if receivers had always followed the senders' recommendations. In the \emph{high bias} treatment, receivers are able to raise their own average payoff by 51.5 points relative to this counterfactual, but at the same time they reduce senders' average payoff by 127 points. Consequently, aggregate welfare is 75.5 points lower than in the counterfactual. Thus, receivers partly protect themselves from the senders’ distortion, shifting the welfare burden to senders. The second benchmark compares observed behavior with the MEU pooling action, which receivers would choose if they ignored the sender’s narrative altogether. In the \emph{high bias} treatment, this pooling action would raise the average payoff by 46.8 points for the receivers and by 92.8 for the senders. Thus both parties experience a loss under this benchmark, but approximately two-thirds of it comes from lower sender payoffs.

\input{results_artifacts/Figure_12_Welfare_and_its_components}

The second observation is that for high bias the welfare falls below the level of the uninformative equilibrium. This is shown in Table \ref{tab:welfare_relative}. It is important to note, however, that the upper bound on attainable welfare itself decreases with bias as incentives become misaligned. The row ``Reference Level'' shows this hypothetical maximum, which is the aggregate welfare if the receiver chose the action that exactly equals the true state. The rest of Table \ref{tab:welfare_relative} shows the average aggregate welfare for the actual decisions in the experiment and for different hypothetical situations relative to this reference level. %
While aggregate welfare in the experiment is slightly higher than in the counterfactual in which the receiver ignores the sender's recommendation (under MEU) in the \emph{no bias} treatment, it is considerably lower than in the counterfactual in which the receiver ignores the sender's recommendation in the \emph{high bias} treatment.

The last row of the welfare table shows the counterfactual results under optimal actions for the MLEU model discussed in Section \ref{sec:comparison_w_MLEU}. It indicates that MLEU behavior without communication would generate low average payoffs at all bias levels, further suggesting that it is a poor decision rule in this particular strategic situation.

\input{results_artifacts/Table_10_Mean_pair_welfare}

\subsection{Other-regarding preferences (SVO)}

The first part of the experiment, the \emph{Slider Task}, allows us to control for social preferences of both senders and receivers. We use a coarser classification than \cite{murphy2011measuring}, combining ``Altruist'' and ``Prosocial'' types into one ``Prosocial'' type, and combining ``Individualist'' and ``Competitive'' types into one ``Proself'' type. With this approach, prosocial subjects comprise $56.3\%$ of the receivers and  $58.8\%$ of the senders. Social value orientation appears to matter more for senders than for receivers, which could be because the former face an apparent choice between telling the truth and distorting the narrative, see Figure \ref{fig:SVO_combined}. In the \emph{no bias} treatment, the variance of proself senders is several times higher than the variance of prosocial senders, who recommend actions at or close to truth. Proself senders in the \emph{low bias} treatment also tend to shift the narratives more than prosocial subjects (Figure \ref{fig:SVO_senders}). This effect is not present in the \emph{high bias} treatment. For receivers, proself subjects adjust downward slightly more than prosocial subjects, but these effects are not statistically significant (Figure \ref{fig:SVO_receivers}). Overall, social preferences do not appear to explain receiver behavior and are more strongly related to sender behavior. 

\input{results_artifacts/Figure_13_Delta_S_and_Delta_R_by_SVO_group}

\section{Conclusion}\label{Section: Conclusion}

We investigate to what extent people in complex and uncertain environments take a narrator's strategic incentives into account. 
Our theoretical model of cheap-talk communication of narratives under model uncertainty is based on \cite{bauch2024strategiccommunicationnarratives}. It predicts that as the sender's bias increases, her action recommendations will deviate more from the receiver's optimal action under the true model. In turn,  the receiver's actions will also deviate more from the sender's recommendations. 

In a laboratory experiment, we find evidence for both predictions. Moreover, we find that, while receivers under-react to low bias relative to senders, they over-react to high bias. 
Additional analyses show that the sender's narrative is, on average, slightly persuasive only when the bias is low.

Finally, a comparison of our MEU approach with an alternative MLEU approach inspired by the literature on narratives \citep{schwartzstein2021using,barron2024narrative} reveals that the MLEU approach neither yields our main hypothesis on receiver behavior nor does it explain receiver behavior well in our experiment. Our results thus complement contemporary work by \cite{barron2024narrative}, who conduct a laboratory experiment to test the model of \cite{schwartzstein2021using}. Using a probabilistic investor-advisor setup in which the conflict of interest is state-independent and (mostly) hidden from the receiver, they find that narratives are persuasive and that the likelihood of a narrative is a key determinant of persuasiveness. This suggests that the set-up, in particular the nature of and information on interest conflicts, has a crucial influence on receiver behavior.

\setlength{\bibsep}{0pt}
\bibliography{references}

\appendix

\newpage

\section{Proofs of Propositions \ref{pro:H2}, \ref{pro:H1} and \ref{pro:H3}}\label{app_proofs}

Consider, first, any $h\in H$ and $b=0$. By Proposition \ref{pro:comp_stat}, $(\sigma,\rho)\in\Gamma(h,b)$ is fully informative. In particular, $\tilde{M} = \sigma^{-1}(r)$ for some $r\in \tilde M$ only if $a(m,h,0)=a(m',h,0)$ for all $m,m'\in \tilde{M}$. Thus, $\rho(\sigma(m))=a(m,h,0)$ for all $m\in M$, which yields
 $\Delta^R(h,0)=0$.
 
Second, consider $b=1$. We frequently write a model $\{x,y\}\in M=2^{\{1, \dots, K\}}$ as $xy$ and proceed by case distinction with respect to the summary statistics $h^\Sigma:=\sum_{k=1}^Kh_k$ associated with history $h$:
\begin{enumerate}
  \item $h^\Sigma=0$. Then $N(h,1)=2$ and $|\Gamma(h,1)|=3$. Specifically, the three equilibria with maximum step size are characterized by 
      $$\big\{\{m\neq \emptyset\},\{\emptyset\}\big\},\ \big\{\{m: |m|\geq 2\},\{m: |m|< 2\}\big\},\text{ and } \big\{\{M\},\{m\neq M\}\big\},$$
      respectively. We obtain $\Delta^R_m(h,1)=1/3$ if $|m|\leq 2$ and $\Delta^R_m(h,1)=0$ if $m=M$, and thus $\Delta^R(h,1)=7/24$.
      
  \item $h^\Sigma=1$. Then $N(h,1)=3$ and $|\Gamma(h,1)|=8$. Specifically, suppose without loss that $h=(1,0,0)$, then the eight equilibria with maximum step size are characterized by
      \begin{align*}       
      &\big\{\{23\},\{2,3,M,\emptyset,12,13\},\{1\}\big\},\ \big\{\{23,2,3\},\{M,\emptyset,12,13\},\{1\}\big\},\\
      &\big\{\{23\},\{2,3,M,\emptyset,12\},\{13,1\}\big\},\ \big\{\{23\},\{2,3,M,\emptyset,13\},\{12,1\}\big\},\\
      &\big\{\{23\},\{2,3,M,\emptyset\},\{12,13,1\}\big\},\ \big\{\{23,2,3\},\{M,\emptyset,12\},\{13,1\}\big\},\\
      &\big\{\{23,2,3\},\{M,\emptyset,13\},\{12,1\}\big\},\ \text{ and } \big\{\{23,2,3\},\{M,\emptyset\},\{12,13,1\}\big\},
      \end{align*}
      respectively. We obtain $\Delta^R_m(h,1)=3/4$ if $m=1$, $\Delta^R_m(h,1)=1/2$ if $m\in \{12,13\}$ and $\Delta^R_m(h,1)=0$ otherwise, and thus $\Delta^R(h,1)=7/32$.      
      
  \item $h^\Sigma=2$. Then $N(h,1)=3$ and $|\Gamma(h,1)|=4$. Specifically, suppose without loss that $h=(1,1,0)$, then the four equilibria with maximum step size are characterized by
      \begin{align*}       
      &\big\{\{3\},\{\emptyset,13,23\},\{M,1,2,12\}\big\},\ \big\{\{3\},\{\emptyset,13\},\{23,M,1,2,12\}\big\},\\
      &\big\{\{3\},\{\emptyset,23\},\{13,M,1,2,12\}\big\},\text{ and } \big\{\{3\},\{\emptyset\},\{13,23,M,1,2,12\}\big\},
      \end{align*}
      respectively. We obtain $\Delta^R_m(h,1)=1$ if $m\in \{M,1,2,12\}$, $\Delta^R_m(h,1)=1/2$ if $m\in \{13,23\}$, and $\Delta^R_m(h,1)=0$ otherwise, and thus $\Delta^R(h,1)=5/8$.
  
  \item $h^\Sigma=3$. Then $N(h,1)=2$ and $|\Gamma(h,1)|=1$. Specifically, the unique equilibrium with maximum step size is characterized by $\big\{\{\emptyset\},\{m\neq \emptyset\}\big\}$.
We obtain $\Delta^R_m(h,1)=1$ if $m\neq \emptyset$ and $\Delta^R_m(h,1)=0$ if $m=\emptyset$, and thus $\Delta^R(h,1)=7/8$.
\end{enumerate}

Third, consider any $h\in H$ and $b=3$. It is easy to check that $N(h,b)=1$. If $h\neq (0,0,0)$, we obtain $\Delta^R_m(h,3)=3$ for all $m$, and thus $\Delta^R(h,3)=3$. If $h= (0,0,0)$, we obtain $\Delta^R_m(h,3)=1$ for all $m$, and thus $\Delta^R(h,3)=1$, which establishes the first claim. The second claim then follows immediately, which establishes Proposition \ref{pro:H2}.

\medskip

Next, we prove Proposition \ref{pro:H1}. Observe that 
\begin{align*}
    \Delta^S(h,b)&=\frac{1}{|M|}\sum_{m\in M}\left(\frac{1}{\lvert \Gamma(h,b)\rvert}\sum_{(\sigma,\rho)\in \Gamma(h,b)} a(\sigma(m),h,0)-a(m,h,0)\right)\\
    &= \Delta^R(h,b)+\frac{1}{|M|}\frac{1}{\lvert \Gamma(h,b)\rvert}\sum_{m\in M}\sum_{(\sigma,\rho)\in \Gamma(h,b)} \rho(\sigma(m))-a(m,h,0).
\end{align*}
We can thus derive the claim from the equilibria and the values for $\Delta^R(h,b)$ derived above. Consider any $h\in H$ and $b=0$. Since $\rho(\sigma(m))=a(m,h,0)$ for all $m\in M$, we obtain $\Delta^S(h,0)=\Delta^R(h,0)=0$.
 
We proceed by case distinction with respect to the summary statistics $h^\Sigma:=\sum_{k=1}^Kh_k$ associated with history $h$:
\begin{enumerate}
  \item $h^\Sigma=0$. We obtain $\Delta^S(h,1)=\Delta^R(h,1)+27/24=7/24+27/24=17/12<\Delta^S(h,3)=\Delta^R(h,3)+17/8=1+17/8=25/8$.
      
  \item $h^\Sigma=1$. We obtain $\Delta^S(h,1)=\Delta^R(h,1)+46/64=7/32+23/32=15/16<\Delta^S(h,3)=\Delta^R(h,3)+8/8=3+8/8=4$.      
     
  \item $h^\Sigma=2$. We obtain $\Delta^S(h,1)=\Delta^R(h,1)-4/32=5/8-1/8=1/2<\Delta^S(h,3)=\Delta^R(h,3)-8/8=3-8/8=2$.
  
  \item $h^\Sigma=3$. We obtain $\Delta^S(h,1)=\Delta^R(h,1)-4/8=7/8-4/8=3/8<\Delta^S(h,3)=\Delta^R(h,3)-17/8=3-17/8=7/8$,
\end{enumerate}
which establishes the first claim. The second claim then follows immediately, which establishes Proposition \ref{pro:H1}.

From the above calculations, we see that $\Delta^S(0) - \Delta^R(0) = 
0$, $\Delta^S(1) - \Delta^R(1) = \tfrac{1}{8}(\tfrac{17}{12} + 3\cdot \tfrac{15}{16} + 3 \cdot \tfrac{1}{2} + \tfrac{3}{8}) - \tfrac{1}{8}(\tfrac{7}{24} + 3\cdot \tfrac{7}{32} + 3 \cdot \tfrac{5}{8} + \tfrac{7}{8}) = \tfrac{77}{256}>0 
$, and $\Delta^S(3) - \Delta^R(3) = \tfrac{1}{8} (\tfrac{25}{8} + 3 \cdot 4+3 \cdot 2 + \tfrac{7}{8}) - \tfrac{1}{8}(1 + 7 \cdot 3) = 0$, proving Proposition \ref{pro:H3}.\hfill\qed

\newpage

\section{Additional analyses}
\label{app:additional_figures}

This section contains additional figures, starting with a more detailed analysis of $\Delta^R$ and $\Delta^S$ by model and history. We then describe possible dynamics and changes in behavior over the course of a session. Finally, we also consider the fit of an alternative model based on maximum-likelihood expected utility (MLEU). In all graphs where values are aggregated by model but not by history, e.g. Figure \ref{fig:hyp2_whiskers}, we weight models according to their frequencies in the experimental sample both for theoretical and observed values for consistent comparisons.

\subsection{\texorpdfstring{$\tilde{\Delta}^R$}{Delta R} and \texorpdfstring{$\tilde{\Delta}^S$}{Delta S} by model and history}\label{app:B1}

The analysis of $\Delta^R$ and $\Delta^S$ by histories and models paints roughly the same picture as the aggregated one in the main body of the article. Receivers tend to adjust their action towards the middle for the more extreme histories. However, the number of history-model combinations where the mean $\tilde{\Delta}_m^R(h,b)$ or $\tilde{\Delta}_m^S(h,b)$ is significantly above zero (shown in green) is increasing in the bias and broadly consistent with the shape of the theoretical shift in equilibrium (shown by triangles). 

Figure \ref{fig:hyp3_whiskers_combined} reports $\tilde{\Delta}_m^S(h,b)-\tilde{\Delta}_m^R(h,b)$, both aggregated by treatment and
disaggregated by history. This quantity measures the net movement of the receiver's action away from the action induced by the true model. The observed values are close to the theoretical predictions. The adjustments are symmetric, which makes the overall metric close to zero when aggregating across histories in \emph{low bias} and \emph{high bias} with a small positive value for \emph{low bias}. This supports the conclusion that receivers largely correct the senders' distortion on average.

\begin{landscape}
\input{results_artifacts/Figure_14_Observed_and_predicted_Delta_R}
\end{landscape}


\begin{landscape}
\input{results_artifacts/Figure_15_Observed_and_predicted_Delta_S}
\end{landscape}


\begin{landscape}
\input{results_artifacts/Figure_16_Observed_persuasiveness_Delta_S_plus_Delta_R}
\end{landscape}


\subsection{Additional comparison of MEU and MLEU}\label{MEU_vs_MLEU}

Figure \ref{fig:MLEU_policy_R} compares the receiver's equilibrium actions under MEU and MLEU for different narratives resulting from the sender's behavior, which are shown on the vertical axis. The difference between models is clearest for \emph{high bias}, where MLEU prescribes only the extreme actions 1 and 9. Figures 
\ref{fig:MLEU_R_whiskers_combined} and \ref{fig:MLEU_S_h_whiskers} show $\Delta^R,~ \tilde{\Delta}^R$ and $\Delta^S,~ \tilde{\Delta}^S$ for the two different ambiguity rules. The MLEU prediction for $\Delta^R$ does not follow Hypothesis \ref{hyp:receiver}:\ $\Delta_m^R(h,b)$ is zero for 3 histories and negative for the fourth one, indicating that the receivers adjust in the opposite direction.

The theoretically-predicted $\Delta^R$ of the MLEU receivers (shown as triangles) goes against the intuition that the receivers' downward adjustment should increase with the bias. When the observed data suggests higher state, i.e., high number of red balls is observed, $\tilde{\Delta}^R$ is not sensitive to the bias:\ the receivers go for the extreme prediction that maximizes the likelihood of this data, which is $9$ red balls.

\begin{landscape}
\input{results_artifacts/Figure_17_Observed_and_predicted_receiver_action_under_MEU_and_MLEU}
\end{landscape}

\begin{landscape}
\input{results_artifacts/Figure_18_Observed_and_predicted_Delta_R_under_MEU_and_MLEU}
\end{landscape}


\begin{landscape}

\input{results_artifacts/Figure_19_Observed_and_predicted_Delta_S_under_MEU_and_MLEU}
\end{landscape}



\subsection{Learning within sessions}

Figure \ref{fig:Learning_combined} separates observations into roughly two halves: the first 6 rounds and the latter rounds. The cutoff of 6 was chosen to equalize the two groups as much as possible given that the sessions had different number of participants and hence different number of rounds. The Figure is a plot of subject-level means to account for subject effects, but only for the early or late rounds in the game (if subject reaches them). Both $\tilde{\Delta}^R$ and $\tilde{\Delta}^S$ are more pronounced in the second half of the \emph{high bias} treatment than in the first half. While this may suggest the existence of some dynamics over time, it is worth reiterating here that subjects received no feedback between rounds. Thus this effect cannot be attributed to learning the behavior of the group over time. Figures \ref{fig:densityLearningR} and \ref{fig:densityLearningS} show a similar picture in terms of distributions of $\tilde{\Delta}^R$ and $\tilde{\Delta}^S$ for the two halves of the game.

\input{results_artifacts/Figure_20_Delta_S_and_Delta_R_over_time}

\input{results_artifacts/Figure_21_Empirical_distribution_of_subject_level_mean_Delta_R_taken_over_first_6_rounds_and_for_rounds_7_plus}

\input{results_artifacts/Figure_22_Empirical_distribution_of_subject_level_mean_Delta_S_taken_over_first_6_rounds_and_for_rounds_7_plus}

\newpage

\section{Experiment Instructions}
\label{app:instructions}

\input{instructions/Instructions}
\end{document}

%% file: results_artifacts/Plot_colors.tex
\definecolor{instructionsblue}{RGB}{0,102,204}
\definecolor{instructionsred}{RGB}{180,0,0}
\definecolor{instructionsblack}{RGB}{0,0,0}
\definecolor{cellgray}{RGB}{245,245,245}
\definecolor{cellblue}{RGB}{220,235,255}
\definecolor{cellred}{RGB}{255,225,225}
\definecolor{captiongray}{RGB}{120,120,120}
\definecolor{borderred}{RGB}{180,0,0}
\definecolor{borderblue}{RGB}{0,102,204}
\definecolor{bordergray}{RGB}{80,80,80}

\definecolor{colorA}{HTML}{D50000}
\definecolor{colorB}{HTML}{0072B2}
\definecolor{series0}{HTML}{D50000}
\definecolor{series1}{HTML}{0072B2}
\definecolor{series2}{HTML}{009E73}
\definecolor{series3}{HTML}{CC79A7}
\definecolor{paircol}{HTML}{009E73}
\definecolor{receiverSeries}{HTML}{D50000}
\definecolor{senderSeries}{HTML}{0072B2}

\definecolor{SVO1}{HTML}{CC79A7}
\definecolor{SVO2}{HTML}{7F7F7F}
\definecolor{biasZero}{HTML}{7AD151}
\definecolor{biasOne}{HTML}{2A788E}
\definecolor{biasThree}{HTML}{440154}

\definecolor{predone}{RGB}{31,119,180}
\definecolor{predtwo}{RGB}{255,127,14}
\definecolor{poscol}{RGB}{44,160,44}
\definecolor{negcol}{RGB}{214,39,40}
\definecolor{neucol}{RGB}{76,76,76}
\definecolor{learningFirst}{HTML}{1F77B4}
\definecolor{learningSecond}{HTML}{D55E00}

%% file: results_artifacts/Figure_4_Truthful_outcomes_by_bias.tex
\begin{figure}[htp!]
\centering
\pgfplotstableread[row sep=\\]{%
bias onlyA inter onlyB\\
0 11.7 33.8 42.1\\
1 11.4 21.4 34.7\\
3 11.7 10.0 27.7\\
}\senderreceiverdata

\begin{tikzpicture}
\begin{axis}[
  width=0.65\textwidth,
  ybar stacked,
  bar width=30pt,
  ymin=0, ymax=100,
  ylabel={\% of matches},
  xlabel={},
  symbolic x coords={0,1,3},
  xticklabels={{No bias (0)},{Low bias (1)},{High bias (3)}},
  xtick=data,
  enlarge x limits=0.25,
  ymajorgrids=true,
  grid style={dashed,gray!40},
  legend style={at={(0.5,-0.38)},anchor=south,legend columns=1},
  legend cell align=left,
  legend image post style={draw=none},
  legend image code/.code={
      \path[#1,draw=none] (0cm,-0.1cm) rectangle (0.6cm,0.1cm);
    },
  nodes near coords,
  point meta=explicit,
  nodes near coords style={font=\scriptsize\bfseries, text=white,
    /pgf/number format/fixed,
    /pgf/number format/precision=1},
]
\addplot+[draw=none, fill=colorB]
  table[x=bias, y=onlyB, meta=onlyB]{\senderreceiverdata};
\addlegendentry{sender truthful, receiver does not follow}
\addplot+[draw=none, fill=colorA!20!white, postaction={pattern=crosshatch, pattern color=colorB}]
  table[x=bias, y=inter, meta=inter]{\senderreceiverdata};
\addlegendentry{sender truthful, receiver follows}
\addplot+[draw=none, fill=colorA]
  table[x=bias, y=onlyA, meta=onlyA]{\senderreceiverdata};
\addlegendentry{sender not truthful, receiver follows}
\end{axis}
\end{tikzpicture}
\caption{Truthful narratives and receiver trust by bias.}
\label{fig:truthfullness}
\end{figure}
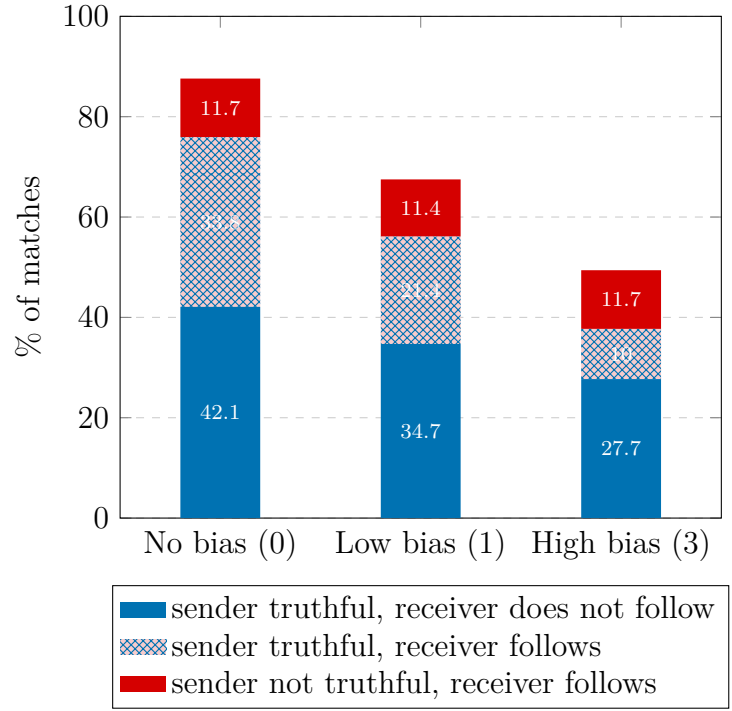

%% file: results_artifacts/Figure_6_Mean_Delta_R_subject_level.tex
\begin{figure}[htp!]
\centering
\pgfplotstableread[row sep=\\]{%
bias mean err_minus err_plus\\
0 0.1512 0.2433 0.2532\\
1 0.3812 0.2371 0.2126\\
3 1.0376 0.2612 0.2707\\
}\treatci

\begin{tikzpicture}
\begin{axis}[
  width=10cm,
  height=6cm,
  ylabel={},
  xlabel={},
  symbolic x coords={0, 1, 3},
  xtick=data,
  xticklabels={{No bias (0)},{Low bias (1)},{High bias (3)}},
  ymajorgrids=true,
  grid style={dashed,gray!40},
  legend style={at={(0.5,1.05)},anchor=south,legend columns=1},
  legend cell align=left,
]
\addplot+[only marks, mark=*, mark options={draw=colorA, fill=colorA},
  error bars/.cd, y dir=both, y explicit, error bar style={black}]
  table[x=bias, y=mean, y error plus=err_plus, y error minus=err_minus]{\treatci};
\end{axis}
\end{tikzpicture}
\vspace{0.4em}
\begin{minipage}{0.9\textwidth}
\footnotesize\centering
Notes: Error bars are subject-level bootstrap 95\% confidence intervals.
\end{minipage}
\caption{Mean $\tilde{\Delta}^R(b)$, subject level.}
\label{fig:hyp2_chart}
\end{figure}

%% file: results_artifacts/Table_2_Delta_R_difference_between_the_senders_recommendation_and_the_receivers_action.tex
\begin{table}[htbp!]
\centering
\begin{threeparttable}
\caption{$\tilde{\Delta}^R$: Difference between the sender's recommendation and the receiver's action.}
\label{tab:receivers}

\begin{tabular}{
    l
    S[table-format=3.0]
    S[table-format=-1.3]
    S[table-format=1.3]
    c
}
\toprule
Treatment & {n} & {Mean} & {Std.\ dev.} & {Bootstrap 95\% conf.\ int.} \\
\midrule
No bias (0) & 24 & 0.151 & 0.628 & (-0.092, 0.404) \\
Low bias (1) & 45 & 0.381 & 0.775 & (0.144, 0.594) \\
High bias (3) & 50 & 1.038 & 0.978 & (0.776, 1.308) \\
\bottomrule
\end{tabular}
\end{threeparttable}
\end{table}

%% file: results_artifacts/Table_3_Pairwise_tests_for_Hypothesis_receiver.tex
\begin{table}[htbp!]
\centering
\begin{threeparttable}
\caption{Pairwise tests for Hypothesis~\ref{hyp:receiver}.}
\label{tab:receivers-tests}

\begin{tabular}{
    l
    S[table-format=-1.3, table-space-text-post={\ensuremath{{}^{***}}}]
    c
    c
    c
}
\toprule
Treat. & {$\tilde{\Delta}^R$ diff.} & {95\% CI} & {$p$ (Welch)} & {$p$ (MW)} \\
\midrule
0 $\to$ 1 & 0.230 { } & (-0.116, 0.576) & 0.188 & 0.331 \\
0 $\to$ 3 & 0.886 {$^{***}$} & (0.510, 1.263) & $< 0.0001$ & $< 0.001$ \\
1 $\to$ 3 & 0.656 {$^{***}$} & (0.298, 1.014) & $< 0.001$ & $< 0.001$ \\
\bottomrule
\end{tabular}

\begin{tablenotes}[flushleft]
\footnotesize
\item Notes:\ The table reports pairwise comparisons of $\tilde{\Delta}^R(b)$ across treatments. The second column reports differences in sample means. Confidence intervals are 95\% confidence intervals for the mean difference. MW denotes the Mann--Whitney $U$ test. Stars following mean differences are based on two-sided Welch tests. ${}^{*}p<0.10$, ${}^{**}p<0.05$, ${}^{***}p<0.01$.
\end{tablenotes}
\end{threeparttable}%
\end{table}

%% file: results_artifacts/Figure_7_Percent_of_trusting_decisions_within_receiver_subjects.tex
\begin{figure}[htp!]
\centering
\begin{tikzpicture}
\begin{axis}[
width=15cm,
height=7cm,
ybar,
bar width=9pt,
xmin=-0.7, xmax=9.7,
ymin=0, ymax=35.0,
xtick={0,1,2,3,4,5,6,7,8,9},
xticklabels={0-10,10-20,20-30,30-40,40-50,50-60,60-70,70-80,80-90,90-100},
xticklabel style={rotate=45,anchor=east},
title={Receivers},
xlabel={Percent of trusting decisions},
ylabel={Percent of subjects},
ymajorgrids=true,
grid style={draw=gray!25},
axis line style={draw=black},
tick style={draw=black},
legend style={at={(0.5,-0.24)},anchor=north,legend columns=4,draw=none},
]
\addplot+[ybar, bar shift=0pt, draw=none, fill=biasZero, fill opacity=0.9, forget plot] coordinates {(-0.25,4.167) (0.75,4.167) (1.75,20.833) (2.75,25.000) (3.75,8.333) (4.75,12.500) (5.75,8.333) (6.75,4.167) (7.75,4.167) (8.75,8.333)};
\addlegendimage{ybar,ybar legend,draw=none,fill=biasZero,fill opacity=0.9}
\addlegendentry{No bias (0)}
\addplot+[ybar, bar shift=0pt, draw=none, fill=biasOne, fill opacity=0.9, forget plot] coordinates {(0.00,22.222) (1.00,20.000) (2.00,22.222) (3.00,4.444) (4.00,11.111) (5.00,0.000) (6.00,4.444) (7.00,2.222) (8.00,6.667) (9.00,6.667)};
\addlegendimage{ybar,ybar legend,draw=none,fill=biasOne,fill opacity=0.9}
\addlegendentry{Low bias (1)}
\addplot+[ybar, bar shift=0pt, draw=none, fill=biasThree, fill opacity=0.9, forget plot] coordinates {(0.25,26.000) (1.25,30.000) (2.25,16.000) (3.25,14.000) (4.25,8.000) (5.25,4.000) (6.25,2.000) (7.25,0.000) (8.25,0.000) (9.25,0.000)};
\addlegendimage{ybar,ybar legend,draw=none,fill=biasThree,fill opacity=0.9}
\addlegendentry{High bias (3)}
\draw[biasZero, dashed, line width=1.1pt] (axis cs:9.500,0) -- (axis cs:9.500,35.0);
\draw[biasOne, dashed, line width=1.1pt] (axis cs:5.124,0) -- (axis cs:5.124,35.0);
\draw[biasThree, dashed, line width=1.1pt] (axis cs:-0.500,0) -- (axis cs:-0.500,35.0);
\addlegendimage{legend image code/.code={\draw[black, dashed, line width=1.1pt] (0cm,0cm) -- (0.35cm,0cm);}}
\addlegendentry{Theory}
\end{axis}
\end{tikzpicture}
\caption{Percent of trusting decisions within receiver subjects.}
\label{fig:trust_R}
\end{figure}

%% file: results_artifacts/Table_4_Effect_of_increasing_bias_on_Delta_R_by_history.tex
\begin{table}[htbp]
\centering
\begin{threeparttable}
\caption{Effect of increasing bias on $\tilde{\Delta}^R(h,b)$, by history $h$.}
\label{tab:by_history_receivers}

\begin{tabular}{
    l
    S[table-format=-1.3]
    @{}c
    S[table-format=1.3]
    S[table-format=-1.3]
    @{}c
    S[table-format=1.3]
    S[table-format=-1.3]
    @{}c
    S[table-format=1.3]
}
\toprule
& \multicolumn{3}{c}{Bias 0} & \multicolumn{3}{c}{0$\to$1} & \multicolumn{3}{c}{1$\to$3} \\
\cmidrule(lr){2-4} \cmidrule(lr){5-7} \cmidrule(lr){8-10}
History & \multicolumn{2}{c}{Coefficient} & {Std.\ err.}
        & \multicolumn{2}{c}{Coefficient} & {Std.\ err.}
        & \multicolumn{2}{c}{Coefficient} & {Std.\ err.} \\
\midrule
$\{\tikz[baseline=-0.6ex]{\fill[black] (0,0) circle (0.07);}, \tikz[baseline=-0.6ex]{\fill[black] (0,0) circle (0.07);}, \tikz[baseline=-0.6ex]{\fill[black] (0,0) circle (0.07);}\}$ & -0.833 & $^{***}$ & 0.230 & 0.522 & $^{*}$ & 0.302 & 0.614 & $^{**}$ & 0.307 \\
$\{\tikz[baseline=-0.6ex]{\fill[red] (0,0) circle (0.07);}, \tikz[baseline=-0.6ex]{\fill[black] (0,0) circle (0.07);}, \tikz[baseline=-0.6ex]{\fill[black] (0,0) circle (0.07);}\}$ & 0.088 &   & 0.204 & 0.123 &   & 0.289 & 1.013 & $^{***}$ & 0.291 \\
$\{\tikz[baseline=-0.6ex]{\fill[red] (0,0) circle (0.07);}, \tikz[baseline=-0.6ex]{\fill[red] (0,0) circle (0.07);}, \tikz[baseline=-0.6ex]{\fill[black] (0,0) circle (0.07);}\}$ & 0.539 & $^{**}$ & 0.216 & 0.104 &   & 0.282 & 0.760 & $^{**}$ & 0.314 \\
$\{\tikz[baseline=-0.6ex]{\fill[red] (0,0) circle (0.07);}, \tikz[baseline=-0.6ex]{\fill[red] (0,0) circle (0.07);}, \tikz[baseline=-0.6ex]{\fill[red] (0,0) circle (0.07);}\}$ & 0.919 & $^{***}$ & 0.213 & -0.004 &   & 0.267 & 0.401 & $^{*}$ & 0.237 \\
\bottomrule
\end{tabular}

\begin{tablenotes}[flushleft]
\footnotesize
\item Notes:\ Entries report coefficient estimates and standard errors clustered by participant. Significance stars are based on two-sided tests of the corresponding linear contrasts using the clustered variance-covariance matrix, where $^{*}p<0.10$, $^{**}p<0.05$, $^{***}p<0.01$.
\end{tablenotes}
\end{threeparttable}
\end{table}

%% file: results_artifacts/Figure_9_Mean_Delta_S_subject_level.tex
\begin{figure}[htp!]
\centering
\pgfplotstableread[row sep=\\]{%
bias mean err_minus err_plus\\
0 -0.0160 0.2267 0.2314\\
1 0.6199 0.3747 0.3468\\
3 0.8342 0.3932 0.3667\\
}\treatci

\begin{tikzpicture}
\begin{axis}[
  width=10cm,
  height=6cm,
  ylabel={},
  xlabel={},
  symbolic x coords={0, 1, 3},
  xtick=data,
  xticklabels={{No bias (0)},{Low bias (1)},{High bias (3)}},
  ymajorgrids=true,
  grid style={dashed,gray!40},
  legend style={at={(0.5,1.05)},anchor=south,legend columns=1},
  legend cell align=left,
]
\addplot+[only marks, mark=square*, mark options={draw=colorB, fill=colorB},
  error bars/.cd, y dir=both, y explicit, error bar style={black}]
  table[x=bias, y=mean, y error plus=err_plus, y error minus=err_minus]{\treatci};
\end{axis}
\end{tikzpicture}
\vspace{0.4em}
\begin{minipage}{0.9\textwidth}
\footnotesize\centering
Notes: Error bars are subject-level bootstrap 95\% confidence intervals.
\end{minipage}
\caption{Mean $\tilde{\Delta}^S(b)$, subject level.}
\label{fig:hyp1_means_chart}
\end{figure}

%% file: results_artifacts/Table_5_Delta_S_difference_between_the_senders_recommendation_and_receivers_optimal_action_for_the_true_model.tex
\begin{table}[htbp]
\centering
\begin{threeparttable}
\caption{$\tilde{\Delta}^S(b)$: Difference between the sender's recommendation and the receiver's optimal action for the true model.}
\label{tab:senders}

\begin{tabular}{
    l
    S[table-format=3.0]
    S[table-format=-1.3]
    S[table-format=1.3]
    c
}
\toprule
Treatment & {n} & {Mean} & {Std.\ dev.} & {Bootstrap 95\% conf.\ int.} \\
\midrule
No bias (0) & 24 & -0.016 & 0.587 & (-0.243, 0.215) \\
Low bias (1) & 45 & 0.620 & 1.254 & (0.245, 0.967) \\
High bias (3) & 50 & 0.834 & 1.405 & (0.441, 1.201) \\
\bottomrule
\end{tabular}
\end{threeparttable}
\end{table}

%% file: results_artifacts/Table_6_Pairwise_tests_for_Hypothesis_sender.tex
\begin{table}[htbp]
\centering
\begin{threeparttable}
\caption{Pairwise tests for Hypothesis~\ref{hyp:sender}.}
\label{tab:senders-tests}

\begin{tabular}{
    l
    S[table-format=-1.3, table-space-text-post={\ensuremath{{}^{***}}}]
    c
    c
    c
}
\toprule
Treat. & {$\tilde{\Delta}^S$ diff.} & {95\% CI} & {$p$ (Welch)} & {$p$ (MW)} \\
\midrule
0 $\to$ 1 & 0.636 {$^{***}$} & (0.193, 1.079) & 0.006 & 0.011 \\
0 $\to$ 3 & 0.850 {$^{***}$} & (0.388, 1.313) & $< 0.001$ & 0.008 \\
1 $\to$ 3 & 0.214 { } & (-0.327, 0.756) & 0.434 & 0.438 \\
\bottomrule
\end{tabular}

\begin{tablenotes}[flushleft]
\footnotesize
\item Notes:\ The table reports pairwise comparisons of $\tilde{\Delta}^S(b)$ across treatments $b$. The second column reports differences in sample means. Confidence intervals are 95\% confidence intervals for the mean difference. MW denotes the Mann--Whitney $U$ test. Stars following mean differences are based on two-sided Welch tests. ${}^{*}p<0.10$, ${}^{**}p<0.05$, ${}^{***}p<0.01$.
\end{tablenotes}
\end{threeparttable}%
\end{table}

%% file: results_artifacts/Figure_10_Percent_of_truthful_decisions_within_sender_subjects.tex
\begin{figure}[htp!]
\centering
\begin{tikzpicture}
\begin{axis}[
width=15cm,
height=7cm,
ybar,
bar width=9pt,
xmin=-0.7, xmax=9.7,
ymin=0, ymax=55.0,
xtick={0,1,2,3,4,5,6,7,8,9},
xticklabels={0-10,10-20,20-30,30-40,40-50,50-60,60-70,70-80,80-90,90-100},
xticklabel style={rotate=45,anchor=east},
title={Senders},
xlabel={Percent of truthful decisions},
ylabel={Percent of subjects},
ymajorgrids=true,
grid style={draw=gray!25},
axis line style={draw=black},
tick style={draw=black},
legend style={at={(0.5,-0.24)},anchor=north,legend columns=4,draw=none},
]
\addplot+[ybar, bar shift=0pt, draw=none, fill=biasZero, fill opacity=0.9, forget plot] coordinates {(-0.25,0.000) (0.75,0.000) (1.75,0.000) (2.75,20.833) (3.75,0.000) (4.75,4.167) (5.75,12.500) (6.75,8.333) (7.75,4.167) (8.75,50.000)};
\addlegendimage{ybar,ybar legend,draw=none,fill=biasZero,fill opacity=0.9}
\addlegendentry{No bias (0)}
\addplot+[ybar, bar shift=0pt, draw=none, fill=biasOne, fill opacity=0.9, forget plot] coordinates {(0.00,4.444) (1.00,4.444) (2.00,11.111) (3.00,13.333) (4.00,20.000) (5.00,4.444) (6.00,11.111) (7.00,6.667) (8.00,6.667) (9.00,17.778)};
\addlegendimage{ybar,ybar legend,draw=none,fill=biasOne,fill opacity=0.9}
\addlegendentry{Low bias (1)}
\addplot+[ybar, bar shift=0pt, draw=none, fill=biasThree, fill opacity=0.9, forget plot] coordinates {(0.25,10.000) (1.25,14.000) (2.25,22.000) (3.25,14.000) (4.25,12.000) (5.25,14.000) (6.25,2.000) (7.25,4.000) (8.25,2.000) (9.25,6.000)};
\addlegendimage{ybar,ybar legend,draw=none,fill=biasThree,fill opacity=0.9}
\addlegendentry{High bias (3)}
\draw[biasZero, dashed, line width=1.1pt] (axis cs:9.500,0) -- (axis cs:9.500,55.0);
\draw[biasOne, dashed, line width=1.1pt] (axis cs:5.878,0) -- (axis cs:5.878,55.0);
\draw[biasThree, dashed, line width=1.1pt] (axis cs:1.972,0) -- (axis cs:1.972,55.0);
\addlegendimage{legend image code/.code={\draw[black, dashed, line width=1.1pt] (0cm,0cm) -- (0.35cm,0cm);}}
\addlegendentry{Theory}
\end{axis}
\end{tikzpicture}
\caption{Percent of truthful decisions within sender subjects.}
\label{fig:truth_S}
\end{figure}

%% file: results_artifacts/Figure_Informativeness_of_the_senders_report_by_history.tex
\begin{figure}[htbp!]
\centering
\begin{tikzpicture}
\begin{axis}[
  width=0.8\textwidth,
  height=7.5cm,
  xlabel={Correlation between $a(r,h,0)$ and $a(m^T,h,0)$},
  ylabel={History $h$},
  xmin=-0.25, xmax=1.15,
  xtick={-0.2,0,0.2,0.4,0.6,0.8,1.0},
  symbolic y coords={{(3,0)},{(2,1)},{(1,2)},{(0,3)}},
  ytick=data,
  yticklabels={\tikz[baseline=-0.55ex]{\fill[red] (0.00em,0) circle (2.2pt);\fill[red] (0.62em,0) circle (2.2pt);\fill[red] (1.24em,0) circle (2.2pt);},\tikz[baseline=-0.55ex]{\fill[red] (0.00em,0) circle (2.2pt);\fill[red] (0.62em,0) circle (2.2pt);\fill[black] (1.24em,0) circle (2.2pt);},\tikz[baseline=-0.55ex]{\fill[red] (0.00em,0) circle (2.2pt);\fill[black] (0.62em,0) circle (2.2pt);\fill[black] (1.24em,0) circle (2.2pt);},\tikz[baseline=-0.55ex]{\fill[black] (0.00em,0) circle (2.2pt);\fill[black] (0.62em,0) circle (2.2pt);\fill[black] (1.24em,0) circle (2.2pt);}},
  yticklabel style={inner sep=3pt},
  enlarge y limits=0.22,
  xmajorgrids=true,
  grid style={dashed,gray!40},
  extra x ticks={0},
  extra x tick labels={},
  extra x tick style={grid=major, grid style={solid, gray!70}},
  legend style={at={(0.5,-0.22)},anchor=north,legend columns=-1, draw=none, font=\small, /tikz/every even column/.append style={column sep=12pt}},
  legend cell align=left,
]
\addplot[only marks, mark=*, mark size=2pt, color=biasZero,
  yshift=9pt, error bars/x dir=both,
  error bars/x explicit, error bars/error bar style={biasZero},
  legend image post style={yshift=-9pt}]
  table[x=mean, y=h, x error minus=lo, x error plus=hi, row sep=\\] {%
  h mean lo hi\\
  (3,0) 0.6300 0.3380 0.2753\\
  (2,1) 0.7355 0.2120 0.1777\\
  (1,2) 0.5324 0.3847 0.3130\\
  (0,3) 0.7591 0.2488 0.1788\\
  };
\addlegendentry{No bias (0)}
\addplot[only marks, mark=triangle*, mark size=3pt,
  mark options={draw=biasZero, fill=biasZero}, yshift=9pt, forget plot]
  table[x=theory, y=h, row sep=\\] {%
  h theory\\
  (3,0) 1.0000\\
  (2,1) 1.0000\\
  (1,2) 1.0000\\
  (0,3) 1.0000\\
  };
\addplot[only marks, mark=*, mark size=2pt, color=biasOne,
  yshift=0pt, error bars/x dir=both,
  error bars/x explicit, error bars/error bar style={biasOne},
  legend image post style={yshift=0pt}]
  table[x=mean, y=h, x error minus=lo, x error plus=hi, row sep=\\] {%
  h mean lo hi\\
  (3,0) 0.3335 0.2343 0.2436\\
  (2,1) 0.4508 0.2639 0.2225\\
  (1,2) 0.5252 0.2324 0.1995\\
  (0,3) 0.3723 0.2054 0.2038\\
  };
\addlegendentry{Low bias (1)}
\addplot[only marks, mark=triangle*, mark size=3pt,
  mark options={draw=biasOne, fill=biasOne}, yshift=0pt, forget plot]
  table[x=theory, y=h, row sep=\\] {%
  h theory\\
  (3,0) 0.9580\\
  (2,1) 0.9200\\
  (1,2) 0.9473\\
  (0,3) 0.8828\\
  };
\addplot[only marks, mark=*, mark size=2pt, color=biasThree,
  yshift=-9pt, error bars/x dir=both,
  error bars/x explicit, error bars/error bar style={biasThree},
  legend image post style={yshift=9pt}]
  table[x=mean, y=h, x error minus=lo, x error plus=hi, row sep=\\] {%
  h mean lo hi\\
  (3,0) 0.3124 0.1952 0.1920\\
  (2,1) 0.1222 0.1921 0.1977\\
  (1,2) 0.3014 0.1364 0.1137\\
  (0,3) 0.1155 0.1793 0.1740\\
  };
\addlegendentry{High bias (3)}
\addplot[only marks, mark=triangle*, mark size=3pt,
  mark options={draw=biasThree, fill=biasThree}, yshift=-9pt, forget plot]
  table[x=theory, y=h, row sep=\\] {%
  h theory\\
  (3,0) 0.0000\\
  (2,1) 0.0000\\
  (1,2) 0.0000\\
  (0,3) 0.0000\\
  };
\end{axis}
\end{tikzpicture}
\begin{minipage}{0.9\textwidth}
\footnotesize\centering
Notes: Dots are observed correlations with sender-clustered 95\% bootstrap intervals, triangles are equilibrium values. Both observed and theoretical values weight models by their empirical frequencies for aggregation rather than uniformly to make them comparable.
\end{minipage}
\caption{Informativeness of the sender's report by history and treatment.  }
\label{fig:report_informativeness_history}
\end{figure}

%% file: results_artifacts/Table_7_Effect_of_increasing_bias_on_Delta_S_by_history.tex
\begin{table}[htbp]
\centering
\begin{threeparttable}
\caption{Effect of increasing bias $b$ on $\tilde{\Delta}^S(h,b)$, by history $h$.}
\label{tab:by_history_senders}

\begin{tabular}{
    l
    S[table-format=-1.3]
    @{}c
    S[table-format=1.3]
    S[table-format=-1.3]
    @{}c
    S[table-format=1.3]
    S[table-format=-1.3]
    @{}c
    S[table-format=1.3]
}
\toprule
& \multicolumn{3}{c}{Bias 0} & \multicolumn{3}{c}{0$\to$1} & \multicolumn{3}{c}{1$\to$3} \\
\cmidrule(lr){2-4} \cmidrule(lr){5-7} \cmidrule(lr){8-10}
History & \multicolumn{2}{c}{Coefficient} & {Std. err.}
        & \multicolumn{2}{c}{Coefficient} & {Std. err.}
        & \multicolumn{2}{c}{Coefficient} & {Std. err.} \\
\midrule
$\{\tikz[baseline=-0.6ex]{\fill[black] (0,0) circle (0.07);}, \tikz[baseline=-0.6ex]{\fill[black] (0,0) circle (0.07);}, \tikz[baseline=-0.6ex]{\fill[black] (0,0) circle (0.07);}\}$ & 0.139 &   & 0.119 & 0.248 &   & 0.226 & 0.734 & $^{**}$ & 0.306 \\
$\{\tikz[baseline=-0.6ex]{\fill[red] (0,0) circle (0.07);}, \tikz[baseline=-0.6ex]{\fill[black] (0,0) circle (0.07);}, \tikz[baseline=-0.6ex]{\fill[black] (0,0) circle (0.07);}\}$ & -0.075 &   & 0.282 & 1.092 & $^{***}$ & 0.381 & 0.556 &   & 0.386 \\
$\{\tikz[baseline=-0.6ex]{\fill[red] (0,0) circle (0.07);}, \tikz[baseline=-0.6ex]{\fill[red] (0,0) circle (0.07);}, \tikz[baseline=-0.6ex]{\fill[black] (0,0) circle (0.07);}\}$ & 0.118 &   & 0.221 & 0.535 &   & 0.338 & -0.002 &   & 0.429 \\
$\{\tikz[baseline=-0.6ex]{\fill[red] (0,0) circle (0.07);}, \tikz[baseline=-0.6ex]{\fill[red] (0,0) circle (0.07);}, \tikz[baseline=-0.6ex]{\fill[red] (0,0) circle (0.07);}\}$ & -0.290 &   & 0.281 & 0.648 & $^{*}$ & 0.360 & -0.441 &   & 0.299 \\
\bottomrule
\end{tabular}

\begin{tablenotes}[flushleft]
\footnotesize
\item Notes: Entries report coefficient estimates and standard errors clustered by participant. Significance stars are based on two-sided tests of the corresponding linear contrasts using the clustered variance-covariance matrix, where $^{*}p<0.10$, $^{**}p<0.05$, $^{***}p<0.01$.
\end{tablenotes}
\end{threeparttable}
\end{table}

%% file: results_artifacts/Figure_11_Mean_observed_persuasiveness_receiver_subject_level.tex
\begin{figure}[htp!]
\centering
\pgfplotstableread[row sep=\\]{%
bias mean err_minus err_plus\\
0 -0.1672 0.2896 0.2937\\
1 0.2387 0.2692 0.2871\\
3 -0.2033 0.2792 0.2755\\
}\treatci

\begin{tikzpicture}
\begin{axis}[
  width=10cm,
  height=6cm,
  ylabel={},
  xlabel={},
  symbolic x coords={0, 1, 3},
  xtick=data,
  xticklabels={{No bias (0)},{Low bias (1)},{High bias (3)}},
  ymajorgrids=true,
  grid style={dashed,gray!40},
  legend style={at={(0.5,1.05)},anchor=south,legend columns=1},
  legend cell align=left,
]
\addplot+[only marks, mark=diamond*, mark options={draw=paircol, fill=paircol},
  error bars/.cd, y dir=both, y explicit, error bar style={black}]
  table[x=bias, y=mean, y error plus=err_plus, y error minus=err_minus]{\treatci};
\end{axis}
\end{tikzpicture}
\vspace{0.4em}
\begin{minipage}{0.9\textwidth}
\footnotesize
Notes: Error bars are subject-level bootstrap 95\% confidence intervals.
\end{minipage}
\caption{Mean $\tilde{\Delta}^S(b)-\tilde{\Delta}^R(b)$, receiver-subject level.}
\label{fig:persuasiveness_subject_level}
\end{figure}

%% file: results_artifacts/Table_8_Pairwise_tests_of_increase_then_decrease_persuasiveness.tex
\begin{table}[htbp]
\centering
\begin{threeparttable}
\caption{Pairwise tests of observed persuasiveness.}
\label{tab:persuasiveness_order_tests}
\begin{tabular}{
    l
    S[table-format=-1.3, table-space-text-post={\ensuremath{{}^{***}}}]
    c
    c
    c
}
\toprule
Treat. & {$\tilde{\Delta}^S-\tilde{\Delta}^R$ diff.} & {95\% CI} & {$p$ (Welch)} & {$p$ (MW)} \\
\midrule
0 $\to$ 1 & 0.406 {$^{*}$} & (-0.010, 0.821) & 0.055 & 0.062 \\
1 $\to$ 3 & -0.442 {$^{**}$} & (-0.845, -0.039) & 0.032 & 0.024 \\
\bottomrule
\end{tabular}
\begin{tablenotes}[flushleft]
\footnotesize
\item Notes:\ The table reports pairwise comparisons of $\tilde{\Delta}^S(b)-\tilde{\Delta}^R(b)$ across treatments. The second column reports differences in sample means. Confidence intervals are 95\% confidence intervals for the mean difference. MW denotes the Mann--Whitney $U$ test. Stars following mean differences are based on two-sided Welch tests. $^{*}p<0.10$, $^{**}p<0.05$, $^{***}p<0.01$.
\end{tablenotes}
\end{threeparttable}%
\end{table}

%% file: results_artifacts/Table_9_Correlations_between_message_likelihood_and_receiver_trust_measures.tex
\begin{table}[htbp]
\centering
\resizebox{\textwidth}{!}{%
\begin{threeparttable}
\caption{Correlations between message likelihood and receiver trust measures.}
\label{tab:likelihood_receiver_correlations}

\begin{tabular}{
    l
    S[table-format=-1.3]
    S[table-format=1.3]
    S[table-format=-1.3]
    S[table-format=-1.3]
    S[table-format=-1.3]
    S[table-format=-1.3]
}
\toprule
& \multicolumn{3}{c}{Observation level} & \multicolumn{3}{c}{Subject level} \\
\cmidrule(lr){2-4} \cmidrule(lr){5-7}
Bias & {Receiver follows} & {$|\tilde{\Delta}^R|$} & {$\tilde{\Delta}^R$}
     & {Receiver follows} & {$|\tilde{\Delta}^R|$} & {$\tilde{\Delta}^R$} \\
\midrule
0 & -0.118 & 0.124 & -0.045 & -0.146 & 0.148 & -0.078 \\
1 & -0.151 & 0.151 & 0.169 & -0.297 & 0.288 & 0.005 \\
3 & -0.057 & 0.053 & 0.079 & 0.242 & -0.218 & -0.041 \\
Total & -0.121 & 0.115 & 0.101 & -0.186 & 0.184 & 0.049 \\
\bottomrule
\end{tabular}

\begin{tablenotes}[flushleft]
\footnotesize
\item Notes: Entries report Pearson correlations between message likelihood and receiver trust measures. Observation-level correlations are computed across individual observations; subject-level correlations are computed using subject-level averages.
\end{tablenotes}
\end{threeparttable}%
}
\end{table}

%% file: results_artifacts/Figure_12_Welfare_and_its_components.tex
\begin{figure}[htb!]
\centering

\pgfplotstableread[row sep=\\]{%
bias m0 e0m e0p m1 e1m e1p m2 e2m e2p m3 e3m e3p\\
0 726.551724 46.206897 45.517241 737.931034 46.551724 44.827586 737.931034 48.620690 48.965517 712.413793 39.310345 40.344828\\
1 729.453682 39.429929 38.954869 713.064133 40.380048 40.855107 743.467933 40.855107 39.667458 724.228029 31.353919 32.066508\\
3 684.386617 34.758364 33.828996 632.899628 37.736989 37.360595 745.167286 35.687732 35.687732 731.226766 28.996283 28.438662\\
}\receiverplotci

\pgfplotstableread[row sep=\\]{%
bias m0 e0m e0p m1 e1m e1p m2 e2m e2p m3 e3m e3p\\
0 726.551724 48.620690 46.896552 737.931034 45.525862 45.862069 737.931034 48.275862 48.965517 712.413793 38.275862 39.310345\\
1 713.776722 39.667458 39.192399 720.902613 39.198337 39.192399 693.586698 39.429929 38.479810 716.389549 37.529691 37.054632\\
3 435.130112 36.802974 35.315985 562.081784 36.988848 37.174721 455.762082 35.878253 35.873606 527.881041 41.454461 41.449814\\
}\senderplotci

\pgfplotstableread[row sep=\\]{%
bias m0 e0m e0p m1 e1m e1p m2 e2m e2p m3 e3m e3p\\
0 1453.103448 93.103448 92.413793 1475.862069 91.724138 92.413793 1475.862069 94.482759 97.931034 1424.827586 76.551724 78.620690\\
1 1443.230404 73.159145 73.871734 1433.966746 76.015439 72.921615 1437.054632 74.109264 74.821853 1440.617577 62.707838 64.133017\\
3 1119.516729 49.256506 50.190520 1194.981413 47.397770 46.840149 1200.929368 47.955390 46.654275 1259.107807 45.539033 44.052045\\
}\pairplotci

\begin{tikzpicture}
\begin{groupplot}[
  group style={group size=3 by 1, horizontal sep=1.6cm},
  width=5.2cm,
  height=5.8cm,
  ymajorgrids=true,
  grid style={dashed,gray!40},
  xlabel={bias},
  symbolic x coords={0, 1, 3},
  xtick=data,
  enlarge x limits=0.25,
]
\nextgroupplot[
  title={Receiver},
  ymin=350, ymax=800,
  ytick={0, 200, 400, 600, 800},
]
\addplot+[only marks, mark=*, mark options={draw=series0, fill=series0}, xshift=-10.00pt, error bars/.cd, y dir=both, y explicit, error bar style={black}]
  table[x=bias, y=m0, y error minus=e0m, y error plus=e0p]{\receiverplotci};
\addplot+[only marks, mark=square*, mark options={draw=series1, fill=series1}, xshift=-3.33pt, error bars/.cd, y dir=both, y explicit, error bar style={black}]
  table[x=bias, y=m1, y error minus=e1m, y error plus=e1p]{\receiverplotci};
\addplot+[only marks, mark=triangle*, mark options={draw=series2, fill=series2}, xshift=3.33pt, error bars/.cd, y dir=both, y explicit, error bar style={black}]
  table[x=bias, y=m2, y error minus=e2m, y error plus=e2p]{\receiverplotci};
\addplot+[only marks, mark=diamond*, mark options={draw=series3, fill=series3}, xshift=10.00pt, error bars/.cd, y dir=both, y explicit, error bar style={black}]
  table[x=bias, y=m3, y error minus=e3m, y error plus=e3p]{\receiverplotci};
\nextgroupplot[
  title={Sender},
  ymin=350, ymax=800,
  ytick={0, 200, 400, 600, 800},
]
\addplot+[only marks, mark=*, mark options={draw=series0, fill=series0}, xshift=-10.00pt, error bars/.cd, y dir=both, y explicit, error bar style={black}]
  table[x=bias, y=m0, y error minus=e0m, y error plus=e0p]{\senderplotci};
\addplot+[only marks, mark=square*, mark options={draw=series1, fill=series1}, xshift=-3.33pt, error bars/.cd, y dir=both, y explicit, error bar style={black}]
  table[x=bias, y=m1, y error minus=e1m, y error plus=e1p]{\senderplotci};
\addplot+[only marks, mark=triangle*, mark options={draw=series2, fill=series2}, xshift=3.33pt, error bars/.cd, y dir=both, y explicit, error bar style={black}]
  table[x=bias, y=m2, y error minus=e2m, y error plus=e2p]{\senderplotci};
\addplot+[only marks, mark=diamond*, mark options={draw=series3, fill=series3}, xshift=10.00pt, error bars/.cd, y dir=both, y explicit, error bar style={black}]
  table[x=bias, y=m3, y error minus=e3m, y error plus=e3p]{\senderplotci};
\nextgroupplot[
  title={Pair},
  ymin=700, ymax=1600,
  ytick={0, 400, 800, 1200, 1600},
]
\addplot+[only marks, mark=*, mark options={draw=series0, fill=series0}, xshift=-10.00pt, error bars/.cd, y dir=both, y explicit, error bar style={black}]
  table[x=bias, y=m0, y error minus=e0m, y error plus=e0p]{\pairplotci};
\addplot+[only marks, mark=square*, mark options={draw=series1, fill=series1}, xshift=-3.33pt, error bars/.cd, y dir=both, y explicit, error bar style={black}]
  table[x=bias, y=m1, y error minus=e1m, y error plus=e1p]{\pairplotci};
\addplot+[only marks, mark=triangle*, mark options={draw=series2, fill=series2}, xshift=3.33pt, error bars/.cd, y dir=both, y explicit, error bar style={black}]
  table[x=bias, y=m2, y error minus=e2m, y error plus=e2p]{\pairplotci};
\addplot+[only marks, mark=diamond*, mark options={draw=series3, fill=series3}, xshift=10.00pt, error bars/.cd, y dir=both, y explicit, error bar style={black}]
  table[x=bias, y=m3, y error minus=e3m, y error plus=e3p]{\pairplotci};
\end{groupplot}
\coordinate (welfarelegendbase) at ($(group c2r1.south)+(0,-1.00cm)$);
\node[anchor=north] at ($(welfarelegendbase)+(0,0.000cm)$) {\tikz[baseline=-0.6ex] \draw plot[only marks, mark=*, mark size=2.6pt, mark options={draw=series0, fill=series0}] coordinates {(0,0)};\hspace{0.45em}Actual receiver action in the experiment};
\node[anchor=north] at ($(welfarelegendbase)+(0,-0.420cm)$) {\tikz[baseline=-0.6ex] \draw plot[only marks, mark=square*, mark size=2.6pt, mark options={draw=series1, fill=series1}] coordinates {(0,0)};\hspace{0.45em}Sender's recommendation in the experiment};
\node[anchor=north] at ($(welfarelegendbase)+(0,-0.840cm)$) {\tikz[baseline=-0.6ex] \draw plot[only marks, mark=triangle*, mark size=2.6pt, mark options={draw=series2, fill=series2}] coordinates {(0,0)};\hspace{0.45em}Optimal action under the true model};
\node[anchor=north] at ($(welfarelegendbase)+(0,-1.260cm)$) {\tikz[baseline=-0.6ex] \draw plot[only marks, mark=diamond*, mark size=2.6pt, mark options={draw=series3, fill=series3}] coordinates {(0,0)};\hspace{0.45em}MEU pooling action (ignore message)};
\end{tikzpicture}
\caption{Welfare (average payoffs) and its components depending on the receiver's choice.}
\label{fig:welfare_triptych}
\end{figure}

%% file: results_artifacts/Table_10_Mean_pair_welfare.tex
\begin{table}[htb!]
\centering
\caption{Aggregate welfare and counterfactuals.}
\label{tab:welfare_relative}

\begin{threeparttable}
\begin{tabularx}{\linewidth}{
    @{}
    Y
    *{3}{S[
        table-format=4.1,
        table-column-width=2.05cm,
        table-number-alignment=center
    ]}
    @{}
}
\toprule
& \multicolumn{1}{C{2.05cm}}{Bias 0}
& \multicolumn{1}{C{2.05cm}}{Bias 1}
& \multicolumn{1}{C{2.05cm}}{Bias 3} \\
\midrule

\multicolumn{4}{@{}l}{\textit{Reference level}} \\
Action equals the true state, in points
& \multicolumn{1}{C{2.05cm}}{2600}
& \multicolumn{1}{C{2.05cm}}{2300}
& \multicolumn{1}{C{2.05cm}}{1700} \\

\addlinespace[4pt]
\midrule
\multicolumn{4}{@{}l}{\textit{Observed welfare, relative to the reference level}} \\
Experimental data
& 56.1 & 63.3 & 65.6 \\
\quad\scriptsize 95\% c.i.
& \ci{(52.2, 60.1)} & \ci{(59.6, 66.8)} & \ci{(62.4, 68.8)} \\

\addlinespace[4pt]
\midrule
\multicolumn{4}{@{}l}{\textit{Counterfactual welfare, relative to the reference level, when the receiver}} \\
knows the true model
& 56.9 & 61.6 & 71.0 \\
\quad\scriptsize 95\% c.i.
& \ci{(53.1, 60.6)} & \ci{(57.7, 65.2)} & \ci{(68.0, 74.1)} \\
\addlinespace[2pt]
follows sender's recommendation
& 56.9 & 60.7 & 70.6 \\
\quad\scriptsize 95\% c.i.
& \ci{(52.6, 60.9)} & \ci{(56.3, 64.8)} & \ci{(67.5, 73.4)} \\
\addlinespace[2pt]
ignores message under MEU
& 54.7 & 63.7 & 73.5 \\
\quad\scriptsize 95\% c.i.
& \ci{(51.5, 57.9)} & \ci{(60.2, 67.1)} & \ci{(70.3, 76.6)} \\
\addlinespace[2pt]
ignores message under MLEU
& 53.0 & 55.9 & 64.2 \\
\quad\scriptsize 95\% c.i.
& \ci{(48.0, 58.1)} & \ci{(51.8, 60.1)} & \ci{(60.4, 67.8)} \\

\bottomrule
\end{tabularx}

\begin{tablenotes}[flushleft]
\footnotesize
\item Notes: The reference level is the aggregate welfare, in points, when the receiver's action exactly equals the true state. All other entries are percentages of this reference level. Counterfactual levels are obtained by replacing the receiver's action with the optimal action under the true model, sender's recommended action in the experiment, or optimal action when all models are in the feasible set under MEU or MLEU preferences.
\end{tablenotes}
\end{threeparttable}
\end{table}

%% file: results_artifacts/Figure_13_Delta_S_and_Delta_R_by_SVO_group.tex
\begin{figure}[htp!]
\centering
\begin{subfigure}[t]{0.48\textwidth}
\centering
\pgfplotstableread[row sep=\\]{%
bias m0 e0m e0p m1 e1m e1p\\
0 0.0336 0.1315 0.1557 -0.0987 0.5688 0.5649\\
1 0.2801 0.3324 0.3489 1.0849 0.6861 0.6100\\
3 1.0460 0.5094 0.4912 0.5418 0.5929 0.5598\\
}\treatciS

\begin{tikzpicture}
\begin{axis}[
  width=\textwidth,
  height=6cm,
  ylabel={Mean $\tilde{\Delta}^S$},
  xlabel={Bias treatment},
  symbolic x coords={0,1,3},
  xtick=data,
  ymajorgrids=true,
  grid style={dashed,gray!40},
  legend to name=SVOlegend,
  legend columns=2,
  legend style={draw=none},
  legend cell align=left,
]
\addplot+[only marks, mark=*, mark options={draw=SVO1, fill=SVO1},
  xshift=-4pt, error bars/.cd, y dir=both, y explicit, error bar style={black}]
  table[x=bias, y=m0, y error minus=e0m, y error plus=e0p]{\treatciS};
\addplot+[only marks, mark=square*, mark options={draw=SVO2, fill=SVO2},
  xshift=4pt, error bars/.cd, y dir=both, y explicit, error bar style={black}]
  table[x=bias, y=m1, y error minus=e1m, y error plus=e1p]{\treatciS};
\addlegendentry{Prosocial}
\addlegendentry{Proself}
\end{axis}
\end{tikzpicture}
\caption{$\tilde{\Delta}^S$ by SVO group.}
\label{fig:SVO_senders}
\end{subfigure}
\hfill
\begin{subfigure}[t]{0.48\textwidth}
\centering
\pgfplotstableread[row sep=\\]{%
bias m0 e0m e0p m1 e1m e1p\\
0 0.2086 0.3421 0.3223 0.0938 0.3421 0.3666\\
1 0.2820 0.2807 0.2512 0.5446 0.3966 0.3501\\
3 0.9891 0.3831 0.3934 1.0944 0.3457 0.3517\\
}\treatciR

\begin{tikzpicture}
\begin{axis}[
  width=\textwidth,
  height=6cm,
  ylabel={Mean $\tilde{\Delta}^R$},
  xlabel={Bias treatment},
  symbolic x coords={0,1,3},
  xtick=data,
  ymajorgrids=true,
  grid style={dashed,gray!40},
]
\addplot+[only marks, mark=*, mark options={draw=SVO1, fill=SVO1},
  xshift=-4pt, error bars/.cd, y dir=both, y explicit, error bar style={black}]
  table[x=bias, y=m0, y error minus=e0m, y error plus=e0p]{\treatciR};
\addplot+[only marks, mark=square*, mark options={draw=SVO2, fill=SVO2},
  xshift=4pt, error bars/.cd, y dir=both, y explicit, error bar style={black}]
  table[x=bias, y=m1, y error minus=e1m, y error plus=e1p]{\treatciR};

\end{axis}
\end{tikzpicture}
\caption{$\tilde{\Delta}^R$ by SVO group.}
\label{fig:SVO_receivers}
\end{subfigure}

\vspace{0.7em}
\pgfplotslegendfromname{SVOlegend}
\vspace{0.4em}
\begin{minipage}{0.9\textwidth}
\footnotesize\centering
Notes: Error bars are subject-level bootstrap 95\% confidence intervals.
\end{minipage}
\caption{$\tilde{\Delta}^S(b)$ and $\tilde{\Delta}^R(b)$ by SVO group and treatment.}
\label{fig:SVO_combined}
\end{figure}

%% file: results_artifacts/Figure_14_Observed_and_predicted_Delta_R.tex
\begin{figure}[p]
\centering
\begin{subfigure}{0.8\linewidth}
\centering
\resizebox{0.8\textwidth}{!}{\begin{tikzpicture}[x=1cm,y=1cm]
\fill[black] (0.450,-0.000) circle (0.10);
\fill[black] (1.100,-0.000) circle (0.10);
\fill[black] (1.750,-0.000) circle (0.10);
\fill[red] (0.450,-0.340) circle (0.10);
\fill[black] (1.100,-0.340) circle (0.10);
\fill[black] (1.750,-0.340) circle (0.10);
\fill[red] (0.450,-0.680) circle (0.10);
\fill[red] (1.100,-0.680) circle (0.10);
\fill[black] (1.750,-0.680) circle (0.10);
\fill[red] (0.450,-1.020) circle (0.10);
\fill[red] (1.100,-1.020) circle (0.10);
\fill[red] (1.750,-1.020) circle (0.10);
\fill[white] (2.840,0.153) rectangle (7.640,-1.120);
\draw[black!75, line width=0.55pt] (2.840,0.153) rectangle (7.640,-1.120);
\draw[black!70, line width=0.3pt] (5.240,0.153) -- (5.240,-1.120);
\draw[gray!30, line width=0.2pt] (3.107,0.153) -- (3.107,-1.120);
\node[font=\scriptsize, anchor=north] at (3.107,-1.200) {-4};
\draw[gray!30, line width=0.2pt] (3.640,0.153) -- (3.640,-1.120);
\node[font=\scriptsize, anchor=north] at (3.640,-1.200) {-3};
\draw[gray!30, line width=0.2pt] (4.173,0.153) -- (4.173,-1.120);
\node[font=\scriptsize, anchor=north] at (4.173,-1.200) {-2};
\draw[gray!30, line width=0.2pt] (4.707,0.153) -- (4.707,-1.120);
\node[font=\scriptsize, anchor=north] at (4.707,-1.200) {-1};
\draw[gray!30, line width=0.2pt] (5.240,0.153) -- (5.240,-1.120);
\node[font=\scriptsize, anchor=north] at (5.240,-1.200) {0};
\draw[gray!30, line width=0.2pt] (5.773,0.153) -- (5.773,-1.120);
\node[font=\scriptsize, anchor=north] at (5.773,-1.200) {1};
\draw[gray!30, line width=0.2pt] (6.307,0.153) -- (6.307,-1.120);
\node[font=\scriptsize, anchor=north] at (6.307,-1.200) {2};
\draw[gray!30, line width=0.2pt] (6.840,0.153) -- (6.840,-1.120);
\node[font=\scriptsize, anchor=north] at (6.840,-1.200) {3};
\draw[gray!30, line width=0.2pt] (7.373,0.153) -- (7.373,-1.120);
\node[font=\scriptsize, anchor=north] at (7.373,-1.200) {4};
\node[font=\small\bfseries, anchor=south] at (5.240,0.340) {No bias (0)};
\draw[negcol, line width=0.8pt] (4.537,-0.000) -- (5.054,-0.000);
\fill[negcol] (4.796,-0.000) circle (0.07);
\filldraw[fill=colorA, draw=black, line width=0.22pt] (5.240,0.085) -- (5.150,-0.075) -- (5.330,-0.075) -- cycle;
\draw[neucol, line width=0.8pt] (5.058,-0.340) -- (5.516,-0.340);
\fill[neucol] (5.287,-0.340) circle (0.07);
\filldraw[fill=colorA, draw=black, line width=0.22pt] (5.240,-0.255) -- (5.150,-0.415) -- (5.330,-0.415) -- cycle;
\draw[poscol, line width=0.8pt] (5.287,-0.680) -- (5.769,-0.680);
\fill[poscol] (5.528,-0.680) circle (0.07);
\filldraw[fill=colorA, draw=black, line width=0.22pt] (5.240,-0.595) -- (5.150,-0.755) -- (5.330,-0.755) -- cycle;
\draw[poscol, line width=0.8pt] (5.490,-1.020) -- (5.970,-1.020);
\fill[poscol] (5.730,-1.020) circle (0.07);
\filldraw[fill=colorA, draw=black, line width=0.22pt] (5.240,-0.935) -- (5.150,-1.095) -- (5.330,-1.095) -- cycle;
\draw[black!75, line width=0.55pt] (2.840,0.153) rectangle (7.640,-1.120);
\fill[white] (8.190,0.153) rectangle (12.990,-1.120);
\draw[black!75, line width=0.55pt] (8.190,0.153) rectangle (12.990,-1.120);
\draw[black!70, line width=0.3pt] (10.590,0.153) -- (10.590,-1.120);
\draw[gray!30, line width=0.2pt] (8.457,0.153) -- (8.457,-1.120);
\node[font=\scriptsize, anchor=north] at (8.457,-1.200) {-4};
\draw[gray!30, line width=0.2pt] (8.990,0.153) -- (8.990,-1.120);
\node[font=\scriptsize, anchor=north] at (8.990,-1.200) {-3};
\draw[gray!30, line width=0.2pt] (9.523,0.153) -- (9.523,-1.120);
\node[font=\scriptsize, anchor=north] at (9.523,-1.200) {-2};
\draw[gray!30, line width=0.2pt] (10.057,0.153) -- (10.057,-1.120);
\node[font=\scriptsize, anchor=north] at (10.057,-1.200) {-1};
\draw[gray!30, line width=0.2pt] (10.590,0.153) -- (10.590,-1.120);
\node[font=\scriptsize, anchor=north] at (10.590,-1.200) {0};
\draw[gray!30, line width=0.2pt] (11.123,0.153) -- (11.123,-1.120);
\node[font=\scriptsize, anchor=north] at (11.123,-1.200) {1};
\draw[gray!30, line width=0.2pt] (11.657,0.153) -- (11.657,-1.120);
\node[font=\scriptsize, anchor=north] at (11.657,-1.200) {2};
\draw[gray!30, line width=0.2pt] (12.190,0.153) -- (12.190,-1.120);
\node[font=\scriptsize, anchor=north] at (12.190,-1.200) {3};
\draw[gray!30, line width=0.2pt] (12.723,0.153) -- (12.723,-1.120);
\node[font=\scriptsize, anchor=north] at (12.723,-1.200) {4};
\node[font=\small\bfseries, anchor=south] at (10.590,0.340) {Low bias (1)};
\draw[neucol, line width=0.8pt] (10.211,-0.000) -- (10.637,-0.000);
\fill[neucol] (10.424,-0.000) circle (0.07);
\filldraw[fill=colorA, draw=black, line width=0.22pt] (10.756,0.085) -- (10.666,-0.075) -- (10.846,-0.075) -- cycle;
\draw[neucol, line width=0.8pt] (10.480,-0.340) -- (10.924,-0.340);
\fill[neucol] (10.702,-0.340) circle (0.07);
\filldraw[fill=colorA, draw=black, line width=0.22pt] (10.694,-0.255) -- (10.604,-0.415) -- (10.784,-0.415) -- cycle;
\draw[poscol, line width=0.8pt] (10.736,-0.680) -- (11.130,-0.680);
\fill[poscol] (10.933,-0.680) circle (0.07);
\filldraw[fill=colorA, draw=black, line width=0.22pt] (10.870,-0.595) -- (10.780,-0.755) -- (10.960,-0.755) -- cycle;
\draw[poscol, line width=0.8pt] (10.905,-1.020) -- (11.252,-1.020);
\fill[poscol] (11.078,-1.020) circle (0.07);
\filldraw[fill=colorA, draw=black, line width=0.22pt] (11.011,-0.935) -- (10.921,-1.095) -- (11.101,-1.095) -- cycle;
\draw[black!75, line width=0.55pt] (8.190,0.153) rectangle (12.990,-1.120);
\fill[white] (13.540,0.153) rectangle (18.340,-1.120);
\draw[black!75, line width=0.55pt] (13.540,0.153) rectangle (18.340,-1.120);
\draw[black!70, line width=0.3pt] (15.940,0.153) -- (15.940,-1.120);
\draw[gray!30, line width=0.2pt] (13.807,0.153) -- (13.807,-1.120);
\node[font=\scriptsize, anchor=north] at (13.807,-1.200) {-4};
\draw[gray!30, line width=0.2pt] (14.340,0.153) -- (14.340,-1.120);
\node[font=\scriptsize, anchor=north] at (14.340,-1.200) {-3};
\draw[gray!30, line width=0.2pt] (14.873,0.153) -- (14.873,-1.120);
\node[font=\scriptsize, anchor=north] at (14.873,-1.200) {-2};
\draw[gray!30, line width=0.2pt] (15.407,0.153) -- (15.407,-1.120);
\node[font=\scriptsize, anchor=north] at (15.407,-1.200) {-1};
\draw[gray!30, line width=0.2pt] (15.940,0.153) -- (15.940,-1.120);
\node[font=\scriptsize, anchor=north] at (15.940,-1.200) {0};
\draw[gray!30, line width=0.2pt] (16.473,0.153) -- (16.473,-1.120);
\node[font=\scriptsize, anchor=north] at (16.473,-1.200) {1};
\draw[gray!30, line width=0.2pt] (17.007,0.153) -- (17.007,-1.120);
\node[font=\scriptsize, anchor=north] at (17.007,-1.200) {2};
\draw[gray!30, line width=0.2pt] (17.540,0.153) -- (17.540,-1.120);
\node[font=\scriptsize, anchor=north] at (17.540,-1.200) {3};
\draw[gray!30, line width=0.2pt] (18.073,0.153) -- (18.073,-1.120);
\node[font=\scriptsize, anchor=north] at (18.073,-1.200) {4};
\node[font=\small\bfseries, anchor=south] at (15.940,0.340) {High bias (3)};
\draw[neucol, line width=0.8pt] (15.847,-0.000) -- (16.356,-0.000);
\fill[neucol] (16.102,-0.000) circle (0.07);
\filldraw[fill=colorA, draw=black, line width=0.22pt] (16.473,0.085) -- (16.383,-0.075) -- (16.563,-0.075) -- cycle;
\draw[poscol, line width=0.8pt] (16.371,-0.340) -- (16.814,-0.340);
\fill[poscol] (16.592,-0.340) circle (0.07);
\filldraw[fill=colorA, draw=black, line width=0.22pt] (17.540,-0.255) -- (17.450,-0.415) -- (17.630,-0.415) -- cycle;
\draw[poscol, line width=0.8pt] (16.413,-0.680) -- (16.964,-0.680);
\fill[poscol] (16.688,-0.680) circle (0.07);
\filldraw[fill=colorA, draw=black, line width=0.22pt] (17.540,-0.595) -- (17.450,-0.755) -- (17.630,-0.755) -- cycle;
\draw[poscol, line width=0.8pt] (16.455,-1.020) -- (16.830,-1.020);
\fill[poscol] (16.642,-1.020) circle (0.07);
\filldraw[fill=colorA, draw=black, line width=0.22pt] (17.540,-0.935) -- (17.450,-1.095) -- (17.630,-1.095) -- cycle;
\draw[black!75, line width=0.55pt] (13.540,0.153) rectangle (18.340,-1.120);
\filldraw[fill=colorA, draw=black, line width=0.22pt] (8.940,-1.835) -- (8.850,-1.995) -- (9.030,-1.995) -- cycle;
\node[font=\scriptsize, anchor=west] at (9.100,-1.920) {theoretical prediction};
\end{tikzpicture}}
\caption{by history.}
\label{fig:hyp2_whiskers}
\end{subfigure}
\vspace{0.8em}
\begin{subfigure}{0.8\linewidth}
\centering
\resizebox{0.8\textwidth}{!}{\begin{tikzpicture}[x=1cm,y=1cm]
\fill[black] (0.450,-0.000) circle (0.10);
\fill[black] (1.100,-0.000) circle (0.10);
\fill[black] (1.750,-0.000) circle (0.10);
\fill[black] (0.450,-0.340) circle (0.10);
\fill[black] (1.100,-0.340) circle (0.10);
\draw[black, line width=0.8pt] (1.630,-0.340) -- (1.870,-0.340);
\draw[red, line width=0.8pt] (0.330,-0.680) -- (0.570,-0.680);
\fill[black] (1.100,-0.680) circle (0.10);
\fill[black] (1.750,-0.680) circle (0.10);
\fill[black] (0.450,-1.020) circle (0.10);
\draw[black, line width=0.8pt] (0.980,-1.020) -- (1.220,-1.020);
\draw[black, line width=0.8pt] (1.630,-1.020) -- (1.870,-1.020);
\draw[red, line width=0.8pt] (0.330,-1.360) -- (0.570,-1.360);
\fill[black] (1.100,-1.360) circle (0.10);
\draw[black, line width=0.8pt] (1.630,-1.360) -- (1.870,-1.360);
\draw[red, line width=0.8pt] (0.330,-1.700) -- (0.570,-1.700);
\draw[red, line width=0.8pt] (0.980,-1.700) -- (1.220,-1.700);
\fill[black] (1.750,-1.700) circle (0.10);
\fill[red] (0.450,-2.040) circle (0.10);
\fill[black] (1.100,-2.040) circle (0.10);
\fill[black] (1.750,-2.040) circle (0.10);
\draw[black, line width=0.8pt] (0.330,-2.380) -- (0.570,-2.380);
\draw[black, line width=0.8pt] (0.980,-2.380) -- (1.220,-2.380);
\draw[black, line width=0.8pt] (1.630,-2.380) -- (1.870,-2.380);
\draw[red, line width=0.8pt] (0.330,-2.720) -- (0.570,-2.720);
\draw[black, line width=0.8pt] (0.980,-2.720) -- (1.220,-2.720);
\draw[black, line width=0.8pt] (1.630,-2.720) -- (1.870,-2.720);
\fill[red] (0.450,-3.060) circle (0.10);
\fill[black] (1.100,-3.060) circle (0.10);
\draw[black, line width=0.8pt] (1.630,-3.060) -- (1.870,-3.060);
\draw[red, line width=0.8pt] (0.330,-3.400) -- (0.570,-3.400);
\draw[red, line width=0.8pt] (0.980,-3.400) -- (1.220,-3.400);
\draw[black, line width=0.8pt] (1.630,-3.400) -- (1.870,-3.400);
\fill[red] (0.450,-3.740) circle (0.10);
\draw[red, line width=0.8pt] (0.980,-3.740) -- (1.220,-3.740);
\fill[black] (1.750,-3.740) circle (0.10);
\draw[red, line width=0.8pt] (0.330,-4.080) -- (0.570,-4.080);
\draw[red, line width=0.8pt] (0.980,-4.080) -- (1.220,-4.080);
\draw[red, line width=0.8pt] (1.630,-4.080) -- (1.870,-4.080);
\fill[red] (0.450,-4.420) circle (0.10);
\fill[red] (1.100,-4.420) circle (0.10);
\fill[black] (1.750,-4.420) circle (0.10);
\fill[red] (0.450,-4.760) circle (0.10);
\draw[black, line width=0.8pt] (0.980,-4.760) -- (1.220,-4.760);
\draw[black, line width=0.8pt] (1.630,-4.760) -- (1.870,-4.760);
\fill[red] (0.450,-5.100) circle (0.10);
\draw[red, line width=0.8pt] (0.980,-5.100) -- (1.220,-5.100);
\draw[black, line width=0.8pt] (1.630,-5.100) -- (1.870,-5.100);
\fill[red] (0.450,-5.440) circle (0.10);
\draw[red, line width=0.8pt] (0.980,-5.440) -- (1.220,-5.440);
\draw[red, line width=0.8pt] (1.630,-5.440) -- (1.870,-5.440);
\fill[red] (0.450,-5.780) circle (0.10);
\fill[red] (1.100,-5.780) circle (0.10);
\draw[black, line width=0.8pt] (1.630,-5.780) -- (1.870,-5.780);
\fill[red] (0.450,-6.120) circle (0.10);
\fill[red] (1.100,-6.120) circle (0.10);
\draw[red, line width=0.8pt] (1.630,-6.120) -- (1.870,-6.120);
\fill[red] (0.450,-6.460) circle (0.10);
\fill[red] (1.100,-6.460) circle (0.10);
\fill[red] (1.750,-6.460) circle (0.10);
\node[font=\small\bfseries, anchor=south] at (1.100,0.340) {};
\fill[white] (2.840,0.153) rectangle (7.640,-6.560);
\draw[black!75, line width=0.55pt] (2.840,0.153) rectangle (7.640,-6.560);
\draw[black!70, line width=0.3pt] (5.240,0.153) -- (5.240,-6.560);
\draw[gray!30, line width=0.2pt] (3.107,0.153) -- (3.107,-6.560);
\node[font=\scriptsize, anchor=north] at (3.107,-6.640) {-4};
\draw[gray!30, line width=0.2pt] (3.640,0.153) -- (3.640,-6.560);
\node[font=\scriptsize, anchor=north] at (3.640,-6.640) {-3};
\draw[gray!30, line width=0.2pt] (4.173,0.153) -- (4.173,-6.560);
\node[font=\scriptsize, anchor=north] at (4.173,-6.640) {-2};
\draw[gray!30, line width=0.2pt] (4.707,0.153) -- (4.707,-6.560);
\node[font=\scriptsize, anchor=north] at (4.707,-6.640) {-1};
\draw[gray!30, line width=0.2pt] (5.240,0.153) -- (5.240,-6.560);
\node[font=\scriptsize, anchor=north] at (5.240,-6.640) {0};
\draw[gray!30, line width=0.2pt] (5.773,0.153) -- (5.773,-6.560);
\node[font=\scriptsize, anchor=north] at (5.773,-6.640) {1};
\draw[gray!30, line width=0.2pt] (6.307,0.153) -- (6.307,-6.560);
\node[font=\scriptsize, anchor=north] at (6.307,-6.640) {2};
\draw[gray!30, line width=0.2pt] (6.840,0.153) -- (6.840,-6.560);
\node[font=\scriptsize, anchor=north] at (6.840,-6.640) {3};
\draw[gray!30, line width=0.2pt] (7.373,0.153) -- (7.373,-6.560);
\node[font=\scriptsize, anchor=north] at (7.373,-6.640) {4};
\node[font=\small\bfseries, anchor=south] at (5.240,0.340) {No bias (0)};
\draw[neucol, line width=0.8pt] (2.505,-0.000) -- (6.553,-0.000);
\fill[neucol] (4.529,-0.000) circle (0.07);
\filldraw[fill=colorA, draw=black, line width=0.22pt] (5.240,0.085) -- (5.150,-0.075) -- (5.330,-0.075) -- cycle;
\draw[negcol, line width=0.8pt] (4.307,-0.340) -- (5.056,-0.340);
\fill[negcol] (4.681,-0.340) circle (0.07);
\filldraw[fill=colorA, draw=black, line width=0.22pt] (5.240,-0.255) -- (5.150,-0.415) -- (5.330,-0.415) -- cycle;
\draw[neucol, line width=0.8pt] (4.471,-0.680) -- (5.440,-0.680);
\fill[neucol] (4.956,-0.680) circle (0.07);
\filldraw[fill=colorA, draw=black, line width=0.22pt] (5.240,-0.595) -- (5.150,-0.755) -- (5.330,-0.755) -- cycle;
\draw[neucol, line width=0.8pt] (4.630,-1.020) -- (5.275,-1.020);
\fill[neucol] (4.953,-1.020) circle (0.07);
\filldraw[fill=colorA, draw=black, line width=0.22pt] (5.240,-0.935) -- (5.150,-1.095) -- (5.330,-1.095) -- cycle;
\draw[neucol, line width=0.8pt] (4.638,-1.360) -- (5.461,-1.360);
\fill[neucol] (5.050,-1.360) circle (0.07);
\filldraw[fill=colorA, draw=black, line width=0.22pt] (5.240,-1.275) -- (5.150,-1.435) -- (5.330,-1.435) -- cycle;
\draw[neucol, line width=0.8pt] (4.916,-1.700) -- (5.831,-1.700);
\fill[neucol] (5.373,-1.700) circle (0.07);
\filldraw[fill=colorA, draw=black, line width=0.22pt] (5.240,-1.615) -- (5.150,-1.775) -- (5.330,-1.775) -- cycle;
\draw[neucol, line width=0.8pt] (5.197,-2.040) -- (5.923,-2.040);
\fill[neucol] (5.560,-2.040) circle (0.07);
\filldraw[fill=colorA, draw=black, line width=0.22pt] (5.240,-1.955) -- (5.150,-2.115) -- (5.330,-2.115) -- cycle;
\draw[neucol, line width=0.8pt] (4.048,-2.380) -- (5.462,-2.380);
\fill[neucol] (4.755,-2.380) circle (0.07);
\filldraw[fill=colorA, draw=black, line width=0.22pt] (5.240,-2.295) -- (5.150,-2.455) -- (5.330,-2.455) -- cycle;
\draw[neucol, line width=0.8pt] (4.966,-2.720) -- (5.457,-2.720);
\fill[neucol] (5.212,-2.720) circle (0.07);
\filldraw[fill=colorA, draw=black, line width=0.22pt] (5.240,-2.635) -- (5.150,-2.795) -- (5.330,-2.795) -- cycle;
\draw[neucol, line width=0.8pt] (5.165,-3.060) -- (5.742,-3.060);
\fill[neucol] (5.453,-3.060) circle (0.07);
\filldraw[fill=colorA, draw=black, line width=0.22pt] (5.240,-2.975) -- (5.150,-3.135) -- (5.330,-3.135) -- cycle;
\draw[neucol, line width=0.8pt] (5.091,-3.400) -- (6.122,-3.400);
\fill[neucol] (5.607,-3.400) circle (0.07);
\filldraw[fill=colorA, draw=black, line width=0.22pt] (5.240,-3.315) -- (5.150,-3.475) -- (5.330,-3.475) -- cycle;
\draw[neucol, line width=0.8pt] (4.884,-3.740) -- (5.738,-3.740);
\fill[neucol] (5.311,-3.740) circle (0.07);
\filldraw[fill=colorA, draw=black, line width=0.22pt] (5.240,-3.655) -- (5.150,-3.815) -- (5.330,-3.815) -- cycle;
\draw[neucol, line width=0.8pt] (4.909,-4.080) -- (6.391,-4.080);
\fill[neucol] (5.650,-4.080) circle (0.07);
\filldraw[fill=colorA, draw=black, line width=0.22pt] (5.240,-3.995) -- (5.150,-4.155) -- (5.330,-4.155) -- cycle;
\draw[neucol, line width=0.8pt] (4.627,-4.420) -- (5.853,-4.420);
\fill[neucol] (5.240,-4.420) circle (0.07);
\filldraw[fill=colorA, draw=black, line width=0.22pt] (5.240,-4.335) -- (5.150,-4.495) -- (5.330,-4.495) -- cycle;
\draw[neucol, line width=0.8pt] (5.005,-4.760) -- (6.542,-4.760);
\fill[neucol] (5.773,-4.760) circle (0.07);
\filldraw[fill=colorA, draw=black, line width=0.22pt] (5.240,-4.675) -- (5.150,-4.835) -- (5.330,-4.835) -- cycle;
\draw[neucol, line width=0.8pt] (4.884,-5.100) -- (6.054,-5.100);
\fill[neucol] (5.469,-5.100) circle (0.07);
\filldraw[fill=colorA, draw=black, line width=0.22pt] (5.240,-5.015) -- (5.150,-5.175) -- (5.330,-5.175) -- cycle;
\draw[poscol, line width=0.8pt] (5.467,-5.440) -- (6.225,-5.440);
\fill[poscol] (5.846,-5.440) circle (0.07);
\filldraw[fill=colorA, draw=black, line width=0.22pt] (5.240,-5.355) -- (5.150,-5.515) -- (5.330,-5.515) -- cycle;
\draw[poscol, line width=0.8pt] (5.570,-5.780) -- (6.469,-5.780);
\fill[poscol] (6.019,-5.780) circle (0.07);
\filldraw[fill=colorA, draw=black, line width=0.22pt] (5.240,-5.695) -- (5.150,-5.855) -- (5.330,-5.855) -- cycle;
\draw[poscol, line width=0.8pt] (5.383,-6.120) -- (6.052,-6.120);
\fill[poscol] (5.717,-6.120) circle (0.07);
\filldraw[fill=colorA, draw=black, line width=0.22pt] (5.240,-6.035) -- (5.150,-6.195) -- (5.330,-6.195) -- cycle;
\draw[neucol, line width=0.8pt] (5.136,-6.460) -- (6.011,-6.460);
\fill[neucol] (5.573,-6.460) circle (0.07);
\filldraw[fill=colorA, draw=black, line width=0.22pt] (5.240,-6.375) -- (5.150,-6.535) -- (5.330,-6.535) -- cycle;
\draw[black!75, line width=0.55pt] (2.840,0.153) rectangle (7.640,-6.560);
\fill[white] (8.190,0.153) rectangle (12.990,-6.560);
\draw[black!75, line width=0.55pt] (8.190,0.153) rectangle (12.990,-6.560);
\draw[black!70, line width=0.3pt] (10.590,0.153) -- (10.590,-6.560);
\draw[gray!30, line width=0.2pt] (8.457,0.153) -- (8.457,-6.560);
\node[font=\scriptsize, anchor=north] at (8.457,-6.640) {-4};
\draw[gray!30, line width=0.2pt] (8.990,0.153) -- (8.990,-6.560);
\node[font=\scriptsize, anchor=north] at (8.990,-6.640) {-3};
\draw[gray!30, line width=0.2pt] (9.523,0.153) -- (9.523,-6.560);
\node[font=\scriptsize, anchor=north] at (9.523,-6.640) {-2};
\draw[gray!30, line width=0.2pt] (10.057,0.153) -- (10.057,-6.560);
\node[font=\scriptsize, anchor=north] at (10.057,-6.640) {-1};
\draw[gray!30, line width=0.2pt] (10.590,0.153) -- (10.590,-6.560);
\node[font=\scriptsize, anchor=north] at (10.590,-6.640) {0};
\draw[gray!30, line width=0.2pt] (11.123,0.153) -- (11.123,-6.560);
\node[font=\scriptsize, anchor=north] at (11.123,-6.640) {1};
\draw[gray!30, line width=0.2pt] (11.657,0.153) -- (11.657,-6.560);
\node[font=\scriptsize, anchor=north] at (11.657,-6.640) {2};
\draw[gray!30, line width=0.2pt] (12.190,0.153) -- (12.190,-6.560);
\node[font=\scriptsize, anchor=north] at (12.190,-6.640) {3};
\draw[gray!30, line width=0.2pt] (12.723,0.153) -- (12.723,-6.560);
\node[font=\scriptsize, anchor=north] at (12.723,-6.640) {4};
\node[font=\small\bfseries, anchor=south] at (10.590,0.340) {Low bias (1)};
\draw[neucol, line width=0.8pt] (9.402,-0.000) -- (11.168,-0.000);
\fill[neucol] (10.285,-0.000) circle (0.07);
\filldraw[fill=colorA, draw=black, line width=0.22pt] (10.590,0.085) -- (10.500,-0.075) -- (10.680,-0.075) -- cycle;
\draw[neucol, line width=0.8pt] (9.901,-0.340) -- (10.674,-0.340);
\fill[neucol] (10.287,-0.340) circle (0.07);
\filldraw[fill=colorA, draw=black, line width=0.22pt] (10.768,-0.255) -- (10.678,-0.415) -- (10.858,-0.415) -- cycle;
\draw[neucol, line width=0.8pt] (9.708,-0.680) -- (10.798,-0.680);
\fill[neucol] (10.253,-0.680) circle (0.07);
\filldraw[fill=colorA, draw=black, line width=0.22pt] (10.590,-0.595) -- (10.500,-0.755) -- (10.680,-0.755) -- cycle;
\draw[neucol, line width=0.8pt] (10.009,-1.020) -- (10.621,-1.020);
\fill[neucol] (10.315,-1.020) circle (0.07);
\filldraw[fill=colorA, draw=black, line width=0.22pt] (10.768,-0.935) -- (10.678,-1.095) -- (10.858,-1.095) -- cycle;
\draw[neucol, line width=0.8pt] (10.201,-1.360) -- (11.072,-1.360);
\fill[neucol] (10.636,-1.360) circle (0.07);
\filldraw[fill=colorA, draw=black, line width=0.22pt] (10.590,-1.275) -- (10.500,-1.435) -- (10.680,-1.435) -- cycle;
\draw[neucol, line width=0.8pt] (10.323,-1.700) -- (11.283,-1.700);
\fill[neucol] (10.803,-1.700) circle (0.07);
\filldraw[fill=colorA, draw=black, line width=0.22pt] (10.590,-1.615) -- (10.500,-1.775) -- (10.680,-1.775) -- cycle;
\draw[neucol, line width=0.8pt] (9.702,-2.040) -- (13.078,-2.040);
\fill[neucol] (11.390,-2.040) circle (0.07);
\filldraw[fill=colorA, draw=black, line width=0.22pt] (10.590,-1.955) -- (10.500,-2.115) -- (10.680,-2.115) -- cycle;
\draw[neucol, line width=0.8pt] (10.494,-2.380) -- (11.017,-2.380);
\fill[neucol] (10.756,-2.380) circle (0.07);
\filldraw[fill=colorA, draw=black, line width=0.22pt] (10.768,-2.295) -- (10.678,-2.455) -- (10.858,-2.455) -- cycle;
\draw[neucol, line width=0.8pt] (10.477,-2.720) -- (11.013,-2.720);
\fill[neucol] (10.745,-2.720) circle (0.07);
\filldraw[fill=colorA, draw=black, line width=0.22pt] (10.590,-2.635) -- (10.500,-2.795) -- (10.680,-2.795) -- cycle;
\draw[neucol, line width=0.8pt] (10.152,-3.060) -- (10.949,-3.060);
\fill[neucol] (10.550,-3.060) circle (0.07);
\filldraw[fill=colorA, draw=black, line width=0.22pt] (10.857,-2.975) -- (10.767,-3.135) -- (10.947,-3.135) -- cycle;
\draw[neucol, line width=0.8pt] (10.504,-3.400) -- (11.186,-3.400);
\fill[neucol] (10.845,-3.400) circle (0.07);
\filldraw[fill=colorA, draw=black, line width=0.22pt] (10.590,-3.315) -- (10.500,-3.475) -- (10.680,-3.475) -- cycle;
\draw[neucol, line width=0.8pt] (10.464,-3.740) -- (11.036,-3.740);
\fill[neucol] (10.750,-3.740) circle (0.07);
\filldraw[fill=colorA, draw=black, line width=0.22pt] (10.857,-3.655) -- (10.767,-3.815) -- (10.947,-3.815) -- cycle;
\draw[poscol, line width=0.8pt] (10.653,-4.080) -- (11.594,-4.080);
\fill[poscol] (11.123,-4.080) circle (0.07);
\filldraw[fill=colorA, draw=black, line width=0.22pt] (10.590,-3.995) -- (10.500,-4.155) -- (10.680,-4.155) -- cycle;
\draw[poscol, line width=0.8pt] (10.705,-4.420) -- (12.075,-4.420);
\fill[poscol] (11.390,-4.420) circle (0.07);
\filldraw[fill=colorA, draw=black, line width=0.22pt] (11.123,-4.335) -- (11.033,-4.495) -- (11.213,-4.495) -- cycle;
\draw[poscol, line width=0.8pt] (10.921,-4.760) -- (11.818,-4.760);
\fill[poscol] (11.369,-4.760) circle (0.07);
\filldraw[fill=colorA, draw=black, line width=0.22pt] (10.990,-4.675) -- (10.900,-4.835) -- (11.080,-4.835) -- cycle;
\draw[poscol, line width=0.8pt] (10.679,-5.100) -- (11.567,-5.100);
\fill[poscol] (11.123,-5.100) circle (0.07);
\filldraw[fill=colorA, draw=black, line width=0.22pt] (11.123,-5.015) -- (11.033,-5.175) -- (11.213,-5.175) -- cycle;
\draw[poscol, line width=0.8pt] (10.799,-5.440) -- (11.241,-5.440);
\fill[poscol] (11.020,-5.440) circle (0.07);
\filldraw[fill=colorA, draw=black, line width=0.22pt] (11.123,-5.355) -- (11.033,-5.515) -- (11.213,-5.515) -- cycle;
\draw[neucol, line width=0.8pt] (10.585,-5.780) -- (11.395,-5.780);
\fill[neucol] (10.990,-5.780) circle (0.07);
\filldraw[fill=colorA, draw=black, line width=0.22pt] (11.123,-5.695) -- (11.033,-5.855) -- (11.213,-5.855) -- cycle;
\draw[poscol, line width=0.8pt] (10.838,-6.120) -- (11.439,-6.120);
\fill[poscol] (11.139,-6.120) circle (0.07);
\filldraw[fill=colorA, draw=black, line width=0.22pt] (11.123,-6.035) -- (11.033,-6.195) -- (11.213,-6.195) -- cycle;
\draw[neucol, line width=0.8pt] (10.331,-6.460) -- (11.560,-6.460);
\fill[neucol] (10.946,-6.460) circle (0.07);
\filldraw[fill=colorA, draw=black, line width=0.22pt] (11.123,-6.375) -- (11.033,-6.535) -- (11.213,-6.535) -- cycle;
\draw[black!75, line width=0.55pt] (8.190,0.153) rectangle (12.990,-6.560);
\fill[white] (13.540,0.153) rectangle (18.340,-6.560);
\draw[black!75, line width=0.55pt] (13.540,0.153) rectangle (18.340,-6.560);
\draw[black!70, line width=0.3pt] (15.940,0.153) -- (15.940,-6.560);
\draw[gray!30, line width=0.2pt] (13.807,0.153) -- (13.807,-6.560);
\node[font=\scriptsize, anchor=north] at (13.807,-6.640) {-4};
\draw[gray!30, line width=0.2pt] (14.340,0.153) -- (14.340,-6.560);
\node[font=\scriptsize, anchor=north] at (14.340,-6.640) {-3};
\draw[gray!30, line width=0.2pt] (14.873,0.153) -- (14.873,-6.560);
\node[font=\scriptsize, anchor=north] at (14.873,-6.640) {-2};
\draw[gray!30, line width=0.2pt] (15.407,0.153) -- (15.407,-6.560);
\node[font=\scriptsize, anchor=north] at (15.407,-6.640) {-1};
\draw[gray!30, line width=0.2pt] (15.940,0.153) -- (15.940,-6.560);
\node[font=\scriptsize, anchor=north] at (15.940,-6.640) {0};
\draw[gray!30, line width=0.2pt] (16.473,0.153) -- (16.473,-6.560);
\node[font=\scriptsize, anchor=north] at (16.473,-6.640) {1};
\draw[gray!30, line width=0.2pt] (17.007,0.153) -- (17.007,-6.560);
\node[font=\scriptsize, anchor=north] at (17.007,-6.640) {2};
\draw[gray!30, line width=0.2pt] (17.540,0.153) -- (17.540,-6.560);
\node[font=\scriptsize, anchor=north] at (17.540,-6.640) {3};
\draw[gray!30, line width=0.2pt] (18.073,0.153) -- (18.073,-6.560);
\node[font=\scriptsize, anchor=north] at (18.073,-6.640) {4};
\node[font=\small\bfseries, anchor=south] at (15.940,0.340) {High bias (3)};
\draw[neucol, line width=0.8pt] (14.774,-0.000) -- (16.718,-0.000);
\fill[neucol] (15.746,-0.000) circle (0.07);
\filldraw[fill=colorA, draw=black, line width=0.22pt] (16.473,0.085) -- (16.383,-0.075) -- (16.563,-0.075) -- cycle;
\draw[neucol, line width=0.8pt] (15.719,-0.340) -- (16.405,-0.340);
\fill[neucol] (16.062,-0.340) circle (0.07);
\filldraw[fill=colorA, draw=black, line width=0.22pt] (16.473,-0.255) -- (16.383,-0.415) -- (16.563,-0.415) -- cycle;
\draw[neucol, line width=0.8pt] (15.686,-0.680) -- (16.533,-0.680);
\fill[neucol] (16.110,-0.680) circle (0.07);
\filldraw[fill=colorA, draw=black, line width=0.22pt] (17.540,-0.595) -- (17.450,-0.755) -- (17.630,-0.755) -- cycle;
\draw[neucol, line width=0.8pt] (15.618,-1.020) -- (16.574,-1.020);
\fill[neucol] (16.096,-1.020) circle (0.07);
\filldraw[fill=colorA, draw=black, line width=0.22pt] (16.473,-0.935) -- (16.383,-1.095) -- (16.563,-1.095) -- cycle;
\draw[poscol, line width=0.8pt] (16.082,-1.360) -- (17.098,-1.360);
\fill[poscol] (16.590,-1.360) circle (0.07);
\filldraw[fill=colorA, draw=black, line width=0.22pt] (17.540,-1.275) -- (17.450,-1.435) -- (17.630,-1.435) -- cycle;
\draw[neucol, line width=0.8pt] (15.684,-1.700) -- (17.130,-1.700);
\fill[neucol] (16.407,-1.700) circle (0.07);
\filldraw[fill=colorA, draw=black, line width=0.22pt] (17.540,-1.615) -- (17.450,-1.775) -- (17.630,-1.775) -- cycle;
\draw[neucol, line width=0.8pt] (15.313,-2.040) -- (17.279,-2.040);
\fill[neucol] (16.296,-2.040) circle (0.07);
\filldraw[fill=colorA, draw=black, line width=0.22pt] (17.540,-1.955) -- (17.450,-2.115) -- (17.630,-2.115) -- cycle;
\draw[poscol, line width=0.8pt] (15.966,-2.380) -- (16.614,-2.380);
\fill[poscol] (16.290,-2.380) circle (0.07);
\filldraw[fill=colorA, draw=black, line width=0.22pt] (16.473,-2.295) -- (16.383,-2.455) -- (16.563,-2.455) -- cycle;
\draw[poscol, line width=0.8pt] (16.228,-2.720) -- (16.973,-2.720);
\fill[poscol] (16.600,-2.720) circle (0.07);
\filldraw[fill=colorA, draw=black, line width=0.22pt] (17.540,-2.635) -- (17.450,-2.795) -- (17.630,-2.795) -- cycle;
\draw[poscol, line width=0.8pt] (16.265,-3.060) -- (17.186,-3.060);
\fill[poscol] (16.725,-3.060) circle (0.07);
\filldraw[fill=colorA, draw=black, line width=0.22pt] (17.540,-2.975) -- (17.450,-3.135) -- (17.630,-3.135) -- cycle;
\draw[poscol, line width=0.8pt] (16.501,-3.400) -- (17.228,-3.400);
\fill[poscol] (16.864,-3.400) circle (0.07);
\filldraw[fill=colorA, draw=black, line width=0.22pt] (17.540,-3.315) -- (17.450,-3.475) -- (17.630,-3.475) -- cycle;
\draw[poscol, line width=0.8pt] (15.995,-3.740) -- (17.116,-3.740);
\fill[poscol] (16.555,-3.740) circle (0.07);
\filldraw[fill=colorA, draw=black, line width=0.22pt] (17.540,-3.655) -- (17.450,-3.815) -- (17.630,-3.815) -- cycle;
\draw[poscol, line width=0.8pt] (16.004,-4.080) -- (16.810,-4.080);
\fill[poscol] (16.407,-4.080) circle (0.07);
\filldraw[fill=colorA, draw=black, line width=0.22pt] (17.540,-3.995) -- (17.450,-4.155) -- (17.630,-4.155) -- cycle;
\draw[poscol, line width=0.8pt] (16.074,-4.420) -- (17.330,-4.420);
\fill[poscol] (16.702,-4.420) circle (0.07);
\filldraw[fill=colorA, draw=black, line width=0.22pt] (17.540,-4.335) -- (17.450,-4.495) -- (17.630,-4.495) -- cycle;
\draw[poscol, line width=0.8pt] (16.488,-4.760) -- (17.725,-4.760);
\fill[poscol] (17.107,-4.760) circle (0.07);
\filldraw[fill=colorA, draw=black, line width=0.22pt] (17.540,-4.675) -- (17.450,-4.835) -- (17.630,-4.835) -- cycle;
\draw[poscol, line width=0.8pt] (16.447,-5.100) -- (17.456,-5.100);
\fill[poscol] (16.951,-5.100) circle (0.07);
\filldraw[fill=colorA, draw=black, line width=0.22pt] (17.540,-5.015) -- (17.450,-5.175) -- (17.630,-5.175) -- cycle;
\draw[poscol, line width=0.8pt] (16.466,-5.440) -- (16.960,-5.440);
\fill[poscol] (16.713,-5.440) circle (0.07);
\filldraw[fill=colorA, draw=black, line width=0.22pt] (17.540,-5.355) -- (17.450,-5.515) -- (17.630,-5.515) -- cycle;
\draw[neucol, line width=0.8pt] (15.686,-5.780) -- (17.210,-5.780);
\fill[neucol] (16.448,-5.780) circle (0.07);
\filldraw[fill=colorA, draw=black, line width=0.22pt] (17.540,-5.695) -- (17.450,-5.855) -- (17.630,-5.855) -- cycle;
\draw[poscol, line width=0.8pt] (16.245,-6.120) -- (16.887,-6.120);
\fill[poscol] (16.566,-6.120) circle (0.07);
\filldraw[fill=colorA, draw=black, line width=0.22pt] (17.540,-6.035) -- (17.450,-6.195) -- (17.630,-6.195) -- cycle;
\draw[poscol, line width=0.8pt] (16.594,-6.460) -- (17.952,-6.460);
\fill[poscol] (17.273,-6.460) circle (0.07);
\filldraw[fill=colorA, draw=black, line width=0.22pt] (17.540,-6.375) -- (17.450,-6.535) -- (17.630,-6.535) -- cycle;
\draw[black!75, line width=0.55pt] (13.540,0.153) rectangle (18.340,-6.560);
\filldraw[fill=colorA, draw=black, line width=0.22pt] (8.940,-7.275) -- (8.850,-7.435) -- (9.030,-7.435) -- cycle;
\node[font=\scriptsize, anchor=west] at (9.100,-7.360) {theoretical prediction};
\end{tikzpicture}}
\caption{by history and model.}
\label{fig:hyp2_whiskers_h}
\end{subfigure}
\vspace{1.0em}
\begin{minipage}{0.95\textwidth}
\footnotesize
Notes: Whiskers indicate pointwise 95\% confidence intervals with standard errors clustered by receiver. Green and red whiskers indicate positive and negative estimates, respectively, whose pointwise 95\% confidence intervals exclude zero.
\end{minipage}
\caption{Observed $\tilde{\Delta}^R(h,b),\tilde{\Delta}_m^R(h,b)$ and predicted $\Delta^R(h,b),\Delta_m^R(h,b)$.}
\label{fig:hyp2_whiskers_combined}
\end{figure}

%% file: results_artifacts/Figure_15_Observed_and_predicted_Delta_S.tex
\begin{figure}[p]
\centering
\begin{subfigure}{0.8\linewidth}
\centering
\resizebox{0.8\textwidth}{!}{\begin{tikzpicture}[x=1cm,y=1cm]
\fill[black] (0.450,-0.000) circle (0.10);
\fill[black] (1.100,-0.000) circle (0.10);
\fill[black] (1.750,-0.000) circle (0.10);
\fill[red] (0.450,-0.340) circle (0.10);
\fill[black] (1.100,-0.340) circle (0.10);
\fill[black] (1.750,-0.340) circle (0.10);
\fill[red] (0.450,-0.680) circle (0.10);
\fill[red] (1.100,-0.680) circle (0.10);
\fill[black] (1.750,-0.680) circle (0.10);
\fill[red] (0.450,-1.020) circle (0.10);
\fill[red] (1.100,-1.020) circle (0.10);
\fill[red] (1.750,-1.020) circle (0.10);
\fill[white] (2.840,0.153) rectangle (7.640,-1.120);
\draw[black!75, line width=0.55pt] (2.840,0.153) rectangle (7.640,-1.120);
\draw[black!70, line width=0.3pt] (5.240,0.153) -- (5.240,-1.120);
\draw[gray!30, line width=0.2pt] (3.107,0.153) -- (3.107,-1.120);
\node[font=\scriptsize, anchor=north] at (3.107,-1.200) {-4};
\draw[gray!30, line width=0.2pt] (3.640,0.153) -- (3.640,-1.120);
\node[font=\scriptsize, anchor=north] at (3.640,-1.200) {-3};
\draw[gray!30, line width=0.2pt] (4.173,0.153) -- (4.173,-1.120);
\node[font=\scriptsize, anchor=north] at (4.173,-1.200) {-2};
\draw[gray!30, line width=0.2pt] (4.707,0.153) -- (4.707,-1.120);
\node[font=\scriptsize, anchor=north] at (4.707,-1.200) {-1};
\draw[gray!30, line width=0.2pt] (5.240,0.153) -- (5.240,-1.120);
\node[font=\scriptsize, anchor=north] at (5.240,-1.200) {0};
\draw[gray!30, line width=0.2pt] (5.773,0.153) -- (5.773,-1.120);
\node[font=\scriptsize, anchor=north] at (5.773,-1.200) {1};
\draw[gray!30, line width=0.2pt] (6.307,0.153) -- (6.307,-1.120);
\node[font=\scriptsize, anchor=north] at (6.307,-1.200) {2};
\draw[gray!30, line width=0.2pt] (6.840,0.153) -- (6.840,-1.120);
\node[font=\scriptsize, anchor=north] at (6.840,-1.200) {3};
\draw[gray!30, line width=0.2pt] (7.373,0.153) -- (7.373,-1.120);
\node[font=\scriptsize, anchor=north] at (7.373,-1.200) {4};
\node[font=\small\bfseries, anchor=south] at (5.240,0.340) {No bias (0)};
\draw[neucol, line width=0.8pt] (5.181,-0.000) -- (5.448,-0.000);
\fill[neucol] (5.314,-0.000) circle (0.07);
\filldraw[fill=colorB, draw=black, line width=0.22pt] (5.240,0.085) -- (5.150,-0.075) -- (5.330,-0.075) -- cycle;
\draw[neucol, line width=0.8pt] (4.884,-0.340) -- (5.516,-0.340);
\fill[neucol] (5.200,-0.340) circle (0.07);
\filldraw[fill=colorB, draw=black, line width=0.22pt] (5.240,-0.255) -- (5.150,-0.415) -- (5.330,-0.415) -- cycle;
\draw[neucol, line width=0.8pt] (5.055,-0.680) -- (5.551,-0.680);
\fill[neucol] (5.303,-0.680) circle (0.07);
\filldraw[fill=colorB, draw=black, line width=0.22pt] (5.240,-0.595) -- (5.150,-0.755) -- (5.330,-0.755) -- cycle;
\draw[neucol, line width=0.8pt] (4.770,-1.020) -- (5.400,-1.020);
\fill[neucol] (5.085,-1.020) circle (0.07);
\filldraw[fill=colorB, draw=black, line width=0.22pt] (5.240,-0.935) -- (5.150,-1.095) -- (5.330,-1.095) -- cycle;
\draw[black!75, line width=0.55pt] (2.840,0.153) rectangle (7.640,-1.120);
\fill[white] (8.190,0.153) rectangle (12.990,-1.120);
\draw[black!75, line width=0.55pt] (8.190,0.153) rectangle (12.990,-1.120);
\draw[black!70, line width=0.3pt] (10.590,0.153) -- (10.590,-1.120);
\draw[gray!30, line width=0.2pt] (8.457,0.153) -- (8.457,-1.120);
\node[font=\scriptsize, anchor=north] at (8.457,-1.200) {-4};
\draw[gray!30, line width=0.2pt] (8.990,0.153) -- (8.990,-1.120);
\node[font=\scriptsize, anchor=north] at (8.990,-1.200) {-3};
\draw[gray!30, line width=0.2pt] (9.523,0.153) -- (9.523,-1.120);
\node[font=\scriptsize, anchor=north] at (9.523,-1.200) {-2};
\draw[gray!30, line width=0.2pt] (10.057,0.153) -- (10.057,-1.120);
\node[font=\scriptsize, anchor=north] at (10.057,-1.200) {-1};
\draw[gray!30, line width=0.2pt] (10.590,0.153) -- (10.590,-1.120);
\node[font=\scriptsize, anchor=north] at (10.590,-1.200) {0};
\draw[gray!30, line width=0.2pt] (11.123,0.153) -- (11.123,-1.120);
\node[font=\scriptsize, anchor=north] at (11.123,-1.200) {1};
\draw[gray!30, line width=0.2pt] (11.657,0.153) -- (11.657,-1.120);
\node[font=\scriptsize, anchor=north] at (11.657,-1.200) {2};
\draw[gray!30, line width=0.2pt] (12.190,0.153) -- (12.190,-1.120);
\node[font=\scriptsize, anchor=north] at (12.190,-1.200) {3};
\draw[gray!30, line width=0.2pt] (12.723,0.153) -- (12.723,-1.120);
\node[font=\scriptsize, anchor=north] at (12.723,-1.200) {4};
\node[font=\small\bfseries, anchor=south] at (10.590,0.340) {Low bias (1)};
\draw[neucol, line width=0.8pt] (10.588,-0.000) -- (11.004,-0.000);
\fill[neucol] (10.796,-0.000) circle (0.07);
\filldraw[fill=colorB, draw=black, line width=0.22pt] (11.244,0.085) -- (11.154,-0.075) -- (11.334,-0.075) -- cycle;
\draw[poscol, line width=0.8pt] (10.855,-0.340) -- (11.409,-0.340);
\fill[poscol] (11.132,-0.340) circle (0.07);
\filldraw[fill=colorB, draw=black, line width=0.22pt] (10.996,-0.255) -- (10.906,-0.415) -- (11.086,-0.415) -- cycle;
\draw[poscol, line width=0.8pt] (10.661,-0.680) -- (11.216,-0.680);
\fill[poscol] (10.939,-0.680) circle (0.07);
\filldraw[fill=colorB, draw=black, line width=0.22pt] (10.727,-0.595) -- (10.637,-0.755) -- (10.817,-0.755) -- cycle;
\draw[neucol, line width=0.8pt] (10.537,-1.020) -- (11.025,-1.020);
\fill[neucol] (10.781,-1.020) circle (0.07);
\filldraw[fill=colorB, draw=black, line width=0.22pt] (10.764,-0.935) -- (10.674,-1.095) -- (10.854,-1.095) -- cycle;
\draw[black!75, line width=0.55pt] (8.190,0.153) rectangle (12.990,-1.120);
\fill[white] (13.540,0.153) rectangle (18.340,-1.120);
\draw[black!75, line width=0.55pt] (13.540,0.153) rectangle (18.340,-1.120);
\draw[black!70, line width=0.3pt] (15.940,0.153) -- (15.940,-1.120);
\draw[gray!30, line width=0.2pt] (13.807,0.153) -- (13.807,-1.120);
\node[font=\scriptsize, anchor=north] at (13.807,-1.200) {-4};
\draw[gray!30, line width=0.2pt] (14.340,0.153) -- (14.340,-1.120);
\node[font=\scriptsize, anchor=north] at (14.340,-1.200) {-3};
\draw[gray!30, line width=0.2pt] (14.873,0.153) -- (14.873,-1.120);
\node[font=\scriptsize, anchor=north] at (14.873,-1.200) {-2};
\draw[gray!30, line width=0.2pt] (15.407,0.153) -- (15.407,-1.120);
\node[font=\scriptsize, anchor=north] at (15.407,-1.200) {-1};
\draw[gray!30, line width=0.2pt] (15.940,0.153) -- (15.940,-1.120);
\node[font=\scriptsize, anchor=north] at (15.940,-1.200) {0};
\draw[gray!30, line width=0.2pt] (16.473,0.153) -- (16.473,-1.120);
\node[font=\scriptsize, anchor=north] at (16.473,-1.200) {1};
\draw[gray!30, line width=0.2pt] (17.007,0.153) -- (17.007,-1.120);
\node[font=\scriptsize, anchor=north] at (17.007,-1.200) {2};
\draw[gray!30, line width=0.2pt] (17.540,0.153) -- (17.540,-1.120);
\node[font=\scriptsize, anchor=north] at (17.540,-1.200) {3};
\draw[gray!30, line width=0.2pt] (18.073,0.153) -- (18.073,-1.120);
\node[font=\scriptsize, anchor=north] at (18.073,-1.200) {4};
\node[font=\small\bfseries, anchor=south] at (15.940,0.340) {High bias (3)};
\draw[poscol, line width=0.8pt] (16.282,-0.000) -- (16.794,-0.000);
\fill[poscol] (16.538,-0.000) circle (0.07);
\filldraw[fill=colorB, draw=black, line width=0.22pt] (17.391,0.085) -- (17.301,-0.075) -- (17.481,-0.075) -- cycle;
\draw[poscol, line width=0.8pt] (16.468,-0.340) -- (17.090,-0.340);
\fill[poscol] (16.779,-0.340) circle (0.07);
\filldraw[fill=colorB, draw=black, line width=0.22pt] (18.033,-0.255) -- (17.943,-0.415) -- (18.123,-0.415) -- cycle;
\draw[neucol, line width=0.8pt] (15.918,-0.680) -- (16.657,-0.680);
\fill[neucol] (16.287,-0.680) circle (0.07);
\filldraw[fill=colorB, draw=black, line width=0.22pt] (17.003,-0.595) -- (16.913,-0.755) -- (17.093,-0.755) -- cycle;
\draw[neucol, line width=0.8pt] (15.683,-1.020) -- (16.108,-1.020);
\fill[neucol] (15.896,-1.020) circle (0.07);
\filldraw[fill=colorB, draw=black, line width=0.22pt] (16.544,-0.935) -- (16.454,-1.095) -- (16.634,-1.095) -- cycle;
\draw[black!75, line width=0.55pt] (13.540,0.153) rectangle (18.340,-1.120);
\filldraw[fill=colorB, draw=black, line width=0.22pt] (8.940,-1.835) -- (8.850,-1.995) -- (9.030,-1.995) -- cycle;
\node[font=\scriptsize, anchor=west] at (9.100,-1.920) {theoretical prediction};
\end{tikzpicture}}
\caption{by history.}
\label{fig:hyp1_whiskers}
\end{subfigure}
\vspace{0.8em}
\begin{subfigure}{0.8\linewidth}
\centering
\resizebox{0.8\textwidth}{!}{\begin{tikzpicture}[x=1cm,y=1cm]
\fill[black] (0.450,-0.000) circle (0.10);
\fill[black] (1.100,-0.000) circle (0.10);
\fill[black] (1.750,-0.000) circle (0.10);
\fill[black] (0.450,-0.340) circle (0.10);
\fill[black] (1.100,-0.340) circle (0.10);
\draw[black, line width=0.8pt] (1.630,-0.340) -- (1.870,-0.340);
\draw[red, line width=0.8pt] (0.330,-0.680) -- (0.570,-0.680);
\fill[black] (1.100,-0.680) circle (0.10);
\fill[black] (1.750,-0.680) circle (0.10);
\fill[black] (0.450,-1.020) circle (0.10);
\draw[black, line width=0.8pt] (0.980,-1.020) -- (1.220,-1.020);
\draw[black, line width=0.8pt] (1.630,-1.020) -- (1.870,-1.020);
\draw[red, line width=0.8pt] (0.330,-1.360) -- (0.570,-1.360);
\fill[black] (1.100,-1.360) circle (0.10);
\draw[black, line width=0.8pt] (1.630,-1.360) -- (1.870,-1.360);
\draw[red, line width=0.8pt] (0.330,-1.700) -- (0.570,-1.700);
\draw[red, line width=0.8pt] (0.980,-1.700) -- (1.220,-1.700);
\fill[black] (1.750,-1.700) circle (0.10);
\fill[red] (0.450,-2.040) circle (0.10);
\fill[black] (1.100,-2.040) circle (0.10);
\fill[black] (1.750,-2.040) circle (0.10);
\draw[black, line width=0.8pt] (0.330,-2.380) -- (0.570,-2.380);
\draw[black, line width=0.8pt] (0.980,-2.380) -- (1.220,-2.380);
\draw[black, line width=0.8pt] (1.630,-2.380) -- (1.870,-2.380);
\draw[red, line width=0.8pt] (0.330,-2.720) -- (0.570,-2.720);
\draw[black, line width=0.8pt] (0.980,-2.720) -- (1.220,-2.720);
\draw[black, line width=0.8pt] (1.630,-2.720) -- (1.870,-2.720);
\fill[red] (0.450,-3.060) circle (0.10);
\fill[black] (1.100,-3.060) circle (0.10);
\draw[black, line width=0.8pt] (1.630,-3.060) -- (1.870,-3.060);
\draw[red, line width=0.8pt] (0.330,-3.400) -- (0.570,-3.400);
\draw[red, line width=0.8pt] (0.980,-3.400) -- (1.220,-3.400);
\draw[black, line width=0.8pt] (1.630,-3.400) -- (1.870,-3.400);
\fill[red] (0.450,-3.740) circle (0.10);
\draw[red, line width=0.8pt] (0.980,-3.740) -- (1.220,-3.740);
\fill[black] (1.750,-3.740) circle (0.10);
\draw[red, line width=0.8pt] (0.330,-4.080) -- (0.570,-4.080);
\draw[red, line width=0.8pt] (0.980,-4.080) -- (1.220,-4.080);
\draw[red, line width=0.8pt] (1.630,-4.080) -- (1.870,-4.080);
\fill[red] (0.450,-4.420) circle (0.10);
\fill[red] (1.100,-4.420) circle (0.10);
\fill[black] (1.750,-4.420) circle (0.10);
\fill[red] (0.450,-4.760) circle (0.10);
\draw[black, line width=0.8pt] (0.980,-4.760) -- (1.220,-4.760);
\draw[black, line width=0.8pt] (1.630,-4.760) -- (1.870,-4.760);
\fill[red] (0.450,-5.100) circle (0.10);
\draw[red, line width=0.8pt] (0.980,-5.100) -- (1.220,-5.100);
\draw[black, line width=0.8pt] (1.630,-5.100) -- (1.870,-5.100);
\fill[red] (0.450,-5.440) circle (0.10);
\draw[red, line width=0.8pt] (0.980,-5.440) -- (1.220,-5.440);
\draw[red, line width=0.8pt] (1.630,-5.440) -- (1.870,-5.440);
\fill[red] (0.450,-5.780) circle (0.10);
\fill[red] (1.100,-5.780) circle (0.10);
\draw[black, line width=0.8pt] (1.630,-5.780) -- (1.870,-5.780);
\fill[red] (0.450,-6.120) circle (0.10);
\fill[red] (1.100,-6.120) circle (0.10);
\draw[red, line width=0.8pt] (1.630,-6.120) -- (1.870,-6.120);
\fill[red] (0.450,-6.460) circle (0.10);
\fill[red] (1.100,-6.460) circle (0.10);
\fill[red] (1.750,-6.460) circle (0.10);
\node[font=\small\bfseries, anchor=south] at (1.100,0.340) {};
\fill[white] (2.840,0.153) rectangle (7.640,-6.560);
\draw[black!75, line width=0.55pt] (2.840,0.153) rectangle (7.640,-6.560);
\draw[black!70, line width=0.3pt] (5.240,0.153) -- (5.240,-6.560);
\draw[gray!30, line width=0.2pt] (3.107,0.153) -- (3.107,-6.560);
\node[font=\scriptsize, anchor=north] at (3.107,-6.640) {-4};
\draw[gray!30, line width=0.2pt] (3.640,0.153) -- (3.640,-6.560);
\node[font=\scriptsize, anchor=north] at (3.640,-6.640) {-3};
\draw[gray!30, line width=0.2pt] (4.173,0.153) -- (4.173,-6.560);
\node[font=\scriptsize, anchor=north] at (4.173,-6.640) {-2};
\draw[gray!30, line width=0.2pt] (4.707,0.153) -- (4.707,-6.560);
\node[font=\scriptsize, anchor=north] at (4.707,-6.640) {-1};
\draw[gray!30, line width=0.2pt] (5.240,0.153) -- (5.240,-6.560);
\node[font=\scriptsize, anchor=north] at (5.240,-6.640) {0};
\draw[gray!30, line width=0.2pt] (5.773,0.153) -- (5.773,-6.560);
\node[font=\scriptsize, anchor=north] at (5.773,-6.640) {1};
\draw[gray!30, line width=0.2pt] (6.307,0.153) -- (6.307,-6.560);
\node[font=\scriptsize, anchor=north] at (6.307,-6.640) {2};
\draw[gray!30, line width=0.2pt] (6.840,0.153) -- (6.840,-6.560);
\node[font=\scriptsize, anchor=north] at (6.840,-6.640) {3};
\draw[gray!30, line width=0.2pt] (7.373,0.153) -- (7.373,-6.560);
\node[font=\scriptsize, anchor=north] at (7.373,-6.640) {4};
\node[font=\small\bfseries, anchor=south] at (5.240,0.340) {No bias (0)};
\draw[neucol, line width=0.8pt] (4.653,-0.000) -- (6.183,-0.000);
\fill[neucol] (5.418,-0.000) circle (0.07);
\filldraw[fill=colorB, draw=black, line width=0.22pt] (5.240,0.085) -- (5.150,-0.075) -- (5.330,-0.075) -- cycle;
\draw[poscol, line width=0.8pt] (5.309,-0.340) -- (5.933,-0.340);
\fill[poscol] (5.621,-0.340) circle (0.07);
\filldraw[fill=colorB, draw=black, line width=0.22pt] (5.240,-0.255) -- (5.150,-0.415) -- (5.330,-0.415) -- cycle;
\draw[poscol, line width=0.8pt] (5.348,-0.680) -- (6.910,-0.680);
\fill[poscol] (6.129,-0.680) circle (0.07);
\filldraw[fill=colorB, draw=black, line width=0.22pt] (5.240,-0.595) -- (5.150,-0.755) -- (5.330,-0.755) -- cycle;
\draw[neucol, line width=0.8pt] (5.123,-1.020) -- (5.521,-1.020);
\fill[neucol] (5.322,-1.020) circle (0.07);
\filldraw[fill=colorB, draw=black, line width=0.22pt] (5.240,-0.935) -- (5.150,-1.095) -- (5.330,-1.095) -- cycle;
\draw[neucol, line width=0.8pt] (5.239,-1.360) -- (6.079,-1.360);
\fill[neucol] (5.659,-1.360) circle (0.07);
\filldraw[fill=colorB, draw=black, line width=0.22pt] (5.240,-1.275) -- (5.150,-1.435) -- (5.330,-1.435) -- cycle;
\draw[poscol, line width=0.8pt] (5.416,-1.700) -- (7.553,-1.700);
\fill[poscol] (6.484,-1.700) circle (0.07);
\filldraw[fill=colorB, draw=black, line width=0.22pt] (5.240,-1.615) -- (5.150,-1.775) -- (5.330,-1.775) -- cycle;
\draw[neucol, line width=0.8pt] (5.051,-2.040) -- (5.643,-2.040);
\fill[neucol] (5.347,-2.040) circle (0.07);
\filldraw[fill=colorB, draw=black, line width=0.22pt] (5.240,-1.955) -- (5.150,-2.115) -- (5.330,-2.115) -- cycle;
\draw[neucol, line width=0.8pt] (4.690,-2.380) -- (5.305,-2.380);
\fill[neucol] (4.998,-2.380) circle (0.07);
\filldraw[fill=colorB, draw=black, line width=0.22pt] (5.240,-2.295) -- (5.150,-2.455) -- (5.330,-2.455) -- cycle;
\draw[neucol, line width=0.8pt] (4.210,-2.720) -- (5.260,-2.720);
\fill[neucol] (4.735,-2.720) circle (0.07);
\filldraw[fill=colorB, draw=black, line width=0.22pt] (5.240,-2.635) -- (5.150,-2.795) -- (5.330,-2.795) -- cycle;
\draw[neucol, line width=0.8pt] (5.240,-3.060) -- (5.240,-3.060);
\fill[neucol] (5.240,-3.060) circle (0.07);
\filldraw[fill=colorB, draw=black, line width=0.22pt] (5.240,-2.975) -- (5.150,-3.135) -- (5.330,-3.135) -- cycle;
\draw[neucol, line width=0.8pt] (5.240,-3.400) -- (5.240,-3.400);
\fill[neucol] (5.240,-3.400) circle (0.07);
\filldraw[fill=colorB, draw=black, line width=0.22pt] (5.240,-3.315) -- (5.150,-3.475) -- (5.330,-3.475) -- cycle;
\draw[neucol, line width=0.8pt] (5.240,-3.740) -- (5.240,-3.740);
\fill[neucol] (5.240,-3.740) circle (0.07);
\filldraw[fill=colorB, draw=black, line width=0.22pt] (5.240,-3.655) -- (5.150,-3.815) -- (5.330,-3.815) -- cycle;
\draw[neucol, line width=0.8pt] (5.052,-4.080) -- (6.002,-4.080);
\fill[neucol] (5.527,-4.080) circle (0.07);
\filldraw[fill=colorB, draw=black, line width=0.22pt] (5.240,-3.995) -- (5.150,-4.155) -- (5.330,-4.155) -- cycle;
\draw[neucol, line width=0.8pt] (4.891,-4.420) -- (5.411,-4.420);
\fill[neucol] (5.151,-4.420) circle (0.07);
\filldraw[fill=colorB, draw=black, line width=0.22pt] (5.240,-4.335) -- (5.150,-4.495) -- (5.330,-4.495) -- cycle;
\draw[neucol, line width=0.8pt] (2.493,-4.760) -- (5.765,-4.760);
\fill[neucol] (4.129,-4.760) circle (0.07);
\filldraw[fill=colorB, draw=black, line width=0.22pt] (5.240,-4.675) -- (5.150,-4.835) -- (5.330,-4.835) -- cycle;
\draw[neucol, line width=0.8pt] (4.504,-5.100) -- (5.291,-5.100);
\fill[neucol] (4.897,-5.100) circle (0.07);
\filldraw[fill=colorB, draw=black, line width=0.22pt] (5.240,-5.015) -- (5.150,-5.175) -- (5.330,-5.175) -- cycle;
\draw[neucol, line width=0.8pt] (4.802,-5.440) -- (5.387,-5.440);
\fill[neucol] (5.095,-5.440) circle (0.07);
\filldraw[fill=colorB, draw=black, line width=0.22pt] (5.240,-5.355) -- (5.150,-5.515) -- (5.330,-5.515) -- cycle;
\draw[neucol, line width=0.8pt] (4.410,-5.780) -- (5.332,-5.780);
\fill[neucol] (4.871,-5.780) circle (0.07);
\filldraw[fill=colorB, draw=black, line width=0.22pt] (5.240,-5.695) -- (5.150,-5.855) -- (5.330,-5.855) -- cycle;
\draw[neucol, line width=0.8pt] (3.903,-6.120) -- (5.566,-6.120);
\fill[neucol] (4.735,-6.120) circle (0.07);
\filldraw[fill=colorB, draw=black, line width=0.22pt] (5.240,-6.035) -- (5.150,-6.195) -- (5.330,-6.195) -- cycle;
\draw[neucol, line width=0.8pt] (5.006,-6.460) -- (5.341,-6.460);
\fill[neucol] (5.173,-6.460) circle (0.07);
\filldraw[fill=colorB, draw=black, line width=0.22pt] (5.240,-6.375) -- (5.150,-6.535) -- (5.330,-6.535) -- cycle;
\draw[black!75, line width=0.55pt] (2.840,0.153) rectangle (7.640,-6.560);
\fill[white] (8.190,0.153) rectangle (12.990,-6.560);
\draw[black!75, line width=0.55pt] (8.190,0.153) rectangle (12.990,-6.560);
\draw[black!70, line width=0.3pt] (10.590,0.153) -- (10.590,-6.560);
\draw[gray!30, line width=0.2pt] (8.457,0.153) -- (8.457,-6.560);
\node[font=\scriptsize, anchor=north] at (8.457,-6.640) {-4};
\draw[gray!30, line width=0.2pt] (8.990,0.153) -- (8.990,-6.560);
\node[font=\scriptsize, anchor=north] at (8.990,-6.640) {-3};
\draw[gray!30, line width=0.2pt] (9.523,0.153) -- (9.523,-6.560);
\node[font=\scriptsize, anchor=north] at (9.523,-6.640) {-2};
\draw[gray!30, line width=0.2pt] (10.057,0.153) -- (10.057,-6.560);
\node[font=\scriptsize, anchor=north] at (10.057,-6.640) {-1};
\draw[gray!30, line width=0.2pt] (10.590,0.153) -- (10.590,-6.560);
\node[font=\scriptsize, anchor=north] at (10.590,-6.640) {0};
\draw[gray!30, line width=0.2pt] (11.123,0.153) -- (11.123,-6.560);
\node[font=\scriptsize, anchor=north] at (11.123,-6.640) {1};
\draw[gray!30, line width=0.2pt] (11.657,0.153) -- (11.657,-6.560);
\node[font=\scriptsize, anchor=north] at (11.657,-6.640) {2};
\draw[gray!30, line width=0.2pt] (12.190,0.153) -- (12.190,-6.560);
\node[font=\scriptsize, anchor=north] at (12.190,-6.640) {3};
\draw[gray!30, line width=0.2pt] (12.723,0.153) -- (12.723,-6.560);
\node[font=\scriptsize, anchor=north] at (12.723,-6.640) {4};
\node[font=\small\bfseries, anchor=south] at (10.590,0.340) {Low bias (1)};
\draw[neucol, line width=0.8pt] (10.577,-0.000) -- (12.584,-0.000);
\fill[neucol] (11.580,-0.000) circle (0.07);
\filldraw[fill=colorB, draw=black, line width=0.22pt] (10.768,0.085) -- (10.678,-0.075) -- (10.858,-0.075) -- cycle;
\draw[poscol, line width=0.8pt] (10.914,-0.340) -- (11.534,-0.340);
\fill[poscol] (11.224,-0.340) circle (0.07);
\filldraw[fill=colorB, draw=black, line width=0.22pt] (11.479,-0.255) -- (11.389,-0.415) -- (11.569,-0.415) -- cycle;
\draw[poscol, line width=0.8pt] (11.238,-0.680) -- (12.581,-0.680);
\fill[poscol] (11.909,-0.680) circle (0.07);
\filldraw[fill=colorB, draw=black, line width=0.22pt] (10.857,-0.595) -- (10.767,-0.755) -- (10.947,-0.755) -- cycle;
\draw[poscol, line width=0.8pt] (10.617,-1.020) -- (11.210,-1.020);
\fill[poscol] (10.913,-1.020) circle (0.07);
\filldraw[fill=colorB, draw=black, line width=0.22pt] (11.657,-0.935) -- (11.567,-1.095) -- (11.747,-1.095) -- cycle;
\draw[poscol, line width=0.8pt] (11.303,-1.360) -- (12.382,-1.360);
\fill[poscol] (11.842,-1.360) circle (0.07);
\filldraw[fill=colorB, draw=black, line width=0.22pt] (11.390,-1.275) -- (11.300,-1.435) -- (11.480,-1.435) -- cycle;
\draw[poscol, line width=0.8pt] (11.607,-1.700) -- (13.271,-1.700);
\fill[poscol] (12.439,-1.700) circle (0.07);
\filldraw[fill=colorB, draw=black, line width=0.22pt] (10.590,-1.615) -- (10.500,-1.775) -- (10.680,-1.775) -- cycle;
\draw[neucol, line width=0.8pt] (9.374,-2.040) -- (13.050,-2.040);
\fill[neucol] (11.212,-2.040) circle (0.07);
\filldraw[fill=colorB, draw=black, line width=0.22pt] (11.123,-1.955) -- (11.033,-2.115) -- (11.213,-2.115) -- cycle;
\draw[negcol, line width=0.8pt] (9.611,-2.380) -- (10.245,-2.380);
\fill[negcol] (9.928,-2.380) circle (0.07);
\filldraw[fill=colorB, draw=black, line width=0.22pt] (10.590,-2.295) -- (10.500,-2.455) -- (10.680,-2.455) -- cycle;
\draw[neucol, line width=0.8pt] (10.427,-2.720) -- (11.234,-2.720);
\fill[neucol] (10.831,-2.720) circle (0.07);
\filldraw[fill=colorB, draw=black, line width=0.22pt] (10.590,-2.635) -- (10.500,-2.795) -- (10.680,-2.795) -- cycle;
\draw[neucol, line width=0.8pt] (10.328,-3.060) -- (10.931,-3.060);
\fill[neucol] (10.630,-3.060) circle (0.07);
\filldraw[fill=colorB, draw=black, line width=0.22pt] (11.390,-2.975) -- (11.300,-3.135) -- (11.480,-3.135) -- cycle;
\draw[poscol, line width=0.8pt] (10.721,-3.400) -- (11.387,-3.400);
\fill[poscol] (11.054,-3.400) circle (0.07);
\filldraw[fill=colorB, draw=black, line width=0.22pt] (10.590,-3.315) -- (10.500,-3.475) -- (10.680,-3.475) -- cycle;
\draw[neucol, line width=0.8pt] (10.583,-3.740) -- (11.077,-3.740);
\fill[neucol] (10.830,-3.740) circle (0.07);
\filldraw[fill=colorB, draw=black, line width=0.22pt] (11.390,-3.655) -- (11.300,-3.815) -- (11.480,-3.815) -- cycle;
\draw[poscol, line width=0.8pt] (11.390,-4.080) -- (12.403,-4.080);
\fill[poscol] (11.897,-4.080) circle (0.07);
\filldraw[fill=colorB, draw=black, line width=0.22pt] (10.590,-3.995) -- (10.500,-4.155) -- (10.680,-4.155) -- cycle;
\draw[neucol, line width=0.8pt] (10.568,-4.420) -- (12.389,-4.420);
\fill[neucol] (11.479,-4.420) circle (0.07);
\filldraw[fill=colorB, draw=black, line width=0.22pt] (11.657,-4.335) -- (11.567,-4.495) -- (11.747,-4.495) -- cycle;
\draw[neucol, line width=0.8pt] (10.180,-4.760) -- (10.754,-4.760);
\fill[neucol] (10.467,-4.760) circle (0.07);
\filldraw[fill=colorB, draw=black, line width=0.22pt] (10.590,-4.675) -- (10.500,-4.835) -- (10.680,-4.835) -- cycle;
\draw[neucol, line width=0.8pt] (10.147,-5.100) -- (10.932,-5.100);
\fill[neucol] (10.539,-5.100) circle (0.07);
\filldraw[fill=colorB, draw=black, line width=0.22pt] (10.590,-5.015) -- (10.500,-5.175) -- (10.680,-5.175) -- cycle;
\draw[neucol, line width=0.8pt] (10.206,-5.440) -- (10.905,-5.440);
\fill[neucol] (10.556,-5.440) circle (0.07);
\filldraw[fill=colorB, draw=black, line width=0.22pt] (11.123,-5.355) -- (11.033,-5.515) -- (11.213,-5.515) -- cycle;
\draw[negcol, line width=0.8pt] (9.130,-5.780) -- (10.517,-5.780);
\fill[negcol] (9.823,-5.780) circle (0.07);
\filldraw[fill=colorB, draw=black, line width=0.22pt] (10.057,-5.695) -- (9.967,-5.855) -- (10.147,-5.855) -- cycle;
\draw[neucol, line width=0.8pt] (10.166,-6.120) -- (10.679,-6.120);
\fill[neucol] (10.422,-6.120) circle (0.07);
\filldraw[fill=colorB, draw=black, line width=0.22pt] (10.590,-6.035) -- (10.500,-6.195) -- (10.680,-6.195) -- cycle;
\draw[neucol, line width=0.8pt] (10.264,-6.460) -- (10.679,-6.460);
\fill[neucol] (10.471,-6.460) circle (0.07);
\filldraw[fill=colorB, draw=black, line width=0.22pt] (10.590,-6.375) -- (10.500,-6.535) -- (10.680,-6.535) -- cycle;
\draw[black!75, line width=0.55pt] (8.190,0.153) rectangle (12.990,-6.560);
\fill[white] (13.540,0.153) rectangle (18.340,-6.560);
\draw[black!75, line width=0.55pt] (13.540,0.153) rectangle (18.340,-6.560);
\draw[black!70, line width=0.3pt] (15.418,0.153) -- (15.418,-6.560);
\draw[gray!30, line width=0.2pt] (13.749,0.153) -- (13.749,-6.560);
\node[font=\scriptsize, anchor=north] at (13.749,-6.640) {-4};
\draw[gray!30, line width=0.2pt] (14.166,0.153) -- (14.166,-6.560);
\node[font=\scriptsize, anchor=north] at (14.166,-6.640) {-3};
\draw[gray!30, line width=0.2pt] (14.583,0.153) -- (14.583,-6.560);
\node[font=\scriptsize, anchor=north] at (14.583,-6.640) {-2};
\draw[gray!30, line width=0.2pt] (15.001,0.153) -- (15.001,-6.560);
\node[font=\scriptsize, anchor=north] at (15.001,-6.640) {-1};
\draw[gray!30, line width=0.2pt] (15.418,0.153) -- (15.418,-6.560);
\node[font=\scriptsize, anchor=north] at (15.418,-6.640) {0};
\draw[gray!30, line width=0.2pt] (15.836,0.153) -- (15.836,-6.560);
\node[font=\scriptsize, anchor=north] at (15.836,-6.640) {1};
\draw[gray!30, line width=0.2pt] (16.253,0.153) -- (16.253,-6.560);
\node[font=\scriptsize, anchor=north] at (16.253,-6.640) {2};
\draw[gray!30, line width=0.2pt] (16.670,0.153) -- (16.670,-6.560);
\node[font=\scriptsize, anchor=north] at (16.670,-6.640) {3};
\draw[gray!30, line width=0.2pt] (17.088,0.153) -- (17.088,-6.560);
\node[font=\scriptsize, anchor=north] at (17.088,-6.640) {4};
\draw[gray!30, line width=0.2pt] (17.505,0.153) -- (17.505,-6.560);
\node[font=\scriptsize, anchor=north] at (17.505,-6.640) {5};
\draw[gray!30, line width=0.2pt] (17.923,0.153) -- (17.923,-6.560);
\node[font=\scriptsize, anchor=north] at (17.923,-6.640) {6};
\draw[gray!30, line width=0.2pt] (18.340,0.153) -- (18.340,-6.560);
\node[font=\scriptsize, anchor=north] at (18.340,-6.640) {7};
\node[font=\small\bfseries, anchor=south] at (15.940,0.340) {High bias (3)};
\draw[poscol, line width=0.8pt] (15.869,-0.000) -- (16.941,-0.000);
\fill[poscol] (16.405,-0.000) circle (0.07);
\filldraw[fill=colorB, draw=black, line width=0.22pt] (17.088,0.085) -- (16.998,-0.075) -- (17.178,-0.075) -- cycle;
\draw[poscol, line width=0.8pt] (16.096,-0.340) -- (16.601,-0.340);
\fill[poscol] (16.349,-0.340) circle (0.07);
\filldraw[fill=colorB, draw=black, line width=0.22pt] (17.088,-0.255) -- (16.998,-0.415) -- (17.178,-0.415) -- cycle;
\draw[poscol, line width=0.8pt] (16.201,-0.680) -- (17.178,-0.680);
\fill[poscol] (16.689,-0.680) circle (0.07);
\filldraw[fill=colorB, draw=black, line width=0.22pt] (18.340,-0.595) -- (18.250,-0.755) -- (18.430,-0.755) -- cycle;
\draw[poscol, line width=0.8pt] (15.741,-1.020) -- (16.216,-1.020);
\fill[poscol] (15.978,-1.020) circle (0.07);
\filldraw[fill=colorB, draw=black, line width=0.22pt] (16.670,-0.935) -- (16.580,-1.095) -- (16.760,-1.095) -- cycle;
\draw[poscol, line width=0.8pt] (16.350,-1.360) -- (17.382,-1.360);
\fill[poscol] (16.866,-1.360) circle (0.07);
\filldraw[fill=colorB, draw=black, line width=0.22pt] (17.923,-1.275) -- (17.833,-1.435) -- (18.013,-1.435) -- cycle;
\draw[poscol, line width=0.8pt] (16.536,-1.700) -- (17.744,-1.700);
\fill[poscol] (17.140,-1.700) circle (0.07);
\filldraw[fill=colorB, draw=black, line width=0.22pt] (17.923,-1.615) -- (17.833,-1.775) -- (18.013,-1.775) -- cycle;
\draw[poscol, line width=0.8pt] (15.624,-2.040) -- (17.160,-2.040);
\fill[poscol] (16.392,-2.040) circle (0.07);
\filldraw[fill=colorB, draw=black, line width=0.22pt] (17.088,-1.955) -- (16.998,-2.115) -- (17.178,-2.115) -- cycle;
\draw[negcol, line width=0.8pt] (14.644,-2.380) -- (15.149,-2.380);
\fill[negcol] (14.897,-2.380) circle (0.07);
\filldraw[fill=colorB, draw=black, line width=0.22pt] (15.418,-2.295) -- (15.328,-2.455) -- (15.508,-2.455) -- cycle;
\draw[neucol, line width=0.8pt] (15.272,-2.720) -- (15.962,-2.720);
\fill[neucol] (15.617,-2.720) circle (0.07);
\filldraw[fill=colorB, draw=black, line width=0.22pt] (16.670,-2.635) -- (16.580,-2.795) -- (16.760,-2.795) -- cycle;
\draw[poscol, line width=0.8pt] (15.462,-3.060) -- (16.232,-3.060);
\fill[poscol] (15.847,-3.060) circle (0.07);
\filldraw[fill=colorB, draw=black, line width=0.22pt] (16.670,-2.975) -- (16.580,-3.135) -- (16.760,-3.135) -- cycle;
\draw[poscol, line width=0.8pt] (15.584,-3.400) -- (16.477,-3.400);
\fill[poscol] (16.030,-3.400) circle (0.07);
\filldraw[fill=colorB, draw=black, line width=0.22pt] (16.670,-3.315) -- (16.580,-3.475) -- (16.760,-3.475) -- cycle;
\draw[poscol, line width=0.8pt] (15.472,-3.740) -- (16.392,-3.740);
\fill[poscol] (15.932,-3.740) circle (0.07);
\filldraw[fill=colorB, draw=black, line width=0.22pt] (16.670,-3.655) -- (16.580,-3.815) -- (16.760,-3.815) -- cycle;
\draw[poscol, line width=0.8pt] (15.781,-4.080) -- (16.482,-4.080);
\fill[poscol] (16.131,-4.080) circle (0.07);
\filldraw[fill=colorB, draw=black, line width=0.22pt] (17.088,-3.995) -- (16.998,-4.155) -- (17.178,-4.155) -- cycle;
\draw[poscol, line width=0.8pt] (15.515,-4.420) -- (16.753,-4.420);
\fill[poscol] (16.134,-4.420) circle (0.07);
\filldraw[fill=colorB, draw=black, line width=0.22pt] (16.253,-4.335) -- (16.163,-4.495) -- (16.343,-4.495) -- cycle;
\draw[neucol, line width=0.8pt] (14.902,-4.760) -- (15.465,-4.760);
\fill[neucol] (15.183,-4.760) circle (0.07);
\filldraw[fill=colorB, draw=black, line width=0.22pt] (15.418,-4.675) -- (15.328,-4.835) -- (15.508,-4.835) -- cycle;
\draw[neucol, line width=0.8pt] (14.672,-5.100) -- (15.503,-5.100);
\fill[neucol] (15.087,-5.100) circle (0.07);
\filldraw[fill=colorB, draw=black, line width=0.22pt] (15.418,-5.015) -- (15.328,-5.175) -- (15.508,-5.175) -- cycle;
\draw[neucol, line width=0.8pt] (15.293,-5.440) -- (15.647,-5.440);
\fill[neucol] (15.470,-5.440) circle (0.07);
\filldraw[fill=colorB, draw=black, line width=0.22pt] (15.836,-5.355) -- (15.746,-5.515) -- (15.926,-5.515) -- cycle;
\draw[negcol, line width=0.8pt] (13.860,-5.780) -- (15.108,-5.780);
\fill[negcol] (14.484,-5.780) circle (0.07);
\filldraw[fill=colorB, draw=black, line width=0.22pt] (15.001,-5.695) -- (14.911,-5.855) -- (15.091,-5.855) -- cycle;
\draw[negcol, line width=0.8pt] (14.708,-6.120) -- (15.148,-6.120);
\fill[negcol] (14.928,-6.120) circle (0.07);
\filldraw[fill=colorB, draw=black, line width=0.22pt] (15.418,-6.035) -- (15.328,-6.195) -- (15.508,-6.195) -- cycle;
\draw[neucol, line width=0.8pt] (15.209,-6.460) -- (15.461,-6.460);
\fill[neucol] (15.335,-6.460) circle (0.07);
\filldraw[fill=colorB, draw=black, line width=0.22pt] (15.418,-6.375) -- (15.328,-6.535) -- (15.508,-6.535) -- cycle;
\draw[black!75, line width=0.55pt] (13.540,0.153) rectangle (18.340,-6.560);
\filldraw[fill=colorB, draw=black, line width=0.22pt] (8.940,-7.275) -- (8.850,-7.435) -- (9.030,-7.435) -- cycle;
\node[font=\scriptsize, anchor=west] at (9.100,-7.360) {theoretical prediction};
\end{tikzpicture}}
\caption{by history and model.}
\label{fig:hyp1_whiskers_h}
\end{subfigure}
\vspace{1.0em}
\begin{minipage}{0.95\textwidth}
\footnotesize
Notes: Whiskers indicate pointwise 95\% confidence intervals with standard errors clustered by sender. Green and red whiskers indicate positive and negative estimates, respectively, whose pointwise 95\% confidence intervals exclude zero.
\end{minipage}
\caption{Observed $\tilde{\Delta}^S(h,b),~ \tilde{\Delta}_m^S(h,b)$ and predicted $\Delta^S(h,b),~ \Delta_m^S(h,b)$.}
\label{fig:hyp1_whiskers_combined}
\end{figure}

%% file: results_artifacts/Figure_16_Observed_persuasiveness_Delta_S_plus_Delta_R.tex
\begin{figure}[p]
\centering
\begin{subfigure}{0.8\linewidth}
\centering
\resizebox{0.8\textwidth}{!}{\begin{tikzpicture}[x=1cm,y=1cm]
\fill[black] (0.450,-0.000) circle (0.10);
\fill[black] (1.100,-0.000) circle (0.10);
\fill[black] (1.750,-0.000) circle (0.10);
\fill[red] (0.450,-0.340) circle (0.10);
\fill[black] (1.100,-0.340) circle (0.10);
\fill[black] (1.750,-0.340) circle (0.10);
\fill[red] (0.450,-0.680) circle (0.10);
\fill[red] (1.100,-0.680) circle (0.10);
\fill[black] (1.750,-0.680) circle (0.10);
\fill[red] (0.450,-1.020) circle (0.10);
\fill[red] (1.100,-1.020) circle (0.10);
\fill[red] (1.750,-1.020) circle (0.10);
\fill[white] (2.840,0.153) rectangle (7.640,-1.120);
\draw[black!75, line width=0.55pt] (2.840,0.153) rectangle (7.640,-1.120);
\draw[black!70, line width=0.3pt] (5.240,0.153) -- (5.240,-1.120);
\draw[gray!30, line width=0.2pt] (3.107,0.153) -- (3.107,-1.120);
\node[font=\scriptsize, anchor=north] at (3.107,-1.200) {-4};
\draw[gray!30, line width=0.2pt] (3.640,0.153) -- (3.640,-1.120);
\node[font=\scriptsize, anchor=north] at (3.640,-1.200) {-3};
\draw[gray!30, line width=0.2pt] (4.173,0.153) -- (4.173,-1.120);
\node[font=\scriptsize, anchor=north] at (4.173,-1.200) {-2};
\draw[gray!30, line width=0.2pt] (4.707,0.153) -- (4.707,-1.120);
\node[font=\scriptsize, anchor=north] at (4.707,-1.200) {-1};
\draw[gray!30, line width=0.2pt] (5.240,0.153) -- (5.240,-1.120);
\node[font=\scriptsize, anchor=north] at (5.240,-1.200) {0};
\draw[gray!30, line width=0.2pt] (5.773,0.153) -- (5.773,-1.120);
\node[font=\scriptsize, anchor=north] at (5.773,-1.200) {1};
\draw[gray!30, line width=0.2pt] (6.307,0.153) -- (6.307,-1.120);
\node[font=\scriptsize, anchor=north] at (6.307,-1.200) {2};
\draw[gray!30, line width=0.2pt] (6.840,0.153) -- (6.840,-1.120);
\node[font=\scriptsize, anchor=north] at (6.840,-1.200) {3};
\draw[gray!30, line width=0.2pt] (7.373,0.153) -- (7.373,-1.120);
\node[font=\scriptsize, anchor=north] at (7.373,-1.200) {4};
\node[font=\small\bfseries, anchor=south] at (5.240,0.340) {No bias (0)};
\draw[poscol, line width=0.8pt] (5.552,-0.000) -- (5.965,-0.000);
\fill[poscol] (5.759,-0.000) circle (0.07);
\filldraw[fill=paircol, draw=black, line width=0.22pt] (5.240,0.085) -- (5.150,-0.075) -- (5.330,-0.075) -- cycle;
\draw[neucol, line width=0.8pt] (4.750,-0.340) -- (5.557,-0.340);
\fill[neucol] (5.153,-0.340) circle (0.07);
\filldraw[fill=paircol, draw=black, line width=0.22pt] (5.240,-0.255) -- (5.150,-0.415) -- (5.330,-0.415) -- cycle;
\draw[neucol, line width=0.8pt] (4.710,-0.680) -- (5.321,-0.680);
\fill[neucol] (5.015,-0.680) circle (0.07);
\filldraw[fill=paircol, draw=black, line width=0.22pt] (5.240,-0.595) -- (5.150,-0.755) -- (5.330,-0.755) -- cycle;
\draw[negcol, line width=0.8pt] (4.285,-1.020) -- (4.905,-1.020);
\fill[negcol] (4.595,-1.020) circle (0.07);
\filldraw[fill=paircol, draw=black, line width=0.22pt] (5.240,-0.935) -- (5.150,-1.095) -- (5.330,-1.095) -- cycle;
\draw[black!75, line width=0.55pt] (2.840,0.153) rectangle (7.640,-1.120);
\fill[white] (8.190,0.153) rectangle (12.990,-1.120);
\draw[black!75, line width=0.55pt] (8.190,0.153) rectangle (12.990,-1.120);
\draw[black!70, line width=0.3pt] (10.590,0.153) -- (10.590,-1.120);
\draw[gray!30, line width=0.2pt] (8.457,0.153) -- (8.457,-1.120);
\node[font=\scriptsize, anchor=north] at (8.457,-1.200) {-4};
\draw[gray!30, line width=0.2pt] (8.990,0.153) -- (8.990,-1.120);
\node[font=\scriptsize, anchor=north] at (8.990,-1.200) {-3};
\draw[gray!30, line width=0.2pt] (9.523,0.153) -- (9.523,-1.120);
\node[font=\scriptsize, anchor=north] at (9.523,-1.200) {-2};
\draw[gray!30, line width=0.2pt] (10.057,0.153) -- (10.057,-1.120);
\node[font=\scriptsize, anchor=north] at (10.057,-1.200) {-1};
\draw[gray!30, line width=0.2pt] (10.590,0.153) -- (10.590,-1.120);
\node[font=\scriptsize, anchor=north] at (10.590,-1.200) {0};
\draw[gray!30, line width=0.2pt] (11.123,0.153) -- (11.123,-1.120);
\node[font=\scriptsize, anchor=north] at (11.123,-1.200) {1};
\draw[gray!30, line width=0.2pt] (11.657,0.153) -- (11.657,-1.120);
\node[font=\scriptsize, anchor=north] at (11.657,-1.200) {2};
\draw[gray!30, line width=0.2pt] (12.190,0.153) -- (12.190,-1.120);
\node[font=\scriptsize, anchor=north] at (12.190,-1.200) {3};
\draw[gray!30, line width=0.2pt] (12.723,0.153) -- (12.723,-1.120);
\node[font=\scriptsize, anchor=north] at (12.723,-1.200) {4};
\node[font=\small\bfseries, anchor=south] at (10.590,0.340) {Low bias (1)};
\draw[poscol, line width=0.8pt] (10.671,-0.000) -- (11.254,-0.000);
\fill[poscol] (10.962,-0.000) circle (0.07);
\filldraw[fill=paircol, draw=black, line width=0.22pt] (11.078,0.085) -- (10.988,-0.075) -- (11.168,-0.075) -- cycle;
\draw[poscol, line width=0.8pt] (10.719,-0.340) -- (11.321,-0.340);
\fill[poscol] (11.020,-0.340) circle (0.07);
\filldraw[fill=paircol, draw=black, line width=0.22pt] (10.891,-0.255) -- (10.801,-0.415) -- (10.981,-0.415) -- cycle;
\draw[neucol, line width=0.8pt] (10.303,-0.680) -- (10.887,-0.680);
\fill[neucol] (10.595,-0.680) circle (0.07);
\filldraw[fill=paircol, draw=black, line width=0.22pt] (10.447,-0.595) -- (10.357,-0.755) -- (10.537,-0.755) -- cycle;
\draw[negcol, line width=0.8pt] (10.120,-1.020) -- (10.465,-1.020);
\fill[negcol] (10.292,-1.020) circle (0.07);
\filldraw[fill=paircol, draw=black, line width=0.22pt] (10.343,-0.935) -- (10.253,-1.095) -- (10.433,-1.095) -- cycle;
\draw[black!75, line width=0.55pt] (8.190,0.153) rectangle (12.990,-1.120);
\fill[white] (13.540,0.153) rectangle (18.340,-1.120);
\draw[black!75, line width=0.55pt] (13.540,0.153) rectangle (18.340,-1.120);
\draw[black!70, line width=0.3pt] (15.940,0.153) -- (15.940,-1.120);
\draw[gray!30, line width=0.2pt] (13.807,0.153) -- (13.807,-1.120);
\node[font=\scriptsize, anchor=north] at (13.807,-1.200) {-4};
\draw[gray!30, line width=0.2pt] (14.340,0.153) -- (14.340,-1.120);
\node[font=\scriptsize, anchor=north] at (14.340,-1.200) {-3};
\draw[gray!30, line width=0.2pt] (14.873,0.153) -- (14.873,-1.120);
\node[font=\scriptsize, anchor=north] at (14.873,-1.200) {-2};
\draw[gray!30, line width=0.2pt] (15.407,0.153) -- (15.407,-1.120);
\node[font=\scriptsize, anchor=north] at (15.407,-1.200) {-1};
\draw[gray!30, line width=0.2pt] (15.940,0.153) -- (15.940,-1.120);
\node[font=\scriptsize, anchor=north] at (15.940,-1.200) {0};
\draw[gray!30, line width=0.2pt] (16.473,0.153) -- (16.473,-1.120);
\node[font=\scriptsize, anchor=north] at (16.473,-1.200) {1};
\draw[gray!30, line width=0.2pt] (17.007,0.153) -- (17.007,-1.120);
\node[font=\scriptsize, anchor=north] at (17.007,-1.200) {2};
\draw[gray!30, line width=0.2pt] (17.540,0.153) -- (17.540,-1.120);
\node[font=\scriptsize, anchor=north] at (17.540,-1.200) {3};
\draw[gray!30, line width=0.2pt] (18.073,0.153) -- (18.073,-1.120);
\node[font=\scriptsize, anchor=north] at (18.073,-1.200) {4};
\node[font=\small\bfseries, anchor=south] at (15.940,0.340) {High bias (3)};
\draw[poscol, line width=0.8pt] (16.140,-0.000) -- (16.612,-0.000);
\fill[poscol] (16.376,-0.000) circle (0.07);
\filldraw[fill=paircol, draw=black, line width=0.22pt] (16.857,0.085) -- (16.767,-0.075) -- (16.947,-0.075) -- cycle;
\draw[neucol, line width=0.8pt] (15.864,-0.340) -- (16.390,-0.340);
\fill[neucol] (16.127,-0.340) circle (0.07);
\filldraw[fill=paircol, draw=black, line width=0.22pt] (16.433,-0.255) -- (16.343,-0.415) -- (16.523,-0.415) -- cycle;
\draw[negcol, line width=0.8pt] (15.289,-0.680) -- (15.789,-0.680);
\fill[negcol] (15.539,-0.680) circle (0.07);
\filldraw[fill=paircol, draw=black, line width=0.22pt] (15.403,-0.595) -- (15.313,-0.755) -- (15.493,-0.755) -- cycle;
\draw[negcol, line width=0.8pt] (14.984,-1.020) -- (15.403,-1.020);
\fill[negcol] (15.193,-1.020) circle (0.07);
\filldraw[fill=paircol, draw=black, line width=0.22pt] (14.944,-0.935) -- (14.854,-1.095) -- (15.034,-1.095) -- cycle;
\draw[black!75, line width=0.55pt] (13.540,0.153) rectangle (18.340,-1.120);
\filldraw[fill=paircol, draw=black, line width=0.22pt] (8.940,-1.835) -- (8.850,-1.995) -- (9.030,-1.995) -- cycle;
\node[font=\scriptsize, anchor=west] at (9.100,-1.920) {theoretical prediction};
\end{tikzpicture}}
\caption{by history.}
\label{fig:hyp3_whiskers}
\end{subfigure}
\vspace{0.8em}
\begin{subfigure}{0.8\linewidth}
\centering
\resizebox{0.8\textwidth}{!}{\begin{tikzpicture}[x=1cm,y=1cm]
\fill[black] (0.450,-0.000) circle (0.10);
\fill[black] (1.100,-0.000) circle (0.10);
\fill[black] (1.750,-0.000) circle (0.10);
\fill[black] (0.450,-0.340) circle (0.10);
\fill[black] (1.100,-0.340) circle (0.10);
\draw[black, line width=0.8pt] (1.630,-0.340) -- (1.870,-0.340);
\draw[red, line width=0.8pt] (0.330,-0.680) -- (0.570,-0.680);
\fill[black] (1.100,-0.680) circle (0.10);
\fill[black] (1.750,-0.680) circle (0.10);
\fill[black] (0.450,-1.020) circle (0.10);
\draw[black, line width=0.8pt] (0.980,-1.020) -- (1.220,-1.020);
\draw[black, line width=0.8pt] (1.630,-1.020) -- (1.870,-1.020);
\draw[red, line width=0.8pt] (0.330,-1.360) -- (0.570,-1.360);
\fill[black] (1.100,-1.360) circle (0.10);
\draw[black, line width=0.8pt] (1.630,-1.360) -- (1.870,-1.360);
\draw[red, line width=0.8pt] (0.330,-1.700) -- (0.570,-1.700);
\draw[red, line width=0.8pt] (0.980,-1.700) -- (1.220,-1.700);
\fill[black] (1.750,-1.700) circle (0.10);
\fill[red] (0.450,-2.040) circle (0.10);
\fill[black] (1.100,-2.040) circle (0.10);
\fill[black] (1.750,-2.040) circle (0.10);
\draw[black, line width=0.8pt] (0.330,-2.380) -- (0.570,-2.380);
\draw[black, line width=0.8pt] (0.980,-2.380) -- (1.220,-2.380);
\draw[black, line width=0.8pt] (1.630,-2.380) -- (1.870,-2.380);
\draw[red, line width=0.8pt] (0.330,-2.720) -- (0.570,-2.720);
\draw[black, line width=0.8pt] (0.980,-2.720) -- (1.220,-2.720);
\draw[black, line width=0.8pt] (1.630,-2.720) -- (1.870,-2.720);
\fill[red] (0.450,-3.060) circle (0.10);
\fill[black] (1.100,-3.060) circle (0.10);
\draw[black, line width=0.8pt] (1.630,-3.060) -- (1.870,-3.060);
\draw[red, line width=0.8pt] (0.330,-3.400) -- (0.570,-3.400);
\draw[red, line width=0.8pt] (0.980,-3.400) -- (1.220,-3.400);
\draw[black, line width=0.8pt] (1.630,-3.400) -- (1.870,-3.400);
\fill[red] (0.450,-3.740) circle (0.10);
\draw[red, line width=0.8pt] (0.980,-3.740) -- (1.220,-3.740);
\fill[black] (1.750,-3.740) circle (0.10);
\draw[red, line width=0.8pt] (0.330,-4.080) -- (0.570,-4.080);
\draw[red, line width=0.8pt] (0.980,-4.080) -- (1.220,-4.080);
\draw[red, line width=0.8pt] (1.630,-4.080) -- (1.870,-4.080);
\fill[red] (0.450,-4.420) circle (0.10);
\fill[red] (1.100,-4.420) circle (0.10);
\fill[black] (1.750,-4.420) circle (0.10);
\fill[red] (0.450,-4.760) circle (0.10);
\draw[black, line width=0.8pt] (0.980,-4.760) -- (1.220,-4.760);
\draw[black, line width=0.8pt] (1.630,-4.760) -- (1.870,-4.760);
\fill[red] (0.450,-5.100) circle (0.10);
\draw[red, line width=0.8pt] (0.980,-5.100) -- (1.220,-5.100);
\draw[black, line width=0.8pt] (1.630,-5.100) -- (1.870,-5.100);
\fill[red] (0.450,-5.440) circle (0.10);
\draw[red, line width=0.8pt] (0.980,-5.440) -- (1.220,-5.440);
\draw[red, line width=0.8pt] (1.630,-5.440) -- (1.870,-5.440);
\fill[red] (0.450,-5.780) circle (0.10);
\fill[red] (1.100,-5.780) circle (0.10);
\draw[black, line width=0.8pt] (1.630,-5.780) -- (1.870,-5.780);
\fill[red] (0.450,-6.120) circle (0.10);
\fill[red] (1.100,-6.120) circle (0.10);
\draw[red, line width=0.8pt] (1.630,-6.120) -- (1.870,-6.120);
\fill[red] (0.450,-6.460) circle (0.10);
\fill[red] (1.100,-6.460) circle (0.10);
\fill[red] (1.750,-6.460) circle (0.10);
\node[font=\small\bfseries, anchor=south] at (1.100,0.340) {};
\fill[white] (2.840,0.153) rectangle (7.640,-6.560);
\draw[black!75, line width=0.55pt] (2.840,0.153) rectangle (7.640,-6.560);
\draw[black!70, line width=0.3pt] (5.240,0.153) -- (5.240,-6.560);
\draw[gray!30, line width=0.2pt] (3.107,0.153) -- (3.107,-6.560);
\node[font=\scriptsize, anchor=north] at (3.107,-6.640) {-4};
\draw[gray!30, line width=0.2pt] (3.640,0.153) -- (3.640,-6.560);
\node[font=\scriptsize, anchor=north] at (3.640,-6.640) {-3};
\draw[gray!30, line width=0.2pt] (4.173,0.153) -- (4.173,-6.560);
\node[font=\scriptsize, anchor=north] at (4.173,-6.640) {-2};
\draw[gray!30, line width=0.2pt] (4.707,0.153) -- (4.707,-6.560);
\node[font=\scriptsize, anchor=north] at (4.707,-6.640) {-1};
\draw[gray!30, line width=0.2pt] (5.240,0.153) -- (5.240,-6.560);
\node[font=\scriptsize, anchor=north] at (5.240,-6.640) {0};
\draw[gray!30, line width=0.2pt] (5.773,0.153) -- (5.773,-6.560);
\node[font=\scriptsize, anchor=north] at (5.773,-6.640) {1};
\draw[gray!30, line width=0.2pt] (6.307,0.153) -- (6.307,-6.560);
\node[font=\scriptsize, anchor=north] at (6.307,-6.640) {2};
\draw[gray!30, line width=0.2pt] (6.840,0.153) -- (6.840,-6.560);
\node[font=\scriptsize, anchor=north] at (6.840,-6.640) {3};
\draw[gray!30, line width=0.2pt] (7.373,0.153) -- (7.373,-6.560);
\node[font=\scriptsize, anchor=north] at (7.373,-6.640) {4};
\node[font=\small\bfseries, anchor=south] at (5.240,0.340) {No bias (0)};
\draw[neucol, line width=0.8pt] (4.105,-0.000) -- (8.153,-0.000);
\fill[neucol] (6.129,-0.000) circle (0.07);
\filldraw[fill=paircol, draw=black, line width=0.22pt] (5.240,0.085) -- (5.150,-0.075) -- (5.330,-0.075) -- cycle;
\draw[poscol, line width=0.8pt] (5.809,-0.340) -- (6.550,-0.340);
\fill[poscol] (6.180,-0.340) circle (0.07);
\filldraw[fill=paircol, draw=black, line width=0.22pt] (5.240,-0.255) -- (5.150,-0.415) -- (5.330,-0.415) -- cycle;
\draw[poscol, line width=0.8pt] (5.731,-0.680) -- (7.095,-0.680);
\fill[poscol] (6.413,-0.680) circle (0.07);
\filldraw[fill=paircol, draw=black, line width=0.22pt] (5.240,-0.595) -- (5.150,-0.755) -- (5.330,-0.755) -- cycle;
\draw[poscol, line width=0.8pt] (5.307,-1.020) -- (5.912,-1.020);
\fill[poscol] (5.609,-1.020) circle (0.07);
\filldraw[fill=paircol, draw=black, line width=0.22pt] (5.240,-0.935) -- (5.150,-1.095) -- (5.330,-1.095) -- cycle;
\draw[poscol, line width=0.8pt] (5.423,-1.360) -- (6.276,-1.360);
\fill[poscol] (5.850,-1.360) circle (0.07);
\filldraw[fill=paircol, draw=black, line width=0.22pt] (5.240,-1.275) -- (5.150,-1.435) -- (5.330,-1.435) -- cycle;
\draw[poscol, line width=0.8pt] (5.538,-1.700) -- (7.164,-1.700);
\fill[poscol] (6.351,-1.700) circle (0.07);
\filldraw[fill=paircol, draw=black, line width=0.22pt] (5.240,-1.615) -- (5.150,-1.775) -- (5.330,-1.775) -- cycle;
\draw[neucol, line width=0.8pt] (4.664,-2.040) -- (5.389,-2.040);
\fill[neucol] (5.027,-2.040) circle (0.07);
\filldraw[fill=paircol, draw=black, line width=0.22pt] (5.240,-1.955) -- (5.150,-2.115) -- (5.330,-2.115) -- cycle;
\draw[neucol, line width=0.8pt] (4.881,-2.380) -- (6.084,-2.380);
\fill[neucol] (5.482,-2.380) circle (0.07);
\filldraw[fill=paircol, draw=black, line width=0.22pt] (5.240,-2.295) -- (5.150,-2.455) -- (5.330,-2.455) -- cycle;
\draw[negcol, line width=0.8pt] (4.357,-2.720) -- (5.169,-2.720);
\fill[negcol] (4.763,-2.720) circle (0.07);
\filldraw[fill=paircol, draw=black, line width=0.22pt] (5.240,-2.635) -- (5.150,-2.795) -- (5.330,-2.795) -- cycle;
\draw[neucol, line width=0.8pt] (4.717,-3.060) -- (5.337,-3.060);
\fill[neucol] (5.027,-3.060) circle (0.07);
\filldraw[fill=paircol, draw=black, line width=0.22pt] (5.240,-2.975) -- (5.150,-3.135) -- (5.330,-3.135) -- cycle;
\draw[neucol, line width=0.8pt] (4.305,-3.400) -- (5.441,-3.400);
\fill[neucol] (4.873,-3.400) circle (0.07);
\filldraw[fill=paircol, draw=black, line width=0.22pt] (5.240,-3.315) -- (5.150,-3.475) -- (5.330,-3.475) -- cycle;
\draw[neucol, line width=0.8pt] (4.740,-3.740) -- (5.598,-3.740);
\fill[neucol] (5.169,-3.740) circle (0.07);
\filldraw[fill=paircol, draw=black, line width=0.22pt] (5.240,-3.655) -- (5.150,-3.815) -- (5.330,-3.815) -- cycle;
\draw[neucol, line width=0.8pt] (4.268,-4.080) -- (5.966,-4.080);
\fill[neucol] (5.117,-4.080) circle (0.07);
\filldraw[fill=paircol, draw=black, line width=0.22pt] (5.240,-3.995) -- (5.150,-4.155) -- (5.330,-4.155) -- cycle;
\draw[neucol, line width=0.8pt] (4.410,-4.420) -- (5.893,-4.420);
\fill[neucol] (5.151,-4.420) circle (0.07);
\filldraw[fill=paircol, draw=black, line width=0.22pt] (5.240,-4.335) -- (5.150,-4.495) -- (5.330,-4.495) -- cycle;
\draw[negcol, line width=0.8pt] (2.378,-4.760) -- (4.813,-4.760);
\fill[negcol] (3.596,-4.760) circle (0.07);
\filldraw[fill=paircol, draw=black, line width=0.22pt] (5.240,-4.675) -- (5.150,-4.835) -- (5.330,-4.835) -- cycle;
\draw[negcol, line width=0.8pt] (4.104,-5.100) -- (5.234,-5.100);
\fill[negcol] (4.669,-5.100) circle (0.07);
\filldraw[fill=paircol, draw=black, line width=0.22pt] (5.240,-5.015) -- (5.150,-5.175) -- (5.330,-5.175) -- cycle;
\draw[negcol, line width=0.8pt] (4.177,-5.440) -- (4.800,-5.440);
\fill[negcol] (4.488,-5.440) circle (0.07);
\filldraw[fill=paircol, draw=black, line width=0.22pt] (5.240,-5.355) -- (5.150,-5.515) -- (5.330,-5.515) -- cycle;
\draw[negcol, line width=0.8pt] (3.539,-5.780) -- (4.643,-5.780);
\fill[negcol] (4.091,-5.780) circle (0.07);
\filldraw[fill=paircol, draw=black, line width=0.22pt] (5.240,-5.695) -- (5.150,-5.855) -- (5.330,-5.855) -- cycle;
\draw[negcol, line width=0.8pt] (3.593,-6.120) -- (4.922,-6.120);
\fill[negcol] (4.258,-6.120) circle (0.07);
\filldraw[fill=paircol, draw=black, line width=0.22pt] (5.240,-6.035) -- (5.150,-6.195) -- (5.330,-6.195) -- cycle;
\draw[neucol, line width=0.8pt] (4.304,-6.460) -- (5.376,-6.460);
\fill[neucol] (4.840,-6.460) circle (0.07);
\filldraw[fill=paircol, draw=black, line width=0.22pt] (5.240,-6.375) -- (5.150,-6.535) -- (5.330,-6.535) -- cycle;
\draw[black!75, line width=0.55pt] (2.840,0.153) rectangle (7.640,-6.560);
\fill[white] (8.190,0.153) rectangle (12.990,-6.560);
\draw[black!75, line width=0.55pt] (8.190,0.153) rectangle (12.990,-6.560);
\draw[black!70, line width=0.3pt] (10.590,0.153) -- (10.590,-6.560);
\draw[gray!30, line width=0.2pt] (8.457,0.153) -- (8.457,-6.560);
\node[font=\scriptsize, anchor=north] at (8.457,-6.640) {-4};
\draw[gray!30, line width=0.2pt] (8.990,0.153) -- (8.990,-6.560);
\node[font=\scriptsize, anchor=north] at (8.990,-6.640) {-3};
\draw[gray!30, line width=0.2pt] (9.523,0.153) -- (9.523,-6.560);
\node[font=\scriptsize, anchor=north] at (9.523,-6.640) {-2};
\draw[gray!30, line width=0.2pt] (10.057,0.153) -- (10.057,-6.560);
\node[font=\scriptsize, anchor=north] at (10.057,-6.640) {-1};
\draw[gray!30, line width=0.2pt] (10.590,0.153) -- (10.590,-6.560);
\node[font=\scriptsize, anchor=north] at (10.590,-6.640) {0};
\draw[gray!30, line width=0.2pt] (11.123,0.153) -- (11.123,-6.560);
\node[font=\scriptsize, anchor=north] at (11.123,-6.640) {1};
\draw[gray!30, line width=0.2pt] (11.657,0.153) -- (11.657,-6.560);
\node[font=\scriptsize, anchor=north] at (11.657,-6.640) {2};
\draw[gray!30, line width=0.2pt] (12.190,0.153) -- (12.190,-6.560);
\node[font=\scriptsize, anchor=north] at (12.190,-6.640) {3};
\draw[gray!30, line width=0.2pt] (12.723,0.153) -- (12.723,-6.560);
\node[font=\scriptsize, anchor=north] at (12.723,-6.640) {4};
\node[font=\small\bfseries, anchor=south] at (10.590,0.340) {Low bias (1)};
\draw[poscol, line width=0.8pt] (11.140,-0.000) -- (12.631,-0.000);
\fill[poscol] (11.885,-0.000) circle (0.07);
\filldraw[fill=paircol, draw=black, line width=0.22pt] (10.768,0.085) -- (10.678,-0.075) -- (10.858,-0.075) -- cycle;
\draw[poscol, line width=0.8pt] (11.094,-0.340) -- (11.960,-0.340);
\fill[poscol] (11.527,-0.340) circle (0.07);
\filldraw[fill=paircol, draw=black, line width=0.22pt] (11.301,-0.255) -- (11.211,-0.415) -- (11.391,-0.415) -- cycle;
\draw[poscol, line width=0.8pt] (11.722,-0.680) -- (12.770,-0.680);
\fill[poscol] (12.246,-0.680) circle (0.07);
\filldraw[fill=paircol, draw=black, line width=0.22pt] (10.857,-0.595) -- (10.767,-0.755) -- (10.947,-0.755) -- cycle;
\draw[poscol, line width=0.8pt] (10.802,-1.020) -- (11.574,-1.020);
\fill[poscol] (11.188,-1.020) circle (0.07);
\filldraw[fill=paircol, draw=black, line width=0.22pt] (11.479,-0.935) -- (11.389,-1.095) -- (11.569,-1.095) -- cycle;
\draw[poscol, line width=0.8pt] (11.245,-1.360) -- (12.347,-1.360);
\fill[poscol] (11.796,-1.360) circle (0.07);
\filldraw[fill=paircol, draw=black, line width=0.22pt] (11.390,-1.275) -- (11.300,-1.435) -- (11.480,-1.435) -- cycle;
\draw[poscol, line width=0.8pt] (11.679,-1.700) -- (12.772,-1.700);
\fill[poscol] (12.226,-1.700) circle (0.07);
\filldraw[fill=paircol, draw=black, line width=0.22pt] (10.590,-1.615) -- (10.500,-1.775) -- (10.680,-1.775) -- cycle;
\draw[neucol, line width=0.8pt] (9.528,-2.040) -- (11.297,-2.040);
\fill[neucol] (10.412,-2.040) circle (0.07);
\filldraw[fill=paircol, draw=black, line width=0.22pt] (11.123,-1.955) -- (11.033,-2.115) -- (11.213,-2.115) -- cycle;
\draw[negcol, line width=0.8pt] (9.394,-2.380) -- (10.131,-2.380);
\fill[negcol] (9.762,-2.380) circle (0.07);
\filldraw[fill=paircol, draw=black, line width=0.22pt] (10.412,-2.295) -- (10.322,-2.455) -- (10.502,-2.455) -- cycle;
\draw[neucol, line width=0.8pt] (10.243,-2.720) -- (11.109,-2.720);
\fill[neucol] (10.676,-2.720) circle (0.07);
\filldraw[fill=paircol, draw=black, line width=0.22pt] (10.590,-2.635) -- (10.500,-2.795) -- (10.680,-2.795) -- cycle;
\draw[neucol, line width=0.8pt] (10.291,-3.060) -- (11.047,-3.060);
\fill[neucol] (10.669,-3.060) circle (0.07);
\filldraw[fill=paircol, draw=black, line width=0.22pt] (11.123,-2.975) -- (11.033,-3.135) -- (11.213,-3.135) -- cycle;
\draw[neucol, line width=0.8pt] (10.440,-3.400) -- (11.158,-3.400);
\fill[neucol] (10.799,-3.400) circle (0.07);
\filldraw[fill=paircol, draw=black, line width=0.22pt] (10.590,-3.315) -- (10.500,-3.475) -- (10.680,-3.475) -- cycle;
\draw[neucol, line width=0.8pt] (10.424,-3.740) -- (10.916,-3.740);
\fill[neucol] (10.670,-3.740) circle (0.07);
\filldraw[fill=paircol, draw=black, line width=0.22pt] (11.123,-3.655) -- (11.033,-3.815) -- (11.213,-3.815) -- cycle;
\draw[poscol, line width=0.8pt] (10.930,-4.080) -- (11.796,-4.080);
\fill[poscol] (11.363,-4.080) circle (0.07);
\filldraw[fill=paircol, draw=black, line width=0.22pt] (10.590,-3.995) -- (10.500,-4.155) -- (10.680,-4.155) -- cycle;
\draw[neucol, line width=0.8pt] (9.302,-4.420) -- (12.055,-4.420);
\fill[neucol] (10.679,-4.420) circle (0.07);
\filldraw[fill=paircol, draw=black, line width=0.22pt] (11.123,-4.335) -- (11.033,-4.495) -- (11.213,-4.495) -- cycle;
\draw[negcol, line width=0.8pt] (9.194,-4.760) -- (10.181,-4.760);
\fill[negcol] (9.687,-4.760) circle (0.07);
\filldraw[fill=paircol, draw=black, line width=0.22pt] (10.190,-4.675) -- (10.100,-4.835) -- (10.280,-4.835) -- cycle;
\draw[negcol, line width=0.8pt] (9.531,-5.100) -- (10.480,-5.100);
\fill[negcol] (10.006,-5.100) circle (0.07);
\filldraw[fill=paircol, draw=black, line width=0.22pt] (10.057,-5.015) -- (9.967,-5.175) -- (10.147,-5.175) -- cycle;
\draw[negcol, line width=0.8pt] (9.773,-5.440) -- (10.478,-5.440);
\fill[negcol] (10.125,-5.440) circle (0.07);
\filldraw[fill=paircol, draw=black, line width=0.22pt] (10.590,-5.355) -- (10.500,-5.515) -- (10.680,-5.515) -- cycle;
\draw[negcol, line width=0.8pt] (9.076,-5.780) -- (9.771,-5.780);
\fill[negcol] (9.423,-5.780) circle (0.07);
\filldraw[fill=paircol, draw=black, line width=0.22pt] (9.523,-5.695) -- (9.433,-5.855) -- (9.613,-5.855) -- cycle;
\draw[negcol, line width=0.8pt] (9.617,-6.120) -- (10.131,-6.120);
\fill[negcol] (9.874,-6.120) circle (0.07);
\filldraw[fill=paircol, draw=black, line width=0.22pt] (10.057,-6.035) -- (9.967,-6.195) -- (10.147,-6.195) -- cycle;
\draw[negcol, line width=0.8pt] (9.659,-6.460) -- (10.572,-6.460);
\fill[negcol] (10.116,-6.460) circle (0.07);
\filldraw[fill=paircol, draw=black, line width=0.22pt] (10.057,-6.375) -- (9.967,-6.535) -- (10.147,-6.535) -- cycle;
\draw[black!75, line width=0.55pt] (8.190,0.153) rectangle (12.990,-6.560);
\fill[white] (13.540,0.153) rectangle (18.340,-6.560);
\draw[black!75, line width=0.55pt] (13.540,0.153) rectangle (18.340,-6.560);
\draw[black!70, line width=0.3pt] (15.940,0.153) -- (15.940,-6.560);
\draw[gray!30, line width=0.2pt] (13.807,0.153) -- (13.807,-6.560);
\node[font=\scriptsize, anchor=north] at (13.807,-6.640) {-4};
\draw[gray!30, line width=0.2pt] (14.340,0.153) -- (14.340,-6.560);
\node[font=\scriptsize, anchor=north] at (14.340,-6.640) {-3};
\draw[gray!30, line width=0.2pt] (14.873,0.153) -- (14.873,-6.560);
\node[font=\scriptsize, anchor=north] at (14.873,-6.640) {-2};
\draw[gray!30, line width=0.2pt] (15.407,0.153) -- (15.407,-6.560);
\node[font=\scriptsize, anchor=north] at (15.407,-6.640) {-1};
\draw[gray!30, line width=0.2pt] (15.940,0.153) -- (15.940,-6.560);
\node[font=\scriptsize, anchor=north] at (15.940,-6.640) {0};
\draw[gray!30, line width=0.2pt] (16.473,0.153) -- (16.473,-6.560);
\node[font=\scriptsize, anchor=north] at (16.473,-6.640) {1};
\draw[gray!30, line width=0.2pt] (17.007,0.153) -- (17.007,-6.560);
\node[font=\scriptsize, anchor=north] at (17.007,-6.640) {2};
\draw[gray!30, line width=0.2pt] (17.540,0.153) -- (17.540,-6.560);
\node[font=\scriptsize, anchor=north] at (17.540,-6.640) {3};
\draw[gray!30, line width=0.2pt] (18.073,0.153) -- (18.073,-6.560);
\node[font=\scriptsize, anchor=north] at (18.073,-6.640) {4};
\node[font=\small\bfseries, anchor=south] at (15.940,0.340) {High bias (3)};
\draw[poscol, line width=0.8pt] (16.795,-0.000) -- (17.994,-0.000);
\fill[poscol] (17.395,-0.000) circle (0.07);
\filldraw[fill=paircol, draw=black, line width=0.22pt] (17.540,0.085) -- (17.450,-0.075) -- (17.630,-0.075) -- cycle;
\draw[poscol, line width=0.8pt] (16.664,-0.340) -- (17.350,-0.340);
\fill[poscol] (17.007,-0.340) circle (0.07);
\filldraw[fill=paircol, draw=black, line width=0.22pt] (17.540,-0.255) -- (17.450,-0.415) -- (17.630,-0.415) -- cycle;
\draw[poscol, line width=0.8pt] (16.983,-0.680) -- (17.806,-0.680);
\fill[poscol] (17.395,-0.680) circle (0.07);
\filldraw[fill=paircol, draw=black, line width=0.22pt] (18.073,-0.595) -- (17.983,-0.755) -- (18.163,-0.755) -- cycle;
\draw[poscol, line width=0.8pt] (16.140,-1.020) -- (16.858,-1.020);
\fill[poscol] (16.499,-1.020) circle (0.07);
\filldraw[fill=paircol, draw=black, line width=0.22pt] (17.007,-0.935) -- (16.917,-1.095) -- (17.097,-1.095) -- cycle;
\draw[poscol, line width=0.8pt] (16.654,-1.360) -- (17.626,-1.360);
\fill[poscol] (17.140,-1.360) circle (0.07);
\filldraw[fill=paircol, draw=black, line width=0.22pt] (17.540,-1.275) -- (17.450,-1.435) -- (17.630,-1.435) -- cycle;
\draw[poscol, line width=0.8pt] (17.195,-1.700) -- (18.152,-1.700);
\fill[poscol] (17.673,-1.700) circle (0.07);
\filldraw[fill=paircol, draw=black, line width=0.22pt] (17.540,-1.615) -- (17.450,-1.775) -- (17.630,-1.775) -- cycle;
\draw[neucol, line width=0.8pt] (15.863,-2.040) -- (17.795,-2.040);
\fill[neucol] (16.829,-2.040) circle (0.07);
\filldraw[fill=paircol, draw=black, line width=0.22pt] (16.473,-1.955) -- (16.383,-2.115) -- (16.563,-2.115) -- cycle;
\draw[negcol, line width=0.8pt] (14.697,-2.380) -- (15.149,-2.380);
\fill[negcol] (14.923,-2.380) circle (0.07);
\filldraw[fill=paircol, draw=black, line width=0.22pt] (15.407,-2.295) -- (15.317,-2.455) -- (15.497,-2.455) -- cycle;
\draw[negcol, line width=0.8pt] (15.257,-2.720) -- (15.810,-2.720);
\fill[negcol] (15.534,-2.720) circle (0.07);
\filldraw[fill=paircol, draw=black, line width=0.22pt] (15.940,-2.635) -- (15.850,-2.795) -- (16.030,-2.795) -- cycle;
\draw[neucol, line width=0.8pt] (15.234,-3.060) -- (16.172,-3.060);
\fill[neucol] (15.703,-3.060) circle (0.07);
\filldraw[fill=paircol, draw=black, line width=0.22pt] (15.940,-2.975) -- (15.850,-3.135) -- (16.030,-3.135) -- cycle;
\draw[neucol, line width=0.8pt] (15.494,-3.400) -- (16.101,-3.400);
\fill[neucol] (15.798,-3.400) circle (0.07);
\filldraw[fill=paircol, draw=black, line width=0.22pt] (15.940,-3.315) -- (15.850,-3.475) -- (16.030,-3.475) -- cycle;
\draw[neucol, line width=0.8pt] (15.622,-3.740) -- (16.340,-3.740);
\fill[neucol] (15.981,-3.740) circle (0.07);
\filldraw[fill=paircol, draw=black, line width=0.22pt] (15.940,-3.655) -- (15.850,-3.815) -- (16.030,-3.815) -- cycle;
\draw[neucol, line width=0.8pt] (15.929,-4.080) -- (16.840,-4.080);
\fill[neucol] (16.384,-4.080) circle (0.07);
\filldraw[fill=paircol, draw=black, line width=0.22pt] (16.473,-3.995) -- (16.383,-4.155) -- (16.563,-4.155) -- cycle;
\draw[neucol, line width=0.8pt] (15.118,-4.420) -- (17.067,-4.420);
\fill[neucol] (16.092,-4.420) circle (0.07);
\filldraw[fill=paircol, draw=black, line width=0.22pt] (15.407,-4.335) -- (15.317,-4.495) -- (15.497,-4.495) -- cycle;
\draw[negcol, line width=0.8pt] (14.066,-4.760) -- (14.880,-4.760);
\fill[negcol] (14.473,-4.760) circle (0.07);
\filldraw[fill=paircol, draw=black, line width=0.22pt] (14.340,-4.675) -- (14.250,-4.835) -- (14.430,-4.835) -- cycle;
\draw[negcol, line width=0.8pt] (13.969,-5.100) -- (15.042,-5.100);
\fill[negcol] (14.506,-5.100) circle (0.07);
\filldraw[fill=paircol, draw=black, line width=0.22pt] (14.340,-5.015) -- (14.250,-5.175) -- (14.430,-5.175) -- cycle;
\draw[negcol, line width=0.8pt] (14.972,-5.440) -- (15.494,-5.440);
\fill[negcol] (15.233,-5.440) circle (0.07);
\filldraw[fill=paircol, draw=black, line width=0.22pt] (14.873,-5.355) -- (14.783,-5.515) -- (14.963,-5.515) -- cycle;
\draw[negcol, line width=0.8pt] (13.722,-5.780) -- (14.755,-5.780);
\fill[negcol] (14.238,-5.780) circle (0.07);
\filldraw[fill=paircol, draw=black, line width=0.22pt] (13.807,-5.695) -- (13.717,-5.855) -- (13.897,-5.855) -- cycle;
\draw[negcol, line width=0.8pt] (14.373,-6.120) -- (15.003,-6.120);
\fill[negcol] (14.688,-6.120) circle (0.07);
\filldraw[fill=paircol, draw=black, line width=0.22pt] (14.340,-6.035) -- (14.250,-6.195) -- (14.430,-6.195) -- cycle;
\draw[negcol, line width=0.8pt] (13.695,-6.460) -- (15.305,-6.460);
\fill[negcol] (14.500,-6.460) circle (0.07);
\filldraw[fill=paircol, draw=black, line width=0.22pt] (14.340,-6.375) -- (14.250,-6.535) -- (14.430,-6.535) -- cycle;
\draw[black!75, line width=0.55pt] (13.540,0.153) rectangle (18.340,-6.560);
\filldraw[fill=paircol, draw=black, line width=0.22pt] (8.940,-7.275) -- (8.850,-7.435) -- (9.030,-7.435) -- cycle;
\node[font=\scriptsize, anchor=west] at (9.100,-7.360) {theoretical prediction};
\end{tikzpicture}}
\caption{by history and model.}
\label{fig:hyp3_whiskers_h}
\end{subfigure}
\vspace{1.0em}
\begin{minipage}{0.95\textwidth}
\footnotesize
Notes: Whiskers indicate pointwise 95\% confidence intervals with two-way standard errors clustered by receiver and sender. Green and red whiskers indicate positive and negative estimates, respectively, whose pointwise 95\% confidence intervals exclude zero.
\end{minipage}
\caption{Observed persuasiveness, $\tilde{\Delta}_m^S(h,b) - \tilde{\Delta}_m^R(h,b)$.}
\label{fig:hyp3_whiskers_combined}
\end{figure}

%% file: results_artifacts/Figure_17_Observed_and_predicted_receiver_action_under_MEU_and_MLEU.tex
\begin{figure}[p]
\centering
\resizebox{0.8\linewidth}{!}{\begin{tikzpicture}[x=1cm,y=1cm]
\fill[black] (0.450,-0.000) circle (0.10);
\fill[black] (1.100,-0.000) circle (0.10);
\fill[black] (1.750,-0.000) circle (0.10);
\fill[black] (0.450,-0.340) circle (0.10);
\fill[black] (1.100,-0.340) circle (0.10);
\draw[black, line width=0.8pt] (1.630,-0.340) -- (1.870,-0.340);
\draw[red, line width=0.8pt] (0.330,-0.680) -- (0.570,-0.680);
\fill[black] (1.100,-0.680) circle (0.10);
\fill[black] (1.750,-0.680) circle (0.10);
\fill[black] (0.450,-1.020) circle (0.10);
\draw[black, line width=0.8pt] (0.980,-1.020) -- (1.220,-1.020);
\draw[black, line width=0.8pt] (1.630,-1.020) -- (1.870,-1.020);
\draw[red, line width=0.8pt] (0.330,-1.360) -- (0.570,-1.360);
\fill[black] (1.100,-1.360) circle (0.10);
\draw[black, line width=0.8pt] (1.630,-1.360) -- (1.870,-1.360);
\draw[red, line width=0.8pt] (0.330,-1.700) -- (0.570,-1.700);
\draw[red, line width=0.8pt] (0.980,-1.700) -- (1.220,-1.700);
\fill[black] (1.750,-1.700) circle (0.10);
\fill[red] (0.450,-2.040) circle (0.10);
\fill[black] (1.100,-2.040) circle (0.10);
\fill[black] (1.750,-2.040) circle (0.10);
\draw[black, line width=0.8pt] (0.330,-2.380) -- (0.570,-2.380);
\draw[black, line width=0.8pt] (0.980,-2.380) -- (1.220,-2.380);
\draw[black, line width=0.8pt] (1.630,-2.380) -- (1.870,-2.380);
\draw[red, line width=0.8pt] (0.330,-2.720) -- (0.570,-2.720);
\draw[black, line width=0.8pt] (0.980,-2.720) -- (1.220,-2.720);
\draw[black, line width=0.8pt] (1.630,-2.720) -- (1.870,-2.720);
\fill[red] (0.450,-3.060) circle (0.10);
\fill[black] (1.100,-3.060) circle (0.10);
\draw[black, line width=0.8pt] (1.630,-3.060) -- (1.870,-3.060);
\draw[red, line width=0.8pt] (0.330,-3.400) -- (0.570,-3.400);
\draw[red, line width=0.8pt] (0.980,-3.400) -- (1.220,-3.400);
\draw[black, line width=0.8pt] (1.630,-3.400) -- (1.870,-3.400);
\fill[red] (0.450,-3.740) circle (0.10);
\draw[red, line width=0.8pt] (0.980,-3.740) -- (1.220,-3.740);
\fill[black] (1.750,-3.740) circle (0.10);
\draw[red, line width=0.8pt] (0.330,-4.080) -- (0.570,-4.080);
\draw[red, line width=0.8pt] (0.980,-4.080) -- (1.220,-4.080);
\draw[red, line width=0.8pt] (1.630,-4.080) -- (1.870,-4.080);
\fill[red] (0.450,-4.420) circle (0.10);
\fill[red] (1.100,-4.420) circle (0.10);
\fill[black] (1.750,-4.420) circle (0.10);
\fill[red] (0.450,-4.760) circle (0.10);
\draw[black, line width=0.8pt] (0.980,-4.760) -- (1.220,-4.760);
\draw[black, line width=0.8pt] (1.630,-4.760) -- (1.870,-4.760);
\fill[red] (0.450,-5.100) circle (0.10);
\draw[red, line width=0.8pt] (0.980,-5.100) -- (1.220,-5.100);
\draw[black, line width=0.8pt] (1.630,-5.100) -- (1.870,-5.100);
\fill[red] (0.450,-5.440) circle (0.10);
\draw[red, line width=0.8pt] (0.980,-5.440) -- (1.220,-5.440);
\draw[red, line width=0.8pt] (1.630,-5.440) -- (1.870,-5.440);
\fill[red] (0.450,-5.780) circle (0.10);
\fill[red] (1.100,-5.780) circle (0.10);
\draw[black, line width=0.8pt] (1.630,-5.780) -- (1.870,-5.780);
\fill[red] (0.450,-6.120) circle (0.10);
\fill[red] (1.100,-6.120) circle (0.10);
\draw[red, line width=0.8pt] (1.630,-6.120) -- (1.870,-6.120);
\fill[red] (0.450,-6.460) circle (0.10);
\fill[red] (1.100,-6.460) circle (0.10);
\fill[red] (1.750,-6.460) circle (0.10);
\node[font=\small\bfseries, anchor=south] at (1.100,0.340) {};
\fill[white] (2.840,0.153) rectangle (7.640,-6.560);
\draw[black!75, line width=0.55pt] (2.840,0.153) rectangle (7.640,-6.560);
\draw[black!70, line width=0.3pt] (2.840,0.153) -- (2.840,-6.560);
\draw[gray!30, line width=0.2pt] (2.840,0.153) -- (2.840,-6.560);
\node[font=\scriptsize, anchor=north] at (2.840,-6.640) {0};
\draw[gray!30, line width=0.2pt] (3.320,0.153) -- (3.320,-6.560);
\node[font=\scriptsize, anchor=north] at (3.320,-6.640) {1};
\draw[gray!30, line width=0.2pt] (3.800,0.153) -- (3.800,-6.560);
\node[font=\scriptsize, anchor=north] at (3.800,-6.640) {2};
\draw[gray!30, line width=0.2pt] (4.280,0.153) -- (4.280,-6.560);
\node[font=\scriptsize, anchor=north] at (4.280,-6.640) {3};
\draw[gray!30, line width=0.2pt] (4.760,0.153) -- (4.760,-6.560);
\node[font=\scriptsize, anchor=north] at (4.760,-6.640) {4};
\draw[gray!30, line width=0.2pt] (5.240,0.153) -- (5.240,-6.560);
\node[font=\scriptsize, anchor=north] at (5.240,-6.640) {5};
\draw[gray!30, line width=0.2pt] (5.720,0.153) -- (5.720,-6.560);
\node[font=\scriptsize, anchor=north] at (5.720,-6.640) {6};
\draw[gray!30, line width=0.2pt] (6.200,0.153) -- (6.200,-6.560);
\node[font=\scriptsize, anchor=north] at (6.200,-6.640) {7};
\draw[gray!30, line width=0.2pt] (6.680,0.153) -- (6.680,-6.560);
\node[font=\scriptsize, anchor=north] at (6.680,-6.640) {8};
\draw[gray!30, line width=0.2pt] (7.160,0.153) -- (7.160,-6.560);
\node[font=\scriptsize, anchor=north] at (7.160,-6.640) {9};
\draw[gray!30, line width=0.2pt] (7.640,0.153) -- (7.640,-6.560);
\node[font=\scriptsize, anchor=north] at (7.640,-6.640) {10};
\node[font=\small\bfseries, anchor=south] at (5.240,0.340) {No bias (0)};
\draw[neucol, line width=0.8pt] (2.299,-0.000) -- (5.941,-0.000);
\fill[neucol] (4.120,-0.000) circle (0.07);
\filldraw[fill=predone, draw=black, line width=0.22pt] (3.320,0.085) -- (3.230,-0.075) -- (3.410,-0.075) -- cycle;
\filldraw[fill=predtwo, draw=black, line width=0.22pt] (3.320,-0.085) -- (3.230,0.075) -- (3.410,0.075) -- cycle;
\draw[neucol, line width=0.8pt] (3.809,-0.340) -- (4.523,-0.340);
\fill[neucol] (4.166,-0.340) circle (0.07);
\filldraw[fill=predone, draw=black, line width=0.22pt] (3.320,-0.255) -- (3.230,-0.415) -- (3.410,-0.415) -- cycle;
\filldraw[fill=predtwo, draw=black, line width=0.22pt] (3.320,-0.425) -- (3.230,-0.265) -- (3.410,-0.265) -- cycle;
\draw[neucol, line width=0.8pt] (3.788,-0.680) -- (4.964,-0.680);
\fill[neucol] (4.376,-0.680) circle (0.07);
\filldraw[fill=predone, draw=black, line width=0.22pt] (3.320,-0.595) -- (3.230,-0.755) -- (3.410,-0.755) -- cycle;
\filldraw[fill=predtwo, draw=black, line width=0.22pt] (3.320,-0.765) -- (3.230,-0.605) -- (3.410,-0.605) -- cycle;
\draw[neucol, line width=0.8pt] (3.873,-1.020) -- (4.392,-1.020);
\fill[neucol] (4.132,-1.020) circle (0.07);
\filldraw[fill=predone, draw=black, line width=0.22pt] (3.800,-0.935) -- (3.710,-1.095) -- (3.890,-1.095) -- cycle;
\filldraw[fill=predtwo, draw=black, line width=0.22pt] (3.800,-1.105) -- (3.710,-0.945) -- (3.890,-0.945) -- cycle;
\draw[neucol, line width=0.8pt] (3.964,-1.360) -- (4.733,-1.360);
\fill[neucol] (4.349,-1.360) circle (0.07);
\filldraw[fill=predone, draw=black, line width=0.22pt] (3.800,-1.275) -- (3.710,-1.435) -- (3.890,-1.435) -- cycle;
\filldraw[fill=predtwo, draw=black, line width=0.22pt] (3.800,-1.445) -- (3.710,-1.285) -- (3.890,-1.285) -- cycle;
\draw[neucol, line width=0.8pt] (4.218,-1.700) -- (5.382,-1.700);
\fill[neucol] (4.800,-1.700) circle (0.07);
\filldraw[fill=predone, draw=black, line width=0.22pt] (3.800,-1.615) -- (3.710,-1.775) -- (3.890,-1.775) -- cycle;
\filldraw[fill=predtwo, draw=black, line width=0.22pt] (3.800,-1.785) -- (3.710,-1.625) -- (3.890,-1.625) -- cycle;
\draw[neucol, line width=0.8pt] (4.242,-2.040) -- (4.894,-2.040);
\fill[neucol] (4.568,-2.040) circle (0.07);
\filldraw[fill=predone, draw=black, line width=0.22pt] (4.760,-1.955) -- (4.670,-2.115) -- (4.850,-2.115) -- cycle;
\filldraw[fill=predtwo, draw=black, line width=0.22pt] (4.760,-2.125) -- (4.670,-1.965) -- (4.850,-1.965) -- cycle;
\draw[neucol, line width=0.8pt] (4.875,-2.380) -- (6.042,-2.380);
\fill[neucol] (5.458,-2.380) circle (0.07);
\filldraw[fill=predone, draw=black, line width=0.22pt] (5.240,-2.295) -- (5.150,-2.455) -- (5.330,-2.455) -- cycle;
\filldraw[fill=predtwo, draw=black, line width=0.22pt] (5.240,-2.465) -- (5.150,-2.305) -- (5.330,-2.305) -- cycle;
\draw[neucol, line width=0.8pt] (4.506,-2.720) -- (5.115,-2.720);
\fill[neucol] (4.811,-2.720) circle (0.07);
\filldraw[fill=predone, draw=black, line width=0.22pt] (5.240,-2.635) -- (5.150,-2.795) -- (5.330,-2.795) -- cycle;
\filldraw[fill=predtwo, draw=black, line width=0.22pt] (5.240,-2.805) -- (5.150,-2.645) -- (5.330,-2.645) -- cycle;
\draw[neucol, line width=0.8pt] (4.788,-3.060) -- (5.308,-3.060);
\fill[neucol] (5.048,-3.060) circle (0.07);
\filldraw[fill=predone, draw=black, line width=0.22pt] (5.240,-2.975) -- (5.150,-3.135) -- (5.330,-3.135) -- cycle;
\filldraw[fill=predtwo, draw=black, line width=0.22pt] (5.240,-3.145) -- (5.150,-2.985) -- (5.330,-2.985) -- cycle;
\draw[neucol, line width=0.8pt] (4.446,-3.400) -- (5.374,-3.400);
\fill[neucol] (4.910,-3.400) circle (0.07);
\filldraw[fill=predone, draw=black, line width=0.22pt] (5.240,-3.315) -- (5.150,-3.475) -- (5.330,-3.475) -- cycle;
\filldraw[fill=predtwo, draw=black, line width=0.22pt] (5.240,-3.485) -- (5.150,-3.325) -- (5.330,-3.325) -- cycle;
\draw[neucol, line width=0.8pt] (4.791,-3.740) -- (5.561,-3.740);
\fill[neucol] (5.176,-3.740) circle (0.07);
\filldraw[fill=predone, draw=black, line width=0.22pt] (5.240,-3.655) -- (5.150,-3.815) -- (5.330,-3.815) -- cycle;
\filldraw[fill=predtwo, draw=black, line width=0.22pt] (5.240,-3.825) -- (5.150,-3.665) -- (5.330,-3.665) -- cycle;
\draw[neucol, line width=0.8pt] (4.442,-4.080) -- (5.817,-4.080);
\fill[neucol] (5.129,-4.080) circle (0.07);
\filldraw[fill=predone, draw=black, line width=0.22pt] (5.240,-3.995) -- (5.150,-4.155) -- (5.330,-4.155) -- cycle;
\filldraw[fill=predtwo, draw=black, line width=0.22pt] (5.240,-4.165) -- (5.150,-4.005) -- (5.330,-4.005) -- cycle;
\draw[neucol, line width=0.8pt] (5.051,-4.420) -- (6.229,-4.420);
\fill[neucol] (5.640,-4.420) circle (0.07);
\filldraw[fill=predone, draw=black, line width=0.22pt] (5.720,-4.335) -- (5.630,-4.495) -- (5.810,-4.495) -- cycle;
\filldraw[fill=predtwo, draw=black, line width=0.22pt] (5.720,-4.505) -- (5.630,-4.345) -- (5.810,-4.345) -- cycle;
\draw[neucol, line width=0.8pt] (4.312,-4.760) -- (6.088,-4.760);
\fill[neucol] (5.200,-4.760) circle (0.07);
\filldraw[fill=predone, draw=black, line width=0.22pt] (6.680,-4.675) -- (6.590,-4.835) -- (6.770,-4.835) -- cycle;
\filldraw[fill=predtwo, draw=black, line width=0.22pt] (6.680,-4.845) -- (6.590,-4.685) -- (6.770,-4.685) -- cycle;
\draw[neucol, line width=0.8pt] (5.657,-5.100) -- (6.674,-5.100);
\fill[neucol] (6.166,-5.100) circle (0.07);
\filldraw[fill=predone, draw=black, line width=0.22pt] (6.680,-5.015) -- (6.590,-5.175) -- (6.770,-5.175) -- cycle;
\filldraw[fill=predtwo, draw=black, line width=0.22pt] (6.680,-5.185) -- (6.590,-5.025) -- (6.770,-5.025) -- cycle;
\draw[neucol, line width=0.8pt] (5.654,-5.440) -- (6.353,-5.440);
\fill[neucol] (6.004,-5.440) circle (0.07);
\filldraw[fill=predone, draw=black, line width=0.22pt] (6.680,-5.355) -- (6.590,-5.515) -- (6.770,-5.515) -- cycle;
\filldraw[fill=predtwo, draw=black, line width=0.22pt] (6.680,-5.525) -- (6.590,-5.365) -- (6.770,-5.365) -- cycle;
\draw[neucol, line width=0.8pt] (5.599,-5.780) -- (6.654,-5.780);
\fill[neucol] (6.126,-5.780) circle (0.07);
\filldraw[fill=predone, draw=black, line width=0.22pt] (7.160,-5.695) -- (7.070,-5.855) -- (7.250,-5.855) -- cycle;
\filldraw[fill=predtwo, draw=black, line width=0.22pt] (7.160,-5.865) -- (7.070,-5.705) -- (7.250,-5.705) -- cycle;
\draw[neucol, line width=0.8pt] (5.871,-6.120) -- (6.681,-6.120);
\fill[neucol] (6.276,-6.120) circle (0.07);
\filldraw[fill=predone, draw=black, line width=0.22pt] (7.160,-6.035) -- (7.070,-6.195) -- (7.250,-6.195) -- cycle;
\filldraw[fill=predtwo, draw=black, line width=0.22pt] (7.160,-6.205) -- (7.070,-6.045) -- (7.250,-6.045) -- cycle;
\draw[neucol, line width=0.8pt] (6.312,-6.460) -- (7.288,-6.460);
\fill[neucol] (6.800,-6.460) circle (0.07);
\filldraw[fill=predone, draw=black, line width=0.22pt] (7.160,-6.375) -- (7.070,-6.535) -- (7.250,-6.535) -- cycle;
\filldraw[fill=predtwo, draw=black, line width=0.22pt] (7.160,-6.545) -- (7.070,-6.385) -- (7.250,-6.385) -- cycle;
\draw[black!75, line width=0.55pt] (2.840,0.153) rectangle (7.640,-6.560);
\fill[white] (8.190,0.153) rectangle (12.990,-6.560);
\draw[black!75, line width=0.55pt] (8.190,0.153) rectangle (12.990,-6.560);
\draw[black!70, line width=0.3pt] (8.190,0.153) -- (8.190,-6.560);
\draw[gray!30, line width=0.2pt] (8.190,0.153) -- (8.190,-6.560);
\node[font=\scriptsize, anchor=north] at (8.190,-6.640) {0};
\draw[gray!30, line width=0.2pt] (8.670,0.153) -- (8.670,-6.560);
\node[font=\scriptsize, anchor=north] at (8.670,-6.640) {1};
\draw[gray!30, line width=0.2pt] (9.150,0.153) -- (9.150,-6.560);
\node[font=\scriptsize, anchor=north] at (9.150,-6.640) {2};
\draw[gray!30, line width=0.2pt] (9.630,0.153) -- (9.630,-6.560);
\node[font=\scriptsize, anchor=north] at (9.630,-6.640) {3};
\draw[gray!30, line width=0.2pt] (10.110,0.153) -- (10.110,-6.560);
\node[font=\scriptsize, anchor=north] at (10.110,-6.640) {4};
\draw[gray!30, line width=0.2pt] (10.590,0.153) -- (10.590,-6.560);
\node[font=\scriptsize, anchor=north] at (10.590,-6.640) {5};
\draw[gray!30, line width=0.2pt] (11.070,0.153) -- (11.070,-6.560);
\node[font=\scriptsize, anchor=north] at (11.070,-6.640) {6};
\draw[gray!30, line width=0.2pt] (11.550,0.153) -- (11.550,-6.560);
\node[font=\scriptsize, anchor=north] at (11.550,-6.640) {7};
\draw[gray!30, line width=0.2pt] (12.030,0.153) -- (12.030,-6.560);
\node[font=\scriptsize, anchor=north] at (12.030,-6.640) {8};
\draw[gray!30, line width=0.2pt] (12.510,0.153) -- (12.510,-6.560);
\node[font=\scriptsize, anchor=north] at (12.510,-6.640) {9};
\draw[gray!30, line width=0.2pt] (12.990,0.153) -- (12.990,-6.560);
\node[font=\scriptsize, anchor=north] at (12.990,-6.640) {10};
\node[font=\small\bfseries, anchor=south] at (10.590,0.340) {Low bias (1)};
\draw[neucol, line width=0.8pt] (9.165,-0.000) -- (10.507,-0.000);
\fill[neucol] (9.836,-0.000) circle (0.07);
\filldraw[fill=predone, draw=black, line width=0.22pt] (8.830,0.085) -- (8.740,-0.075) -- (8.920,-0.075) -- cycle;
\filldraw[fill=predtwo, draw=black, line width=0.22pt] (8.670,-0.085) -- (8.580,0.075) -- (8.760,0.075) -- cycle;
\draw[neucol, line width=0.8pt] (9.149,-0.340) -- (9.877,-0.340);
\fill[neucol] (9.513,-0.340) circle (0.07);
\filldraw[fill=predone, draw=black, line width=0.22pt] (9.310,-0.255) -- (9.220,-0.415) -- (9.400,-0.415) -- cycle;
\filldraw[fill=predtwo, draw=black, line width=0.22pt] (8.670,-0.425) -- (8.580,-0.265) -- (8.760,-0.265) -- cycle;
\draw[neucol, line width=0.8pt] (9.693,-0.680) -- (10.628,-0.680);
\fill[neucol] (10.161,-0.680) circle (0.07);
\filldraw[fill=predone, draw=black, line width=0.22pt] (8.910,-0.595) -- (8.820,-0.755) -- (9.000,-0.755) -- cycle;
\filldraw[fill=predtwo, draw=black, line width=0.22pt] (8.670,-0.765) -- (8.580,-0.605) -- (8.760,-0.605) -- cycle;
\draw[neucol, line width=0.8pt] (9.390,-1.020) -- (9.986,-1.020);
\fill[neucol] (9.688,-1.020) circle (0.07);
\filldraw[fill=predone, draw=black, line width=0.22pt] (9.950,-0.935) -- (9.860,-1.095) -- (10.040,-1.095) -- cycle;
\filldraw[fill=predtwo, draw=black, line width=0.22pt] (10.590,-1.105) -- (10.500,-0.945) -- (10.680,-0.945) -- cycle;
\draw[neucol, line width=0.8pt] (9.772,-1.360) -- (10.698,-1.360);
\fill[neucol] (10.235,-1.360) circle (0.07);
\filldraw[fill=predone, draw=black, line width=0.22pt] (9.870,-1.275) -- (9.780,-1.435) -- (9.960,-1.435) -- cycle;
\filldraw[fill=predtwo, draw=black, line width=0.22pt] (10.590,-1.445) -- (10.500,-1.285) -- (10.680,-1.285) -- cycle;
\draw[neucol, line width=0.8pt] (10.133,-1.700) -- (11.111,-1.700);
\fill[neucol] (10.622,-1.700) circle (0.07);
\filldraw[fill=predone, draw=black, line width=0.22pt] (9.150,-1.615) -- (9.060,-1.775) -- (9.240,-1.775) -- cycle;
\filldraw[fill=predtwo, draw=black, line width=0.22pt] (9.150,-1.785) -- (9.060,-1.625) -- (9.240,-1.625) -- cycle;
\draw[neucol, line width=0.8pt] (9.068,-2.040) -- (10.832,-2.040);
\fill[neucol] (9.950,-2.040) circle (0.07);
\filldraw[fill=predone, draw=black, line width=0.22pt] (10.590,-1.955) -- (10.500,-2.115) -- (10.680,-2.115) -- cycle;
\filldraw[fill=predtwo, draw=black, line width=0.22pt] (10.590,-2.125) -- (10.500,-1.965) -- (10.680,-1.965) -- cycle;
\draw[neucol, line width=0.8pt] (9.484,-2.380) -- (10.207,-2.380);
\fill[neucol] (9.845,-2.380) circle (0.07);
\filldraw[fill=predone, draw=black, line width=0.22pt] (10.430,-2.295) -- (10.340,-2.455) -- (10.520,-2.455) -- cycle;
\filldraw[fill=predtwo, draw=black, line width=0.22pt] (10.590,-2.465) -- (10.500,-2.305) -- (10.680,-2.305) -- cycle;
\draw[neucol, line width=0.8pt] (10.350,-2.720) -- (10.985,-2.720);
\fill[neucol] (10.667,-2.720) circle (0.07);
\filldraw[fill=predone, draw=black, line width=0.22pt] (10.590,-2.635) -- (10.500,-2.795) -- (10.680,-2.795) -- cycle;
\filldraw[fill=predtwo, draw=black, line width=0.22pt] (10.590,-2.805) -- (10.500,-2.645) -- (10.680,-2.645) -- cycle;
\draw[neucol, line width=0.8pt] (10.338,-3.060) -- (10.984,-3.060);
\fill[neucol] (10.661,-3.060) circle (0.07);
\filldraw[fill=predone, draw=black, line width=0.22pt] (11.070,-2.975) -- (10.980,-3.135) -- (11.160,-3.135) -- cycle;
\filldraw[fill=predtwo, draw=black, line width=0.22pt] (10.590,-3.145) -- (10.500,-2.985) -- (10.680,-2.985) -- cycle;
\draw[neucol, line width=0.8pt] (10.470,-3.400) -- (11.085,-3.400);
\fill[neucol] (10.778,-3.400) circle (0.07);
\filldraw[fill=predone, draw=black, line width=0.22pt] (10.590,-3.315) -- (10.500,-3.475) -- (10.680,-3.475) -- cycle;
\filldraw[fill=predtwo, draw=black, line width=0.22pt] (10.590,-3.485) -- (10.500,-3.325) -- (10.680,-3.325) -- cycle;
\draw[neucol, line width=0.8pt] (10.454,-3.740) -- (10.870,-3.740);
\fill[neucol] (10.662,-3.740) circle (0.07);
\filldraw[fill=predone, draw=black, line width=0.22pt] (11.070,-3.655) -- (10.980,-3.815) -- (11.160,-3.815) -- cycle;
\filldraw[fill=predtwo, draw=black, line width=0.22pt] (10.590,-3.825) -- (10.500,-3.665) -- (10.680,-3.665) -- cycle;
\draw[neucol, line width=0.8pt] (10.894,-4.080) -- (11.678,-4.080);
\fill[neucol] (11.286,-4.080) circle (0.07);
\filldraw[fill=predone, draw=black, line width=0.22pt] (10.590,-3.995) -- (10.500,-4.155) -- (10.680,-4.155) -- cycle;
\filldraw[fill=predtwo, draw=black, line width=0.22pt] (10.590,-4.165) -- (10.500,-4.005) -- (10.680,-4.005) -- cycle;
\draw[neucol, line width=0.8pt] (10.027,-4.420) -- (12.273,-4.420);
\fill[neucol] (11.150,-4.420) circle (0.07);
\filldraw[fill=predone, draw=black, line width=0.22pt] (11.550,-4.335) -- (11.460,-4.495) -- (11.640,-4.495) -- cycle;
\filldraw[fill=predtwo, draw=black, line width=0.22pt] (12.510,-4.505) -- (12.420,-4.345) -- (12.600,-4.345) -- cycle;
\draw[neucol, line width=0.8pt] (10.818,-4.760) -- (11.617,-4.760);
\fill[neucol] (11.218,-4.760) circle (0.07);
\filldraw[fill=predone, draw=black, line width=0.22pt] (11.670,-4.675) -- (11.580,-4.835) -- (11.760,-4.835) -- cycle;
\filldraw[fill=predtwo, draw=black, line width=0.22pt] (12.030,-4.845) -- (11.940,-4.685) -- (12.120,-4.685) -- cycle;
\draw[neucol, line width=0.8pt] (11.060,-5.100) -- (11.949,-5.100);
\fill[neucol] (11.504,-5.100) circle (0.07);
\filldraw[fill=predone, draw=black, line width=0.22pt] (11.550,-5.015) -- (11.460,-5.175) -- (11.640,-5.175) -- cycle;
\filldraw[fill=predtwo, draw=black, line width=0.22pt] (12.510,-5.185) -- (12.420,-5.025) -- (12.600,-5.025) -- cycle;
\draw[neucol, line width=0.8pt] (11.361,-5.440) -- (11.863,-5.440);
\fill[neucol] (11.612,-5.440) circle (0.07);
\filldraw[fill=predone, draw=black, line width=0.22pt] (12.030,-5.355) -- (11.940,-5.515) -- (12.120,-5.515) -- cycle;
\filldraw[fill=predtwo, draw=black, line width=0.22pt] (12.510,-5.525) -- (12.420,-5.365) -- (12.600,-5.365) -- cycle;
\draw[neucol, line width=0.8pt] (11.110,-5.780) -- (11.810,-5.780);
\fill[neucol] (11.460,-5.780) circle (0.07);
\filldraw[fill=predone, draw=black, line width=0.22pt] (11.550,-5.695) -- (11.460,-5.855) -- (11.640,-5.855) -- cycle;
\filldraw[fill=predtwo, draw=black, line width=0.22pt] (12.510,-5.865) -- (12.420,-5.705) -- (12.600,-5.705) -- cycle;
\draw[neucol, line width=0.8pt] (11.622,-6.120) -- (12.109,-6.120);
\fill[neucol] (11.865,-6.120) circle (0.07);
\filldraw[fill=predone, draw=black, line width=0.22pt] (12.030,-6.035) -- (11.940,-6.195) -- (12.120,-6.195) -- cycle;
\filldraw[fill=predtwo, draw=black, line width=0.22pt] (12.510,-6.205) -- (12.420,-6.045) -- (12.600,-6.045) -- cycle;
\draw[neucol, line width=0.8pt] (11.615,-6.460) -- (12.552,-6.460);
\fill[neucol] (12.083,-6.460) circle (0.07);
\filldraw[fill=predone, draw=black, line width=0.22pt] (12.030,-6.375) -- (11.940,-6.535) -- (12.120,-6.535) -- cycle;
\filldraw[fill=predtwo, draw=black, line width=0.22pt] (12.510,-6.545) -- (12.420,-6.385) -- (12.600,-6.385) -- cycle;
\draw[black!75, line width=0.55pt] (8.190,0.153) rectangle (12.990,-6.560);
\fill[white] (13.540,0.153) rectangle (18.340,-6.560);
\draw[black!75, line width=0.55pt] (13.540,0.153) rectangle (18.340,-6.560);
\draw[black!70, line width=0.3pt] (13.540,0.153) -- (13.540,-6.560);
\draw[gray!30, line width=0.2pt] (13.540,0.153) -- (13.540,-6.560);
\node[font=\scriptsize, anchor=north] at (13.540,-6.640) {0};
\draw[gray!30, line width=0.2pt] (14.020,0.153) -- (14.020,-6.560);
\node[font=\scriptsize, anchor=north] at (14.020,-6.640) {1};
\draw[gray!30, line width=0.2pt] (14.500,0.153) -- (14.500,-6.560);
\node[font=\scriptsize, anchor=north] at (14.500,-6.640) {2};
\draw[gray!30, line width=0.2pt] (14.980,0.153) -- (14.980,-6.560);
\node[font=\scriptsize, anchor=north] at (14.980,-6.640) {3};
\draw[gray!30, line width=0.2pt] (15.460,0.153) -- (15.460,-6.560);
\node[font=\scriptsize, anchor=north] at (15.460,-6.640) {4};
\draw[gray!30, line width=0.2pt] (15.940,0.153) -- (15.940,-6.560);
\node[font=\scriptsize, anchor=north] at (15.940,-6.640) {5};
\draw[gray!30, line width=0.2pt] (16.420,0.153) -- (16.420,-6.560);
\node[font=\scriptsize, anchor=north] at (16.420,-6.640) {6};
\draw[gray!30, line width=0.2pt] (16.900,0.153) -- (16.900,-6.560);
\node[font=\scriptsize, anchor=north] at (16.900,-6.640) {7};
\draw[gray!30, line width=0.2pt] (17.380,0.153) -- (17.380,-6.560);
\node[font=\scriptsize, anchor=north] at (17.380,-6.640) {8};
\draw[gray!30, line width=0.2pt] (17.860,0.153) -- (17.860,-6.560);
\node[font=\scriptsize, anchor=north] at (17.860,-6.640) {9};
\draw[gray!30, line width=0.2pt] (18.340,0.153) -- (18.340,-6.560);
\node[font=\scriptsize, anchor=north] at (18.340,-6.640) {10};
\node[font=\small\bfseries, anchor=south] at (15.940,0.340) {High bias (3)};
\draw[neucol, line width=0.8pt] (14.790,-0.000) -- (15.869,-0.000);
\fill[neucol] (15.329,-0.000) circle (0.07);
\filldraw[fill=predone, draw=black, line width=0.22pt] (15.460,0.085) -- (15.370,-0.075) -- (15.550,-0.075) -- cycle;
\filldraw[fill=predtwo, draw=black, line width=0.22pt] (14.020,-0.085) -- (13.930,0.075) -- (14.110,0.075) -- cycle;
\draw[neucol, line width=0.8pt] (14.659,-0.340) -- (15.301,-0.340);
\fill[neucol] (14.980,-0.340) circle (0.07);
\filldraw[fill=predone, draw=black, line width=0.22pt] (15.460,-0.255) -- (15.370,-0.415) -- (15.550,-0.415) -- cycle;
\filldraw[fill=predtwo, draw=black, line width=0.22pt] (14.020,-0.425) -- (13.930,-0.265) -- (14.110,-0.265) -- cycle;
\draw[neucol, line width=0.8pt] (14.972,-0.680) -- (15.686,-0.680);
\fill[neucol] (15.329,-0.680) circle (0.07);
\filldraw[fill=predone, draw=black, line width=0.22pt] (15.940,-0.595) -- (15.850,-0.755) -- (16.030,-0.755) -- cycle;
\filldraw[fill=predtwo, draw=black, line width=0.22pt] (14.020,-0.765) -- (13.930,-0.605) -- (14.110,-0.605) -- cycle;
\draw[neucol, line width=0.8pt] (14.686,-1.020) -- (15.321,-1.020);
\fill[neucol] (15.003,-1.020) circle (0.07);
\filldraw[fill=predone, draw=black, line width=0.22pt] (15.460,-0.935) -- (15.370,-1.095) -- (15.550,-1.095) -- cycle;
\filldraw[fill=predtwo, draw=black, line width=0.22pt] (14.020,-1.105) -- (13.930,-0.945) -- (14.110,-0.945) -- cycle;
\draw[neucol, line width=0.8pt] (15.156,-1.360) -- (16.004,-1.360);
\fill[neucol] (15.580,-1.360) circle (0.07);
\filldraw[fill=predone, draw=black, line width=0.22pt] (15.940,-1.275) -- (15.850,-1.435) -- (16.030,-1.435) -- cycle;
\filldraw[fill=predtwo, draw=black, line width=0.22pt] (14.020,-1.445) -- (13.930,-1.285) -- (14.110,-1.285) -- cycle;
\draw[neucol, line width=0.8pt] (15.631,-1.700) -- (16.489,-1.700);
\fill[neucol] (16.060,-1.700) circle (0.07);
\filldraw[fill=predone, draw=black, line width=0.22pt] (15.940,-1.615) -- (15.850,-1.775) -- (16.030,-1.775) -- cycle;
\filldraw[fill=predtwo, draw=black, line width=0.22pt] (17.860,-1.785) -- (17.770,-1.625) -- (17.950,-1.625) -- cycle;
\draw[neucol, line width=0.8pt] (15.395,-2.040) -- (17.125,-2.040);
\fill[neucol] (16.260,-2.040) circle (0.07);
\filldraw[fill=predone, draw=black, line width=0.22pt] (15.940,-1.955) -- (15.850,-2.115) -- (16.030,-2.115) -- cycle;
\filldraw[fill=predtwo, draw=black, line width=0.22pt] (14.020,-2.125) -- (13.930,-1.965) -- (14.110,-1.965) -- cycle;
\draw[neucol, line width=0.8pt] (14.793,-2.380) -- (15.257,-2.380);
\fill[neucol] (15.025,-2.380) circle (0.07);
\filldraw[fill=predone, draw=black, line width=0.22pt] (15.460,-2.295) -- (15.370,-2.455) -- (15.550,-2.455) -- cycle;
\filldraw[fill=predtwo, draw=black, line width=0.22pt] (14.020,-2.465) -- (13.930,-2.305) -- (14.110,-2.305) -- cycle;
\draw[neucol, line width=0.8pt] (15.334,-2.720) -- (15.815,-2.720);
\fill[neucol] (15.574,-2.720) circle (0.07);
\filldraw[fill=predone, draw=black, line width=0.22pt] (15.940,-2.635) -- (15.850,-2.795) -- (16.030,-2.795) -- cycle;
\filldraw[fill=predtwo, draw=black, line width=0.22pt] (14.020,-2.805) -- (13.930,-2.645) -- (14.110,-2.645) -- cycle;
\draw[neucol, line width=0.8pt] (15.332,-3.060) -- (16.121,-3.060);
\fill[neucol] (15.727,-3.060) circle (0.07);
\filldraw[fill=predone, draw=black, line width=0.22pt] (15.940,-2.975) -- (15.850,-3.135) -- (16.030,-3.135) -- cycle;
\filldraw[fill=predtwo, draw=black, line width=0.22pt] (14.020,-3.145) -- (13.930,-2.985) -- (14.110,-2.985) -- cycle;
\draw[neucol, line width=0.8pt] (15.562,-3.400) -- (16.062,-3.400);
\fill[neucol] (15.812,-3.400) circle (0.07);
\filldraw[fill=predone, draw=black, line width=0.22pt] (15.940,-3.315) -- (15.850,-3.475) -- (16.030,-3.475) -- cycle;
\filldraw[fill=predtwo, draw=black, line width=0.22pt] (17.860,-3.485) -- (17.770,-3.325) -- (17.950,-3.325) -- cycle;
\draw[neucol, line width=0.8pt] (15.619,-3.740) -- (16.335,-3.740);
\fill[neucol] (15.977,-3.740) circle (0.07);
\filldraw[fill=predone, draw=black, line width=0.22pt] (15.940,-3.655) -- (15.850,-3.815) -- (16.030,-3.815) -- cycle;
\filldraw[fill=predtwo, draw=black, line width=0.22pt] (17.860,-3.825) -- (17.770,-3.665) -- (17.950,-3.665) -- cycle;
\draw[neucol, line width=0.8pt] (15.922,-4.080) -- (16.758,-4.080);
\fill[neucol] (16.340,-4.080) circle (0.07);
\filldraw[fill=predone, draw=black, line width=0.22pt] (16.420,-3.995) -- (16.330,-4.155) -- (16.510,-4.155) -- cycle;
\filldraw[fill=predtwo, draw=black, line width=0.22pt] (17.860,-4.165) -- (17.770,-4.005) -- (17.950,-4.005) -- cycle;
\draw[neucol, line width=0.8pt] (15.680,-4.420) -- (17.434,-4.420);
\fill[neucol] (16.557,-4.420) circle (0.07);
\filldraw[fill=predone, draw=black, line width=0.22pt] (15.940,-4.335) -- (15.850,-4.495) -- (16.030,-4.495) -- cycle;
\filldraw[fill=predtwo, draw=black, line width=0.22pt] (17.860,-4.505) -- (17.770,-4.345) -- (17.950,-4.345) -- cycle;
\draw[neucol, line width=0.8pt] (15.681,-4.760) -- (16.439,-4.760);
\fill[neucol] (16.060,-4.760) circle (0.07);
\filldraw[fill=predone, draw=black, line width=0.22pt] (15.940,-4.675) -- (15.850,-4.835) -- (16.030,-4.835) -- cycle;
\filldraw[fill=predtwo, draw=black, line width=0.22pt] (14.020,-4.845) -- (13.930,-4.685) -- (14.110,-4.685) -- cycle;
\draw[neucol, line width=0.8pt] (15.639,-5.100) -- (16.538,-5.100);
\fill[neucol] (16.089,-5.100) circle (0.07);
\filldraw[fill=predone, draw=black, line width=0.22pt] (15.940,-5.015) -- (15.850,-5.175) -- (16.030,-5.175) -- cycle;
\filldraw[fill=predtwo, draw=black, line width=0.22pt] (17.860,-5.185) -- (17.770,-5.025) -- (17.950,-5.025) -- cycle;
\draw[neucol, line width=0.8pt] (16.495,-5.440) -- (16.993,-5.440);
\fill[neucol] (16.744,-5.440) circle (0.07);
\filldraw[fill=predone, draw=black, line width=0.22pt] (16.420,-5.355) -- (16.330,-5.515) -- (16.510,-5.515) -- cycle;
\filldraw[fill=predtwo, draw=black, line width=0.22pt] (17.860,-5.525) -- (17.770,-5.365) -- (17.950,-5.365) -- cycle;
\draw[neucol, line width=0.8pt] (15.932,-5.780) -- (16.725,-5.780);
\fill[neucol] (16.329,-5.780) circle (0.07);
\filldraw[fill=predone, draw=black, line width=0.22pt] (15.940,-5.695) -- (15.850,-5.855) -- (16.030,-5.855) -- cycle;
\filldraw[fill=predtwo, draw=black, line width=0.22pt] (17.860,-5.865) -- (17.770,-5.705) -- (17.950,-5.705) -- cycle;
\draw[neucol, line width=0.8pt] (16.434,-6.120) -- (17.032,-6.120);
\fill[neucol] (16.733,-6.120) circle (0.07);
\filldraw[fill=predone, draw=black, line width=0.22pt] (16.420,-6.035) -- (16.330,-6.195) -- (16.510,-6.195) -- cycle;
\filldraw[fill=predtwo, draw=black, line width=0.22pt] (17.860,-6.205) -- (17.770,-6.045) -- (17.950,-6.045) -- cycle;
\draw[neucol, line width=0.8pt] (15.839,-6.460) -- (17.289,-6.460);
\fill[neucol] (16.564,-6.460) circle (0.07);
\filldraw[fill=predone, draw=black, line width=0.22pt] (16.420,-6.375) -- (16.330,-6.535) -- (16.510,-6.535) -- cycle;
\filldraw[fill=predtwo, draw=black, line width=0.22pt] (17.860,-6.545) -- (17.770,-6.385) -- (17.950,-6.385) -- cycle;
\draw[black!75, line width=0.55pt] (13.540,0.153) rectangle (18.340,-6.560);
\filldraw[fill=predone, draw=black, line width=0.22pt] (5.390,-7.275) -- (5.300,-7.435) -- (5.480,-7.435) -- cycle;
\node[font=\scriptsize, anchor=west] at (5.550,-7.360) {receiver's eq. action (MEU)};
\filldraw[fill=predtwo, draw=black, line width=0.22pt] (10.590,-7.445) -- (10.500,-7.285) -- (10.680,-7.285) -- cycle;
\node[font=\scriptsize, anchor=west] at (10.750,-7.360) {receiver's eq. action (MLEU)};
\end{tikzpicture}}
\vspace{1.0em}
\begin{minipage}{0.95\textwidth}
\footnotesize
Notes:\ Whiskers indicate pointwise 95\% confidence intervals with standard errors clustered by receiver.
\end{minipage}
\caption{Predicted action of the receiver under MEU and MLEU and the observed actions in the experiment.}
\label{fig:MLEU_policy_R}
\end{figure}

%% file: results_artifacts/Figure_18_Observed_and_predicted_Delta_R_under_MEU_and_MLEU.tex
\begin{figure}[p]
\centering
\begin{subfigure}{0.8\linewidth}
\centering
\resizebox{0.8\textwidth}{!}{\begin{tikzpicture}[x=1cm,y=1cm]
\fill[black] (0.450,-0.000) circle (0.10);
\fill[black] (1.100,-0.000) circle (0.10);
\fill[black] (1.750,-0.000) circle (0.10);
\fill[red] (0.450,-0.340) circle (0.10);
\fill[black] (1.100,-0.340) circle (0.10);
\fill[black] (1.750,-0.340) circle (0.10);
\fill[red] (0.450,-0.680) circle (0.10);
\fill[red] (1.100,-0.680) circle (0.10);
\fill[black] (1.750,-0.680) circle (0.10);
\fill[red] (0.450,-1.020) circle (0.10);
\fill[red] (1.100,-1.020) circle (0.10);
\fill[red] (1.750,-1.020) circle (0.10);
\fill[white] (2.840,0.153) rectangle (7.640,-1.120);
\draw[black!75, line width=0.55pt] (2.840,0.153) rectangle (7.640,-1.120);
\draw[black!70, line width=0.3pt] (5.240,0.153) -- (5.240,-1.120);
\draw[gray!30, line width=0.2pt] (3.107,0.153) -- (3.107,-1.120);
\node[font=\scriptsize, anchor=north] at (3.107,-1.200) {-4};
\draw[gray!30, line width=0.2pt] (3.640,0.153) -- (3.640,-1.120);
\node[font=\scriptsize, anchor=north] at (3.640,-1.200) {-3};
\draw[gray!30, line width=0.2pt] (4.173,0.153) -- (4.173,-1.120);
\node[font=\scriptsize, anchor=north] at (4.173,-1.200) {-2};
\draw[gray!30, line width=0.2pt] (4.707,0.153) -- (4.707,-1.120);
\node[font=\scriptsize, anchor=north] at (4.707,-1.200) {-1};
\draw[gray!30, line width=0.2pt] (5.240,0.153) -- (5.240,-1.120);
\node[font=\scriptsize, anchor=north] at (5.240,-1.200) {0};
\draw[gray!30, line width=0.2pt] (5.773,0.153) -- (5.773,-1.120);
\node[font=\scriptsize, anchor=north] at (5.773,-1.200) {1};
\draw[gray!30, line width=0.2pt] (6.307,0.153) -- (6.307,-1.120);
\node[font=\scriptsize, anchor=north] at (6.307,-1.200) {2};
\draw[gray!30, line width=0.2pt] (6.840,0.153) -- (6.840,-1.120);
\node[font=\scriptsize, anchor=north] at (6.840,-1.200) {3};
\draw[gray!30, line width=0.2pt] (7.373,0.153) -- (7.373,-1.120);
\node[font=\scriptsize, anchor=north] at (7.373,-1.200) {4};
\node[font=\small\bfseries, anchor=south] at (5.240,0.340) {No bias (0)};
\draw[negcol, line width=0.8pt] (4.537,-0.000) -- (5.054,-0.000);
\fill[negcol] (4.796,-0.000) circle (0.07);
\filldraw[fill=predone, draw=black, line width=0.22pt] (5.240,0.085) -- (5.150,-0.075) -- (5.330,-0.075) -- cycle;
\filldraw[fill=predtwo, draw=black, line width=0.22pt] (5.240,-0.085) -- (5.150,0.075) -- (5.330,0.075) -- cycle;
\draw[neucol, line width=0.8pt] (5.058,-0.340) -- (5.516,-0.340);
\fill[neucol] (5.287,-0.340) circle (0.07);
\filldraw[fill=predone, draw=black, line width=0.22pt] (5.240,-0.255) -- (5.150,-0.415) -- (5.330,-0.415) -- cycle;
\filldraw[fill=predtwo, draw=black, line width=0.22pt] (5.240,-0.425) -- (5.150,-0.265) -- (5.330,-0.265) -- cycle;
\draw[poscol, line width=0.8pt] (5.287,-0.680) -- (5.769,-0.680);
\fill[poscol] (5.528,-0.680) circle (0.07);
\filldraw[fill=predone, draw=black, line width=0.22pt] (5.240,-0.595) -- (5.150,-0.755) -- (5.330,-0.755) -- cycle;
\filldraw[fill=predtwo, draw=black, line width=0.22pt] (5.240,-0.765) -- (5.150,-0.605) -- (5.330,-0.605) -- cycle;
\draw[poscol, line width=0.8pt] (5.490,-1.020) -- (5.970,-1.020);
\fill[poscol] (5.730,-1.020) circle (0.07);
\filldraw[fill=predone, draw=black, line width=0.22pt] (5.240,-0.935) -- (5.150,-1.095) -- (5.330,-1.095) -- cycle;
\filldraw[fill=predtwo, draw=black, line width=0.22pt] (5.240,-1.105) -- (5.150,-0.945) -- (5.330,-0.945) -- cycle;
\draw[black!75, line width=0.55pt] (2.840,0.153) rectangle (7.640,-1.120);
\fill[white] (8.190,0.153) rectangle (12.990,-1.120);
\draw[black!75, line width=0.55pt] (8.190,0.153) rectangle (12.990,-1.120);
\draw[black!70, line width=0.3pt] (10.590,0.153) -- (10.590,-1.120);
\draw[gray!30, line width=0.2pt] (8.457,0.153) -- (8.457,-1.120);
\node[font=\scriptsize, anchor=north] at (8.457,-1.200) {-4};
\draw[gray!30, line width=0.2pt] (8.990,0.153) -- (8.990,-1.120);
\node[font=\scriptsize, anchor=north] at (8.990,-1.200) {-3};
\draw[gray!30, line width=0.2pt] (9.523,0.153) -- (9.523,-1.120);
\node[font=\scriptsize, anchor=north] at (9.523,-1.200) {-2};
\draw[gray!30, line width=0.2pt] (10.057,0.153) -- (10.057,-1.120);
\node[font=\scriptsize, anchor=north] at (10.057,-1.200) {-1};
\draw[gray!30, line width=0.2pt] (10.590,0.153) -- (10.590,-1.120);
\node[font=\scriptsize, anchor=north] at (10.590,-1.200) {0};
\draw[gray!30, line width=0.2pt] (11.123,0.153) -- (11.123,-1.120);
\node[font=\scriptsize, anchor=north] at (11.123,-1.200) {1};
\draw[gray!30, line width=0.2pt] (11.657,0.153) -- (11.657,-1.120);
\node[font=\scriptsize, anchor=north] at (11.657,-1.200) {2};
\draw[gray!30, line width=0.2pt] (12.190,0.153) -- (12.190,-1.120);
\node[font=\scriptsize, anchor=north] at (12.190,-1.200) {3};
\draw[gray!30, line width=0.2pt] (12.723,0.153) -- (12.723,-1.120);
\node[font=\scriptsize, anchor=north] at (12.723,-1.200) {4};
\node[font=\small\bfseries, anchor=south] at (10.590,0.340) {Low bias (1)};
\draw[neucol, line width=0.8pt] (10.211,-0.000) -- (10.637,-0.000);
\fill[neucol] (10.424,-0.000) circle (0.07);
\filldraw[fill=predone, draw=black, line width=0.22pt] (10.756,0.085) -- (10.666,-0.075) -- (10.846,-0.075) -- cycle;
\filldraw[fill=predtwo, draw=black, line width=0.22pt] (10.590,-0.085) -- (10.500,0.075) -- (10.680,0.075) -- cycle;
\draw[neucol, line width=0.8pt] (10.480,-0.340) -- (10.924,-0.340);
\fill[neucol] (10.702,-0.340) circle (0.07);
\filldraw[fill=predone, draw=black, line width=0.22pt] (10.694,-0.255) -- (10.604,-0.415) -- (10.784,-0.415) -- cycle;
\filldraw[fill=predtwo, draw=black, line width=0.22pt] (10.590,-0.425) -- (10.500,-0.265) -- (10.680,-0.265) -- cycle;
\draw[poscol, line width=0.8pt] (10.736,-0.680) -- (11.130,-0.680);
\fill[poscol] (10.933,-0.680) circle (0.07);
\filldraw[fill=predone, draw=black, line width=0.22pt] (10.870,-0.595) -- (10.780,-0.755) -- (10.960,-0.755) -- cycle;
\filldraw[fill=predtwo, draw=black, line width=0.22pt] (10.363,-0.765) -- (10.273,-0.605) -- (10.453,-0.605) -- cycle;
\draw[poscol, line width=0.8pt] (10.905,-1.020) -- (11.252,-1.020);
\fill[poscol] (11.078,-1.020) circle (0.07);
\filldraw[fill=predone, draw=black, line width=0.22pt] (11.011,-0.935) -- (10.921,-1.095) -- (11.101,-1.095) -- cycle;
\filldraw[fill=predtwo, draw=black, line width=0.22pt] (10.590,-1.105) -- (10.500,-0.945) -- (10.680,-0.945) -- cycle;
\draw[black!75, line width=0.55pt] (8.190,0.153) rectangle (12.990,-1.120);
\fill[white] (13.540,0.153) rectangle (18.340,-1.120);
\draw[black!75, line width=0.55pt] (13.540,0.153) rectangle (18.340,-1.120);
\draw[black!70, line width=0.3pt] (15.340,0.153) -- (15.340,-1.120);
\draw[gray!30, line width=0.2pt] (13.740,0.153) -- (13.740,-1.120);
\node[font=\scriptsize, anchor=north] at (13.740,-1.200) {-4};
\draw[gray!30, line width=0.2pt] (14.140,0.153) -- (14.140,-1.120);
\node[font=\scriptsize, anchor=north] at (14.140,-1.200) {-3};
\draw[gray!30, line width=0.2pt] (14.540,0.153) -- (14.540,-1.120);
\node[font=\scriptsize, anchor=north] at (14.540,-1.200) {-2};
\draw[gray!30, line width=0.2pt] (14.940,0.153) -- (14.940,-1.120);
\node[font=\scriptsize, anchor=north] at (14.940,-1.200) {-1};
\draw[gray!30, line width=0.2pt] (15.340,0.153) -- (15.340,-1.120);
\node[font=\scriptsize, anchor=north] at (15.340,-1.200) {0};
\draw[gray!30, line width=0.2pt] (15.740,0.153) -- (15.740,-1.120);
\node[font=\scriptsize, anchor=north] at (15.740,-1.200) {1};
\draw[gray!30, line width=0.2pt] (16.140,0.153) -- (16.140,-1.120);
\node[font=\scriptsize, anchor=north] at (16.140,-1.200) {2};
\draw[gray!30, line width=0.2pt] (16.540,0.153) -- (16.540,-1.120);
\node[font=\scriptsize, anchor=north] at (16.540,-1.200) {3};
\draw[gray!30, line width=0.2pt] (16.940,0.153) -- (16.940,-1.120);
\node[font=\scriptsize, anchor=north] at (16.940,-1.200) {4};
\draw[gray!30, line width=0.2pt] (17.340,0.153) -- (17.340,-1.120);
\node[font=\scriptsize, anchor=north] at (17.340,-1.200) {5};
\draw[gray!30, line width=0.2pt] (17.740,0.153) -- (17.740,-1.120);
\node[font=\scriptsize, anchor=north] at (17.740,-1.200) {6};
\draw[gray!30, line width=0.2pt] (18.140,0.153) -- (18.140,-1.120);
\node[font=\scriptsize, anchor=north] at (18.140,-1.200) {7};
\node[font=\small\bfseries, anchor=south] at (15.940,0.340) {High bias (3)};
\draw[neucol, line width=0.8pt] (15.271,-0.000) -- (15.652,-0.000);
\fill[neucol] (15.461,-0.000) circle (0.07);
\filldraw[fill=predone, draw=black, line width=0.22pt] (15.740,0.085) -- (15.650,-0.075) -- (15.830,-0.075) -- cycle;
\filldraw[fill=predtwo, draw=black, line width=0.22pt] (16.940,-0.085) -- (16.850,0.075) -- (17.030,0.075) -- cycle;
\draw[poscol, line width=0.8pt] (15.663,-0.340) -- (15.995,-0.340);
\fill[poscol] (15.829,-0.340) circle (0.07);
\filldraw[fill=predone, draw=black, line width=0.22pt] (16.540,-0.255) -- (16.450,-0.415) -- (16.630,-0.415) -- cycle;
\filldraw[fill=predtwo, draw=black, line width=0.22pt] (18.140,-0.425) -- (18.050,-0.265) -- (18.230,-0.265) -- cycle;
\draw[poscol, line width=0.8pt] (15.695,-0.680) -- (16.108,-0.680);
\fill[poscol] (15.901,-0.680) circle (0.07);
\filldraw[fill=predone, draw=black, line width=0.22pt] (16.540,-0.595) -- (16.450,-0.755) -- (16.630,-0.755) -- cycle;
\filldraw[fill=predtwo, draw=black, line width=0.22pt] (14.940,-0.765) -- (14.850,-0.605) -- (15.030,-0.605) -- cycle;
\draw[poscol, line width=0.8pt] (15.726,-1.020) -- (16.007,-1.020);
\fill[poscol] (15.867,-1.020) circle (0.07);
\filldraw[fill=predone, draw=black, line width=0.22pt] (16.540,-0.935) -- (16.450,-1.095) -- (16.630,-1.095) -- cycle;
\filldraw[fill=predtwo, draw=black, line width=0.22pt] (15.340,-1.105) -- (15.250,-0.945) -- (15.430,-0.945) -- cycle;
\draw[black!75, line width=0.55pt] (13.540,0.153) rectangle (18.340,-1.120);
\filldraw[fill=predone, draw=black, line width=0.22pt] (5.390,-1.835) -- (5.300,-1.995) -- (5.480,-1.995) -- cycle;
\node[font=\scriptsize, anchor=west] at (5.550,-1.920) {MEU};
\filldraw[fill=predtwo, draw=black, line width=0.22pt] (10.590,-2.005) -- (10.500,-1.845) -- (10.680,-1.845) -- cycle;
\node[font=\scriptsize, anchor=west] at (10.750,-1.920) {MLEU};
\end{tikzpicture}}
\caption{by history.}
\label{fig:MLEU_R_whiskers}
\end{subfigure}
\vspace{0.8em}
\begin{subfigure}{0.8\linewidth}
\centering
\resizebox{0.8\textwidth}{!}{\begin{tikzpicture}[x=1cm,y=1cm]
\fill[black] (0.450,-0.000) circle (0.10);
\fill[black] (1.100,-0.000) circle (0.10);
\fill[black] (1.750,-0.000) circle (0.10);
\fill[black] (0.450,-0.340) circle (0.10);
\fill[black] (1.100,-0.340) circle (0.10);
\draw[black, line width=0.8pt] (1.630,-0.340) -- (1.870,-0.340);
\draw[red, line width=0.8pt] (0.330,-0.680) -- (0.570,-0.680);
\fill[black] (1.100,-0.680) circle (0.10);
\fill[black] (1.750,-0.680) circle (0.10);
\fill[black] (0.450,-1.020) circle (0.10);
\draw[black, line width=0.8pt] (0.980,-1.020) -- (1.220,-1.020);
\draw[black, line width=0.8pt] (1.630,-1.020) -- (1.870,-1.020);
\draw[red, line width=0.8pt] (0.330,-1.360) -- (0.570,-1.360);
\fill[black] (1.100,-1.360) circle (0.10);
\draw[black, line width=0.8pt] (1.630,-1.360) -- (1.870,-1.360);
\draw[red, line width=0.8pt] (0.330,-1.700) -- (0.570,-1.700);
\draw[red, line width=0.8pt] (0.980,-1.700) -- (1.220,-1.700);
\fill[black] (1.750,-1.700) circle (0.10);
\fill[red] (0.450,-2.040) circle (0.10);
\fill[black] (1.100,-2.040) circle (0.10);
\fill[black] (1.750,-2.040) circle (0.10);
\draw[black, line width=0.8pt] (0.330,-2.380) -- (0.570,-2.380);
\draw[black, line width=0.8pt] (0.980,-2.380) -- (1.220,-2.380);
\draw[black, line width=0.8pt] (1.630,-2.380) -- (1.870,-2.380);
\draw[red, line width=0.8pt] (0.330,-2.720) -- (0.570,-2.720);
\draw[black, line width=0.8pt] (0.980,-2.720) -- (1.220,-2.720);
\draw[black, line width=0.8pt] (1.630,-2.720) -- (1.870,-2.720);
\fill[red] (0.450,-3.060) circle (0.10);
\fill[black] (1.100,-3.060) circle (0.10);
\draw[black, line width=0.8pt] (1.630,-3.060) -- (1.870,-3.060);
\draw[red, line width=0.8pt] (0.330,-3.400) -- (0.570,-3.400);
\draw[red, line width=0.8pt] (0.980,-3.400) -- (1.220,-3.400);
\draw[black, line width=0.8pt] (1.630,-3.400) -- (1.870,-3.400);
\fill[red] (0.450,-3.740) circle (0.10);
\draw[red, line width=0.8pt] (0.980,-3.740) -- (1.220,-3.740);
\fill[black] (1.750,-3.740) circle (0.10);
\draw[red, line width=0.8pt] (0.330,-4.080) -- (0.570,-4.080);
\draw[red, line width=0.8pt] (0.980,-4.080) -- (1.220,-4.080);
\draw[red, line width=0.8pt] (1.630,-4.080) -- (1.870,-4.080);
\fill[red] (0.450,-4.420) circle (0.10);
\fill[red] (1.100,-4.420) circle (0.10);
\fill[black] (1.750,-4.420) circle (0.10);
\fill[red] (0.450,-4.760) circle (0.10);
\draw[black, line width=0.8pt] (0.980,-4.760) -- (1.220,-4.760);
\draw[black, line width=0.8pt] (1.630,-4.760) -- (1.870,-4.760);
\fill[red] (0.450,-5.100) circle (0.10);
\draw[red, line width=0.8pt] (0.980,-5.100) -- (1.220,-5.100);
\draw[black, line width=0.8pt] (1.630,-5.100) -- (1.870,-5.100);
\fill[red] (0.450,-5.440) circle (0.10);
\draw[red, line width=0.8pt] (0.980,-5.440) -- (1.220,-5.440);
\draw[red, line width=0.8pt] (1.630,-5.440) -- (1.870,-5.440);
\fill[red] (0.450,-5.780) circle (0.10);
\fill[red] (1.100,-5.780) circle (0.10);
\draw[black, line width=0.8pt] (1.630,-5.780) -- (1.870,-5.780);
\fill[red] (0.450,-6.120) circle (0.10);
\fill[red] (1.100,-6.120) circle (0.10);
\draw[red, line width=0.8pt] (1.630,-6.120) -- (1.870,-6.120);
\fill[red] (0.450,-6.460) circle (0.10);
\fill[red] (1.100,-6.460) circle (0.10);
\fill[red] (1.750,-6.460) circle (0.10);
\node[font=\small\bfseries, anchor=south] at (1.100,0.340) {};
\fill[white] (2.840,0.153) rectangle (7.640,-6.560);
\draw[black!75, line width=0.55pt] (2.840,0.153) rectangle (7.640,-6.560);
\draw[black!70, line width=0.3pt] (5.240,0.153) -- (5.240,-6.560);
\draw[gray!30, line width=0.2pt] (3.107,0.153) -- (3.107,-6.560);
\node[font=\scriptsize, anchor=north] at (3.107,-6.640) {-4};
\draw[gray!30, line width=0.2pt] (3.640,0.153) -- (3.640,-6.560);
\node[font=\scriptsize, anchor=north] at (3.640,-6.640) {-3};
\draw[gray!30, line width=0.2pt] (4.173,0.153) -- (4.173,-6.560);
\node[font=\scriptsize, anchor=north] at (4.173,-6.640) {-2};
\draw[gray!30, line width=0.2pt] (4.707,0.153) -- (4.707,-6.560);
\node[font=\scriptsize, anchor=north] at (4.707,-6.640) {-1};
\draw[gray!30, line width=0.2pt] (5.240,0.153) -- (5.240,-6.560);
\node[font=\scriptsize, anchor=north] at (5.240,-6.640) {0};
\draw[gray!30, line width=0.2pt] (5.773,0.153) -- (5.773,-6.560);
\node[font=\scriptsize, anchor=north] at (5.773,-6.640) {1};
\draw[gray!30, line width=0.2pt] (6.307,0.153) -- (6.307,-6.560);
\node[font=\scriptsize, anchor=north] at (6.307,-6.640) {2};
\draw[gray!30, line width=0.2pt] (6.840,0.153) -- (6.840,-6.560);
\node[font=\scriptsize, anchor=north] at (6.840,-6.640) {3};
\draw[gray!30, line width=0.2pt] (7.373,0.153) -- (7.373,-6.560);
\node[font=\scriptsize, anchor=north] at (7.373,-6.640) {4};
\node[font=\small\bfseries, anchor=south] at (5.240,0.340) {No bias (0)};
\draw[neucol, line width=0.8pt] (2.505,-0.000) -- (6.553,-0.000);
\fill[neucol] (4.529,-0.000) circle (0.07);
\filldraw[fill=predone, draw=black, line width=0.22pt] (5.240,0.085) -- (5.150,-0.075) -- (5.330,-0.075) -- cycle;
\filldraw[fill=predtwo, draw=black, line width=0.22pt] (5.240,-0.085) -- (5.150,0.075) -- (5.330,0.075) -- cycle;
\draw[negcol, line width=0.8pt] (4.307,-0.340) -- (5.056,-0.340);
\fill[negcol] (4.681,-0.340) circle (0.07);
\filldraw[fill=predone, draw=black, line width=0.22pt] (5.240,-0.255) -- (5.150,-0.415) -- (5.330,-0.415) -- cycle;
\filldraw[fill=predtwo, draw=black, line width=0.22pt] (5.240,-0.425) -- (5.150,-0.265) -- (5.330,-0.265) -- cycle;
\draw[neucol, line width=0.8pt] (4.471,-0.680) -- (5.440,-0.680);
\fill[neucol] (4.956,-0.680) circle (0.07);
\filldraw[fill=predone, draw=black, line width=0.22pt] (5.240,-0.595) -- (5.150,-0.755) -- (5.330,-0.755) -- cycle;
\filldraw[fill=predtwo, draw=black, line width=0.22pt] (5.240,-0.765) -- (5.150,-0.605) -- (5.330,-0.605) -- cycle;
\draw[neucol, line width=0.8pt] (4.630,-1.020) -- (5.275,-1.020);
\fill[neucol] (4.953,-1.020) circle (0.07);
\filldraw[fill=predone, draw=black, line width=0.22pt] (5.240,-0.935) -- (5.150,-1.095) -- (5.330,-1.095) -- cycle;
\filldraw[fill=predtwo, draw=black, line width=0.22pt] (5.240,-1.105) -- (5.150,-0.945) -- (5.330,-0.945) -- cycle;
\draw[neucol, line width=0.8pt] (4.638,-1.360) -- (5.461,-1.360);
\fill[neucol] (5.050,-1.360) circle (0.07);
\filldraw[fill=predone, draw=black, line width=0.22pt] (5.240,-1.275) -- (5.150,-1.435) -- (5.330,-1.435) -- cycle;
\filldraw[fill=predtwo, draw=black, line width=0.22pt] (5.240,-1.445) -- (5.150,-1.285) -- (5.330,-1.285) -- cycle;
\draw[neucol, line width=0.8pt] (4.916,-1.700) -- (5.831,-1.700);
\fill[neucol] (5.373,-1.700) circle (0.07);
\filldraw[fill=predone, draw=black, line width=0.22pt] (5.240,-1.615) -- (5.150,-1.775) -- (5.330,-1.775) -- cycle;
\filldraw[fill=predtwo, draw=black, line width=0.22pt] (5.240,-1.785) -- (5.150,-1.625) -- (5.330,-1.625) -- cycle;
\draw[neucol, line width=0.8pt] (5.197,-2.040) -- (5.923,-2.040);
\fill[neucol] (5.560,-2.040) circle (0.07);
\filldraw[fill=predone, draw=black, line width=0.22pt] (5.240,-1.955) -- (5.150,-2.115) -- (5.330,-2.115) -- cycle;
\filldraw[fill=predtwo, draw=black, line width=0.22pt] (5.240,-2.125) -- (5.150,-1.965) -- (5.330,-1.965) -- cycle;
\draw[neucol, line width=0.8pt] (4.048,-2.380) -- (5.462,-2.380);
\fill[neucol] (4.755,-2.380) circle (0.07);
\filldraw[fill=predone, draw=black, line width=0.22pt] (5.240,-2.295) -- (5.150,-2.455) -- (5.330,-2.455) -- cycle;
\filldraw[fill=predtwo, draw=black, line width=0.22pt] (5.240,-2.465) -- (5.150,-2.305) -- (5.330,-2.305) -- cycle;
\draw[neucol, line width=0.8pt] (4.966,-2.720) -- (5.457,-2.720);
\fill[neucol] (5.212,-2.720) circle (0.07);
\filldraw[fill=predone, draw=black, line width=0.22pt] (5.240,-2.635) -- (5.150,-2.795) -- (5.330,-2.795) -- cycle;
\filldraw[fill=predtwo, draw=black, line width=0.22pt] (5.240,-2.805) -- (5.150,-2.645) -- (5.330,-2.645) -- cycle;
\draw[neucol, line width=0.8pt] (5.165,-3.060) -- (5.742,-3.060);
\fill[neucol] (5.453,-3.060) circle (0.07);
\filldraw[fill=predone, draw=black, line width=0.22pt] (5.240,-2.975) -- (5.150,-3.135) -- (5.330,-3.135) -- cycle;
\filldraw[fill=predtwo, draw=black, line width=0.22pt] (5.240,-3.145) -- (5.150,-2.985) -- (5.330,-2.985) -- cycle;
\draw[neucol, line width=0.8pt] (5.091,-3.400) -- (6.122,-3.400);
\fill[neucol] (5.607,-3.400) circle (0.07);
\filldraw[fill=predone, draw=black, line width=0.22pt] (5.240,-3.315) -- (5.150,-3.475) -- (5.330,-3.475) -- cycle;
\filldraw[fill=predtwo, draw=black, line width=0.22pt] (5.240,-3.485) -- (5.150,-3.325) -- (5.330,-3.325) -- cycle;
\draw[neucol, line width=0.8pt] (4.884,-3.740) -- (5.738,-3.740);
\fill[neucol] (5.311,-3.740) circle (0.07);
\filldraw[fill=predone, draw=black, line width=0.22pt] (5.240,-3.655) -- (5.150,-3.815) -- (5.330,-3.815) -- cycle;
\filldraw[fill=predtwo, draw=black, line width=0.22pt] (5.240,-3.825) -- (5.150,-3.665) -- (5.330,-3.665) -- cycle;
\draw[neucol, line width=0.8pt] (4.909,-4.080) -- (6.391,-4.080);
\fill[neucol] (5.650,-4.080) circle (0.07);
\filldraw[fill=predone, draw=black, line width=0.22pt] (5.240,-3.995) -- (5.150,-4.155) -- (5.330,-4.155) -- cycle;
\filldraw[fill=predtwo, draw=black, line width=0.22pt] (5.240,-4.165) -- (5.150,-4.005) -- (5.330,-4.005) -- cycle;
\draw[neucol, line width=0.8pt] (4.627,-4.420) -- (5.853,-4.420);
\fill[neucol] (5.240,-4.420) circle (0.07);
\filldraw[fill=predone, draw=black, line width=0.22pt] (5.240,-4.335) -- (5.150,-4.495) -- (5.330,-4.495) -- cycle;
\filldraw[fill=predtwo, draw=black, line width=0.22pt] (5.240,-4.505) -- (5.150,-4.345) -- (5.330,-4.345) -- cycle;
\draw[neucol, line width=0.8pt] (5.005,-4.760) -- (6.542,-4.760);
\fill[neucol] (5.773,-4.760) circle (0.07);
\filldraw[fill=predone, draw=black, line width=0.22pt] (5.240,-4.675) -- (5.150,-4.835) -- (5.330,-4.835) -- cycle;
\filldraw[fill=predtwo, draw=black, line width=0.22pt] (5.240,-4.845) -- (5.150,-4.685) -- (5.330,-4.685) -- cycle;
\draw[neucol, line width=0.8pt] (4.884,-5.100) -- (6.054,-5.100);
\fill[neucol] (5.469,-5.100) circle (0.07);
\filldraw[fill=predone, draw=black, line width=0.22pt] (5.240,-5.015) -- (5.150,-5.175) -- (5.330,-5.175) -- cycle;
\filldraw[fill=predtwo, draw=black, line width=0.22pt] (5.240,-5.185) -- (5.150,-5.025) -- (5.330,-5.025) -- cycle;
\draw[poscol, line width=0.8pt] (5.467,-5.440) -- (6.225,-5.440);
\fill[poscol] (5.846,-5.440) circle (0.07);
\filldraw[fill=predone, draw=black, line width=0.22pt] (5.240,-5.355) -- (5.150,-5.515) -- (5.330,-5.515) -- cycle;
\filldraw[fill=predtwo, draw=black, line width=0.22pt] (5.240,-5.525) -- (5.150,-5.365) -- (5.330,-5.365) -- cycle;
\draw[poscol, line width=0.8pt] (5.570,-5.780) -- (6.469,-5.780);
\fill[poscol] (6.019,-5.780) circle (0.07);
\filldraw[fill=predone, draw=black, line width=0.22pt] (5.240,-5.695) -- (5.150,-5.855) -- (5.330,-5.855) -- cycle;
\filldraw[fill=predtwo, draw=black, line width=0.22pt] (5.240,-5.865) -- (5.150,-5.705) -- (5.330,-5.705) -- cycle;
\draw[poscol, line width=0.8pt] (5.383,-6.120) -- (6.052,-6.120);
\fill[poscol] (5.717,-6.120) circle (0.07);
\filldraw[fill=predone, draw=black, line width=0.22pt] (5.240,-6.035) -- (5.150,-6.195) -- (5.330,-6.195) -- cycle;
\filldraw[fill=predtwo, draw=black, line width=0.22pt] (5.240,-6.205) -- (5.150,-6.045) -- (5.330,-6.045) -- cycle;
\draw[neucol, line width=0.8pt] (5.136,-6.460) -- (6.011,-6.460);
\fill[neucol] (5.573,-6.460) circle (0.07);
\filldraw[fill=predone, draw=black, line width=0.22pt] (5.240,-6.375) -- (5.150,-6.535) -- (5.330,-6.535) -- cycle;
\filldraw[fill=predtwo, draw=black, line width=0.22pt] (5.240,-6.545) -- (5.150,-6.385) -- (5.330,-6.385) -- cycle;
\draw[black!75, line width=0.55pt] (2.840,0.153) rectangle (7.640,-6.560);
\fill[white] (8.190,0.153) rectangle (12.990,-6.560);
\draw[black!75, line width=0.55pt] (8.190,0.153) rectangle (12.990,-6.560);
\draw[black!70, line width=0.3pt] (10.590,0.153) -- (10.590,-6.560);
\draw[gray!30, line width=0.2pt] (8.457,0.153) -- (8.457,-6.560);
\node[font=\scriptsize, anchor=north] at (8.457,-6.640) {-4};
\draw[gray!30, line width=0.2pt] (8.990,0.153) -- (8.990,-6.560);
\node[font=\scriptsize, anchor=north] at (8.990,-6.640) {-3};
\draw[gray!30, line width=0.2pt] (9.523,0.153) -- (9.523,-6.560);
\node[font=\scriptsize, anchor=north] at (9.523,-6.640) {-2};
\draw[gray!30, line width=0.2pt] (10.057,0.153) -- (10.057,-6.560);
\node[font=\scriptsize, anchor=north] at (10.057,-6.640) {-1};
\draw[gray!30, line width=0.2pt] (10.590,0.153) -- (10.590,-6.560);
\node[font=\scriptsize, anchor=north] at (10.590,-6.640) {0};
\draw[gray!30, line width=0.2pt] (11.123,0.153) -- (11.123,-6.560);
\node[font=\scriptsize, anchor=north] at (11.123,-6.640) {1};
\draw[gray!30, line width=0.2pt] (11.657,0.153) -- (11.657,-6.560);
\node[font=\scriptsize, anchor=north] at (11.657,-6.640) {2};
\draw[gray!30, line width=0.2pt] (12.190,0.153) -- (12.190,-6.560);
\node[font=\scriptsize, anchor=north] at (12.190,-6.640) {3};
\draw[gray!30, line width=0.2pt] (12.723,0.153) -- (12.723,-6.560);
\node[font=\scriptsize, anchor=north] at (12.723,-6.640) {4};
\node[font=\small\bfseries, anchor=south] at (10.590,0.340) {Low bias (1)};
\draw[neucol, line width=0.8pt] (9.402,-0.000) -- (11.168,-0.000);
\fill[neucol] (10.285,-0.000) circle (0.07);
\filldraw[fill=predone, draw=black, line width=0.22pt] (10.590,0.085) -- (10.500,-0.075) -- (10.680,-0.075) -- cycle;
\filldraw[fill=predtwo, draw=black, line width=0.22pt] (10.590,-0.085) -- (10.500,0.075) -- (10.680,0.075) -- cycle;
\draw[neucol, line width=0.8pt] (9.901,-0.340) -- (10.674,-0.340);
\fill[neucol] (10.287,-0.340) circle (0.07);
\filldraw[fill=predone, draw=black, line width=0.22pt] (10.768,-0.255) -- (10.678,-0.415) -- (10.858,-0.415) -- cycle;
\filldraw[fill=predtwo, draw=black, line width=0.22pt] (10.590,-0.425) -- (10.500,-0.265) -- (10.680,-0.265) -- cycle;
\draw[neucol, line width=0.8pt] (9.708,-0.680) -- (10.798,-0.680);
\fill[neucol] (10.253,-0.680) circle (0.07);
\filldraw[fill=predone, draw=black, line width=0.22pt] (10.590,-0.595) -- (10.500,-0.755) -- (10.680,-0.755) -- cycle;
\filldraw[fill=predtwo, draw=black, line width=0.22pt] (10.590,-0.765) -- (10.500,-0.605) -- (10.680,-0.605) -- cycle;
\draw[neucol, line width=0.8pt] (10.009,-1.020) -- (10.621,-1.020);
\fill[neucol] (10.315,-1.020) circle (0.07);
\filldraw[fill=predone, draw=black, line width=0.22pt] (10.768,-0.935) -- (10.678,-1.095) -- (10.858,-1.095) -- cycle;
\filldraw[fill=predtwo, draw=black, line width=0.22pt] (10.590,-1.105) -- (10.500,-0.945) -- (10.680,-0.945) -- cycle;
\draw[neucol, line width=0.8pt] (10.201,-1.360) -- (11.072,-1.360);
\fill[neucol] (10.636,-1.360) circle (0.07);
\filldraw[fill=predone, draw=black, line width=0.22pt] (10.590,-1.275) -- (10.500,-1.435) -- (10.680,-1.435) -- cycle;
\filldraw[fill=predtwo, draw=black, line width=0.22pt] (10.590,-1.445) -- (10.500,-1.285) -- (10.680,-1.285) -- cycle;
\draw[neucol, line width=0.8pt] (10.323,-1.700) -- (11.283,-1.700);
\fill[neucol] (10.803,-1.700) circle (0.07);
\filldraw[fill=predone, draw=black, line width=0.22pt] (10.590,-1.615) -- (10.500,-1.775) -- (10.680,-1.775) -- cycle;
\filldraw[fill=predtwo, draw=black, line width=0.22pt] (10.590,-1.785) -- (10.500,-1.625) -- (10.680,-1.625) -- cycle;
\draw[neucol, line width=0.8pt] (9.702,-2.040) -- (13.078,-2.040);
\fill[neucol] (11.390,-2.040) circle (0.07);
\filldraw[fill=predone, draw=black, line width=0.22pt] (10.590,-1.955) -- (10.500,-2.115) -- (10.680,-2.115) -- cycle;
\filldraw[fill=predtwo, draw=black, line width=0.22pt] (10.590,-2.125) -- (10.500,-1.965) -- (10.680,-1.965) -- cycle;
\draw[neucol, line width=0.8pt] (10.494,-2.380) -- (11.017,-2.380);
\fill[neucol] (10.756,-2.380) circle (0.07);
\filldraw[fill=predone, draw=black, line width=0.22pt] (10.768,-2.295) -- (10.678,-2.455) -- (10.858,-2.455) -- cycle;
\filldraw[fill=predtwo, draw=black, line width=0.22pt] (10.590,-2.465) -- (10.500,-2.305) -- (10.680,-2.305) -- cycle;
\draw[neucol, line width=0.8pt] (10.477,-2.720) -- (11.013,-2.720);
\fill[neucol] (10.745,-2.720) circle (0.07);
\filldraw[fill=predone, draw=black, line width=0.22pt] (10.590,-2.635) -- (10.500,-2.795) -- (10.680,-2.795) -- cycle;
\filldraw[fill=predtwo, draw=black, line width=0.22pt] (10.590,-2.805) -- (10.500,-2.645) -- (10.680,-2.645) -- cycle;
\draw[neucol, line width=0.8pt] (10.152,-3.060) -- (10.949,-3.060);
\fill[neucol] (10.550,-3.060) circle (0.07);
\filldraw[fill=predone, draw=black, line width=0.22pt] (10.857,-2.975) -- (10.767,-3.135) -- (10.947,-3.135) -- cycle;
\filldraw[fill=predtwo, draw=black, line width=0.22pt] (10.590,-3.145) -- (10.500,-2.985) -- (10.680,-2.985) -- cycle;
\draw[neucol, line width=0.8pt] (10.504,-3.400) -- (11.186,-3.400);
\fill[neucol] (10.845,-3.400) circle (0.07);
\filldraw[fill=predone, draw=black, line width=0.22pt] (10.590,-3.315) -- (10.500,-3.475) -- (10.680,-3.475) -- cycle;
\filldraw[fill=predtwo, draw=black, line width=0.22pt] (10.590,-3.485) -- (10.500,-3.325) -- (10.680,-3.325) -- cycle;
\draw[neucol, line width=0.8pt] (10.464,-3.740) -- (11.036,-3.740);
\fill[neucol] (10.750,-3.740) circle (0.07);
\filldraw[fill=predone, draw=black, line width=0.22pt] (10.857,-3.655) -- (10.767,-3.815) -- (10.947,-3.815) -- cycle;
\filldraw[fill=predtwo, draw=black, line width=0.22pt] (10.590,-3.825) -- (10.500,-3.665) -- (10.680,-3.665) -- cycle;
\draw[poscol, line width=0.8pt] (10.653,-4.080) -- (11.594,-4.080);
\fill[poscol] (11.123,-4.080) circle (0.07);
\filldraw[fill=predone, draw=black, line width=0.22pt] (10.590,-3.995) -- (10.500,-4.155) -- (10.680,-4.155) -- cycle;
\filldraw[fill=predtwo, draw=black, line width=0.22pt] (10.590,-4.165) -- (10.500,-4.005) -- (10.680,-4.005) -- cycle;
\draw[poscol, line width=0.8pt] (10.705,-4.420) -- (12.075,-4.420);
\fill[poscol] (11.390,-4.420) circle (0.07);
\filldraw[fill=predone, draw=black, line width=0.22pt] (11.123,-4.335) -- (11.033,-4.495) -- (11.213,-4.495) -- cycle;
\filldraw[fill=predtwo, draw=black, line width=0.22pt] (10.057,-4.505) -- (9.967,-4.345) -- (10.147,-4.345) -- cycle;
\draw[poscol, line width=0.8pt] (10.921,-4.760) -- (11.818,-4.760);
\fill[poscol] (11.369,-4.760) circle (0.07);
\filldraw[fill=predone, draw=black, line width=0.22pt] (10.990,-4.675) -- (10.900,-4.835) -- (11.080,-4.835) -- cycle;
\filldraw[fill=predtwo, draw=black, line width=0.22pt] (10.590,-4.845) -- (10.500,-4.685) -- (10.680,-4.685) -- cycle;
\draw[poscol, line width=0.8pt] (10.679,-5.100) -- (11.567,-5.100);
\fill[poscol] (11.123,-5.100) circle (0.07);
\filldraw[fill=predone, draw=black, line width=0.22pt] (11.123,-5.015) -- (11.033,-5.175) -- (11.213,-5.175) -- cycle;
\filldraw[fill=predtwo, draw=black, line width=0.22pt] (10.057,-5.185) -- (9.967,-5.025) -- (10.147,-5.025) -- cycle;
\draw[poscol, line width=0.8pt] (10.799,-5.440) -- (11.241,-5.440);
\fill[poscol] (11.020,-5.440) circle (0.07);
\filldraw[fill=predone, draw=black, line width=0.22pt] (11.123,-5.355) -- (11.033,-5.515) -- (11.213,-5.515) -- cycle;
\filldraw[fill=predtwo, draw=black, line width=0.22pt] (10.590,-5.525) -- (10.500,-5.365) -- (10.680,-5.365) -- cycle;
\draw[neucol, line width=0.8pt] (10.585,-5.780) -- (11.395,-5.780);
\fill[neucol] (10.990,-5.780) circle (0.07);
\filldraw[fill=predone, draw=black, line width=0.22pt] (11.123,-5.695) -- (11.033,-5.855) -- (11.213,-5.855) -- cycle;
\filldraw[fill=predtwo, draw=black, line width=0.22pt] (10.057,-5.865) -- (9.967,-5.705) -- (10.147,-5.705) -- cycle;
\draw[poscol, line width=0.8pt] (10.838,-6.120) -- (11.439,-6.120);
\fill[poscol] (11.139,-6.120) circle (0.07);
\filldraw[fill=predone, draw=black, line width=0.22pt] (11.123,-6.035) -- (11.033,-6.195) -- (11.213,-6.195) -- cycle;
\filldraw[fill=predtwo, draw=black, line width=0.22pt] (10.590,-6.205) -- (10.500,-6.045) -- (10.680,-6.045) -- cycle;
\draw[neucol, line width=0.8pt] (10.331,-6.460) -- (11.560,-6.460);
\fill[neucol] (10.946,-6.460) circle (0.07);
\filldraw[fill=predone, draw=black, line width=0.22pt] (11.123,-6.375) -- (11.033,-6.535) -- (11.213,-6.535) -- cycle;
\filldraw[fill=predtwo, draw=black, line width=0.22pt] (10.590,-6.545) -- (10.500,-6.385) -- (10.680,-6.385) -- cycle;
\draw[black!75, line width=0.55pt] (8.190,0.153) rectangle (12.990,-6.560);
\fill[white] (13.540,0.153) rectangle (18.340,-6.560);
\draw[black!75, line width=0.55pt] (13.540,0.153) rectangle (18.340,-6.560);
\draw[black!70, line width=0.3pt] (15.340,0.153) -- (15.340,-6.560);
\draw[gray!30, line width=0.2pt] (13.740,0.153) -- (13.740,-6.560);
\node[font=\scriptsize, anchor=north] at (13.740,-6.640) {-4};
\draw[gray!30, line width=0.2pt] (14.140,0.153) -- (14.140,-6.560);
\node[font=\scriptsize, anchor=north] at (14.140,-6.640) {-3};
\draw[gray!30, line width=0.2pt] (14.540,0.153) -- (14.540,-6.560);
\node[font=\scriptsize, anchor=north] at (14.540,-6.640) {-2};
\draw[gray!30, line width=0.2pt] (14.940,0.153) -- (14.940,-6.560);
\node[font=\scriptsize, anchor=north] at (14.940,-6.640) {-1};
\draw[gray!30, line width=0.2pt] (15.340,0.153) -- (15.340,-6.560);
\node[font=\scriptsize, anchor=north] at (15.340,-6.640) {0};
\draw[gray!30, line width=0.2pt] (15.740,0.153) -- (15.740,-6.560);
\node[font=\scriptsize, anchor=north] at (15.740,-6.640) {1};
\draw[gray!30, line width=0.2pt] (16.140,0.153) -- (16.140,-6.560);
\node[font=\scriptsize, anchor=north] at (16.140,-6.640) {2};
\draw[gray!30, line width=0.2pt] (16.540,0.153) -- (16.540,-6.560);
\node[font=\scriptsize, anchor=north] at (16.540,-6.640) {3};
\draw[gray!30, line width=0.2pt] (16.940,0.153) -- (16.940,-6.560);
\node[font=\scriptsize, anchor=north] at (16.940,-6.640) {4};
\draw[gray!30, line width=0.2pt] (17.340,0.153) -- (17.340,-6.560);
\node[font=\scriptsize, anchor=north] at (17.340,-6.640) {5};
\draw[gray!30, line width=0.2pt] (17.740,0.153) -- (17.740,-6.560);
\node[font=\scriptsize, anchor=north] at (17.740,-6.640) {6};
\draw[gray!30, line width=0.2pt] (18.140,0.153) -- (18.140,-6.560);
\node[font=\scriptsize, anchor=north] at (18.140,-6.640) {7};
\node[font=\small\bfseries, anchor=south] at (15.940,0.340) {High bias (3)};
\draw[neucol, line width=0.8pt] (14.466,-0.000) -- (15.923,-0.000);
\fill[neucol] (15.195,-0.000) circle (0.07);
\filldraw[fill=predone, draw=black, line width=0.22pt] (15.740,0.085) -- (15.650,-0.075) -- (15.830,-0.075) -- cycle;
\filldraw[fill=predtwo, draw=black, line width=0.22pt] (16.940,-0.085) -- (16.850,0.075) -- (17.030,0.075) -- cycle;
\draw[neucol, line width=0.8pt] (15.174,-0.340) -- (15.689,-0.340);
\fill[neucol] (15.432,-0.340) circle (0.07);
\filldraw[fill=predone, draw=black, line width=0.22pt] (15.740,-0.255) -- (15.650,-0.415) -- (15.830,-0.415) -- cycle;
\filldraw[fill=predtwo, draw=black, line width=0.22pt] (16.940,-0.425) -- (16.850,-0.265) -- (17.030,-0.265) -- cycle;
\draw[neucol, line width=0.8pt] (15.150,-0.680) -- (15.785,-0.680);
\fill[neucol] (15.467,-0.680) circle (0.07);
\filldraw[fill=predone, draw=black, line width=0.22pt] (16.540,-0.595) -- (16.450,-0.755) -- (16.630,-0.755) -- cycle;
\filldraw[fill=predtwo, draw=black, line width=0.22pt] (18.140,-0.765) -- (18.050,-0.605) -- (18.230,-0.605) -- cycle;
\draw[neucol, line width=0.8pt] (15.099,-1.020) -- (15.815,-1.020);
\fill[neucol] (15.457,-1.020) circle (0.07);
\filldraw[fill=predone, draw=black, line width=0.22pt] (15.740,-0.935) -- (15.650,-1.095) -- (15.830,-1.095) -- cycle;
\filldraw[fill=predtwo, draw=black, line width=0.22pt] (16.940,-1.105) -- (16.850,-0.945) -- (17.030,-0.945) -- cycle;
\draw[poscol, line width=0.8pt] (15.446,-1.360) -- (16.209,-1.360);
\fill[poscol] (15.827,-1.360) circle (0.07);
\filldraw[fill=predone, draw=black, line width=0.22pt] (16.540,-1.275) -- (16.450,-1.435) -- (16.630,-1.435) -- cycle;
\filldraw[fill=predtwo, draw=black, line width=0.22pt] (18.140,-1.445) -- (18.050,-1.285) -- (18.230,-1.285) -- cycle;
\draw[neucol, line width=0.8pt] (15.148,-1.700) -- (16.232,-1.700);
\fill[neucol] (15.690,-1.700) circle (0.07);
\filldraw[fill=predone, draw=black, line width=0.22pt] (16.540,-1.615) -- (16.450,-1.775) -- (16.630,-1.775) -- cycle;
\filldraw[fill=predtwo, draw=black, line width=0.22pt] (14.940,-1.785) -- (14.850,-1.625) -- (15.030,-1.625) -- cycle;
\draw[neucol, line width=0.8pt] (14.869,-2.040) -- (16.344,-2.040);
\fill[neucol] (15.607,-2.040) circle (0.07);
\filldraw[fill=predone, draw=black, line width=0.22pt] (16.540,-1.955) -- (16.450,-2.115) -- (16.630,-2.115) -- cycle;
\filldraw[fill=predtwo, draw=black, line width=0.22pt] (18.140,-2.125) -- (18.050,-1.965) -- (18.230,-1.965) -- cycle;
\draw[poscol, line width=0.8pt] (15.359,-2.380) -- (15.846,-2.380);
\fill[poscol] (15.602,-2.380) circle (0.07);
\filldraw[fill=predone, draw=black, line width=0.22pt] (15.740,-2.295) -- (15.650,-2.455) -- (15.830,-2.455) -- cycle;
\filldraw[fill=predtwo, draw=black, line width=0.22pt] (16.940,-2.465) -- (16.850,-2.305) -- (17.030,-2.305) -- cycle;
\draw[poscol, line width=0.8pt] (15.556,-2.720) -- (16.114,-2.720);
\fill[poscol] (15.835,-2.720) circle (0.07);
\filldraw[fill=predone, draw=black, line width=0.22pt] (16.540,-2.635) -- (16.450,-2.795) -- (16.630,-2.795) -- cycle;
\filldraw[fill=predtwo, draw=black, line width=0.22pt] (18.140,-2.805) -- (18.050,-2.645) -- (18.230,-2.645) -- cycle;
\draw[poscol, line width=0.8pt] (15.584,-3.060) -- (16.274,-3.060);
\fill[poscol] (15.929,-3.060) circle (0.07);
\filldraw[fill=predone, draw=black, line width=0.22pt] (16.540,-2.975) -- (16.450,-3.135) -- (16.630,-3.135) -- cycle;
\filldraw[fill=predtwo, draw=black, line width=0.22pt] (18.140,-3.145) -- (18.050,-2.985) -- (18.230,-2.985) -- cycle;
\draw[poscol, line width=0.8pt] (15.761,-3.400) -- (16.306,-3.400);
\fill[poscol] (16.033,-3.400) circle (0.07);
\filldraw[fill=predone, draw=black, line width=0.22pt] (16.540,-3.315) -- (16.450,-3.475) -- (16.630,-3.475) -- cycle;
\filldraw[fill=predtwo, draw=black, line width=0.22pt] (14.940,-3.485) -- (14.850,-3.325) -- (15.030,-3.325) -- cycle;
\draw[poscol, line width=0.8pt] (15.381,-3.740) -- (16.222,-3.740);
\fill[poscol] (15.802,-3.740) circle (0.07);
\filldraw[fill=predone, draw=black, line width=0.22pt] (16.540,-3.655) -- (16.450,-3.815) -- (16.630,-3.815) -- cycle;
\filldraw[fill=predtwo, draw=black, line width=0.22pt] (14.940,-3.825) -- (14.850,-3.665) -- (15.030,-3.665) -- cycle;
\draw[poscol, line width=0.8pt] (15.388,-4.080) -- (15.992,-4.080);
\fill[poscol] (15.690,-4.080) circle (0.07);
\filldraw[fill=predone, draw=black, line width=0.22pt] (16.540,-3.995) -- (16.450,-4.155) -- (16.630,-4.155) -- cycle;
\filldraw[fill=predtwo, draw=black, line width=0.22pt] (15.340,-4.165) -- (15.250,-4.005) -- (15.430,-4.005) -- cycle;
\draw[poscol, line width=0.8pt] (15.441,-4.420) -- (16.382,-4.420);
\fill[poscol] (15.911,-4.420) circle (0.07);
\filldraw[fill=predone, draw=black, line width=0.22pt] (16.540,-4.335) -- (16.450,-4.495) -- (16.630,-4.495) -- cycle;
\filldraw[fill=predtwo, draw=black, line width=0.22pt] (14.940,-4.505) -- (14.850,-4.345) -- (15.030,-4.345) -- cycle;
\draw[poscol, line width=0.8pt] (15.751,-4.760) -- (16.679,-4.760);
\fill[poscol] (16.215,-4.760) circle (0.07);
\filldraw[fill=predone, draw=black, line width=0.22pt] (16.540,-4.675) -- (16.450,-4.835) -- (16.630,-4.835) -- cycle;
\filldraw[fill=predtwo, draw=black, line width=0.22pt] (18.140,-4.845) -- (18.050,-4.685) -- (18.230,-4.685) -- cycle;
\draw[poscol, line width=0.8pt] (15.720,-5.100) -- (16.477,-5.100);
\fill[poscol] (16.099,-5.100) circle (0.07);
\filldraw[fill=predone, draw=black, line width=0.22pt] (16.540,-5.015) -- (16.450,-5.175) -- (16.630,-5.175) -- cycle;
\filldraw[fill=predtwo, draw=black, line width=0.22pt] (14.940,-5.185) -- (14.850,-5.025) -- (15.030,-5.025) -- cycle;
\draw[poscol, line width=0.8pt] (15.735,-5.440) -- (16.105,-5.440);
\fill[poscol] (15.920,-5.440) circle (0.07);
\filldraw[fill=predone, draw=black, line width=0.22pt] (16.540,-5.355) -- (16.450,-5.515) -- (16.630,-5.515) -- cycle;
\filldraw[fill=predtwo, draw=black, line width=0.22pt] (15.340,-5.525) -- (15.250,-5.365) -- (15.430,-5.365) -- cycle;
\draw[neucol, line width=0.8pt] (15.149,-5.780) -- (16.293,-5.780);
\fill[neucol] (15.721,-5.780) circle (0.07);
\filldraw[fill=predone, draw=black, line width=0.22pt] (16.540,-5.695) -- (16.450,-5.855) -- (16.630,-5.855) -- cycle;
\filldraw[fill=predtwo, draw=black, line width=0.22pt] (14.940,-5.865) -- (14.850,-5.705) -- (15.030,-5.705) -- cycle;
\draw[poscol, line width=0.8pt] (15.569,-6.120) -- (16.050,-6.120);
\fill[poscol] (15.810,-6.120) circle (0.07);
\filldraw[fill=predone, draw=black, line width=0.22pt] (16.540,-6.035) -- (16.450,-6.195) -- (16.630,-6.195) -- cycle;
\filldraw[fill=predtwo, draw=black, line width=0.22pt] (15.340,-6.205) -- (15.250,-6.045) -- (15.430,-6.045) -- cycle;
\draw[poscol, line width=0.8pt] (15.831,-6.460) -- (16.849,-6.460);
\fill[poscol] (16.340,-6.460) circle (0.07);
\filldraw[fill=predone, draw=black, line width=0.22pt] (16.540,-6.375) -- (16.450,-6.535) -- (16.630,-6.535) -- cycle;
\filldraw[fill=predtwo, draw=black, line width=0.22pt] (15.340,-6.545) -- (15.250,-6.385) -- (15.430,-6.385) -- cycle;
\draw[black!75, line width=0.55pt] (13.540,0.153) rectangle (18.340,-6.560);
\filldraw[fill=predone, draw=black, line width=0.22pt] (5.390,-7.275) -- (5.300,-7.435) -- (5.480,-7.435) -- cycle;
\node[font=\scriptsize, anchor=west] at (5.550,-7.360) {MEU};
\filldraw[fill=predtwo, draw=black, line width=0.22pt] (10.590,-7.445) -- (10.500,-7.285) -- (10.680,-7.285) -- cycle;
\node[font=\scriptsize, anchor=west] at (10.750,-7.360) {MLEU};
\end{tikzpicture}}
\caption{by history and model.}
\label{fig:MLEU_R_h_whiskers}
\end{subfigure}
\vspace{1.0em}
\begin{minipage}{0.95\textwidth}
\footnotesize
Notes: Whiskers indicate pointwise 95\% confidence intervals with standard errors clustered by receiver. Green and red whiskers indicate positive and negative estimates, respectively, whose pointwise 95\% confidence intervals exclude zero.
\end{minipage}
\caption{Predicted $\Delta^R(h,b),~\Delta_m^R(h,b)$ under MEU and MLEU and observed $\tilde{\Delta}^R(h,b),~ \tilde{\Delta}_m^R(h,b)$.}
\label{fig:MLEU_R_whiskers_combined}
\end{figure}

%% file: results_artifacts/Figure_19_Observed_and_predicted_Delta_S_under_MEU_and_MLEU.tex
\begin{figure}[p]
\centering
\resizebox{0.9\linewidth}{!}{\begin{tikzpicture}[x=1cm,y=1cm]
\fill[black] (0.450,-0.000) circle (0.10);
\fill[black] (1.100,-0.000) circle (0.10);
\fill[black] (1.750,-0.000) circle (0.10);
\fill[red] (0.450,-0.340) circle (0.10);
\fill[black] (1.100,-0.340) circle (0.10);
\fill[black] (1.750,-0.340) circle (0.10);
\fill[red] (0.450,-0.680) circle (0.10);
\fill[red] (1.100,-0.680) circle (0.10);
\fill[black] (1.750,-0.680) circle (0.10);
\fill[red] (0.450,-1.020) circle (0.10);
\fill[red] (1.100,-1.020) circle (0.10);
\fill[red] (1.750,-1.020) circle (0.10);
\fill[white] (2.840,0.153) rectangle (7.640,-1.120);
\draw[black!75, line width=0.55pt] (2.840,0.153) rectangle (7.640,-1.120);
\draw[black!70, line width=0.3pt] (5.240,0.153) -- (5.240,-1.120);
\draw[gray!30, line width=0.2pt] (3.107,0.153) -- (3.107,-1.120);
\node[font=\scriptsize, anchor=north] at (3.107,-1.200) {-4};
\draw[gray!30, line width=0.2pt] (3.640,0.153) -- (3.640,-1.120);
\node[font=\scriptsize, anchor=north] at (3.640,-1.200) {-3};
\draw[gray!30, line width=0.2pt] (4.173,0.153) -- (4.173,-1.120);
\node[font=\scriptsize, anchor=north] at (4.173,-1.200) {-2};
\draw[gray!30, line width=0.2pt] (4.707,0.153) -- (4.707,-1.120);
\node[font=\scriptsize, anchor=north] at (4.707,-1.200) {-1};
\draw[gray!30, line width=0.2pt] (5.240,0.153) -- (5.240,-1.120);
\node[font=\scriptsize, anchor=north] at (5.240,-1.200) {0};
\draw[gray!30, line width=0.2pt] (5.773,0.153) -- (5.773,-1.120);
\node[font=\scriptsize, anchor=north] at (5.773,-1.200) {1};
\draw[gray!30, line width=0.2pt] (6.307,0.153) -- (6.307,-1.120);
\node[font=\scriptsize, anchor=north] at (6.307,-1.200) {2};
\draw[gray!30, line width=0.2pt] (6.840,0.153) -- (6.840,-1.120);
\node[font=\scriptsize, anchor=north] at (6.840,-1.200) {3};
\draw[gray!30, line width=0.2pt] (7.373,0.153) -- (7.373,-1.120);
\node[font=\scriptsize, anchor=north] at (7.373,-1.200) {4};
\node[font=\small\bfseries, anchor=south] at (5.240,0.340) {No bias (0)};
\draw[neucol, line width=0.8pt] (5.181,-0.000) -- (5.448,-0.000);
\fill[neucol] (5.314,-0.000) circle (0.07);
\filldraw[fill=predone, draw=black, line width=0.22pt] (5.240,0.085) -- (5.150,-0.075) -- (5.330,-0.075) -- cycle;
\filldraw[fill=predtwo, draw=black, line width=0.22pt] (5.240,-0.085) -- (5.150,0.075) -- (5.330,0.075) -- cycle;
\draw[neucol, line width=0.8pt] (4.884,-0.340) -- (5.516,-0.340);
\fill[neucol] (5.200,-0.340) circle (0.07);
\filldraw[fill=predone, draw=black, line width=0.22pt] (5.240,-0.255) -- (5.150,-0.415) -- (5.330,-0.415) -- cycle;
\filldraw[fill=predtwo, draw=black, line width=0.22pt] (5.240,-0.425) -- (5.150,-0.265) -- (5.330,-0.265) -- cycle;
\draw[neucol, line width=0.8pt] (5.055,-0.680) -- (5.551,-0.680);
\fill[neucol] (5.303,-0.680) circle (0.07);
\filldraw[fill=predone, draw=black, line width=0.22pt] (5.240,-0.595) -- (5.150,-0.755) -- (5.330,-0.755) -- cycle;
\filldraw[fill=predtwo, draw=black, line width=0.22pt] (5.240,-0.765) -- (5.150,-0.605) -- (5.330,-0.605) -- cycle;
\draw[neucol, line width=0.8pt] (4.770,-1.020) -- (5.400,-1.020);
\fill[neucol] (5.085,-1.020) circle (0.07);
\filldraw[fill=predone, draw=black, line width=0.22pt] (5.240,-0.935) -- (5.150,-1.095) -- (5.330,-1.095) -- cycle;
\filldraw[fill=predtwo, draw=black, line width=0.22pt] (5.240,-1.105) -- (5.150,-0.945) -- (5.330,-0.945) -- cycle;
\draw[black!75, line width=0.55pt] (2.840,0.153) rectangle (7.640,-1.120);
\fill[white] (8.190,0.153) rectangle (12.990,-1.120);
\draw[black!75, line width=0.55pt] (8.190,0.153) rectangle (12.990,-1.120);
\draw[black!70, line width=0.3pt] (10.590,0.153) -- (10.590,-1.120);
\draw[gray!30, line width=0.2pt] (8.457,0.153) -- (8.457,-1.120);
\node[font=\scriptsize, anchor=north] at (8.457,-1.200) {-4};
\draw[gray!30, line width=0.2pt] (8.990,0.153) -- (8.990,-1.120);
\node[font=\scriptsize, anchor=north] at (8.990,-1.200) {-3};
\draw[gray!30, line width=0.2pt] (9.523,0.153) -- (9.523,-1.120);
\node[font=\scriptsize, anchor=north] at (9.523,-1.200) {-2};
\draw[gray!30, line width=0.2pt] (10.057,0.153) -- (10.057,-1.120);
\node[font=\scriptsize, anchor=north] at (10.057,-1.200) {-1};
\draw[gray!30, line width=0.2pt] (10.590,0.153) -- (10.590,-1.120);
\node[font=\scriptsize, anchor=north] at (10.590,-1.200) {0};
\draw[gray!30, line width=0.2pt] (11.123,0.153) -- (11.123,-1.120);
\node[font=\scriptsize, anchor=north] at (11.123,-1.200) {1};
\draw[gray!30, line width=0.2pt] (11.657,0.153) -- (11.657,-1.120);
\node[font=\scriptsize, anchor=north] at (11.657,-1.200) {2};
\draw[gray!30, line width=0.2pt] (12.190,0.153) -- (12.190,-1.120);
\node[font=\scriptsize, anchor=north] at (12.190,-1.200) {3};
\draw[gray!30, line width=0.2pt] (12.723,0.153) -- (12.723,-1.120);
\node[font=\scriptsize, anchor=north] at (12.723,-1.200) {4};
\node[font=\small\bfseries, anchor=south] at (10.590,0.340) {Low bias (1)};
\draw[neucol, line width=0.8pt] (10.588,-0.000) -- (11.004,-0.000);
\fill[neucol] (10.796,-0.000) circle (0.07);
\filldraw[fill=predone, draw=black, line width=0.22pt] (11.244,0.085) -- (11.154,-0.075) -- (11.334,-0.075) -- cycle;
\filldraw[fill=predtwo, draw=black, line width=0.22pt] (11.088,-0.085) -- (10.998,0.075) -- (11.178,0.075) -- cycle;
\draw[poscol, line width=0.8pt] (10.855,-0.340) -- (11.409,-0.340);
\fill[poscol] (11.132,-0.340) circle (0.07);
\filldraw[fill=predone, draw=black, line width=0.22pt] (10.996,-0.255) -- (10.906,-0.415) -- (11.086,-0.415) -- cycle;
\filldraw[fill=predtwo, draw=black, line width=0.22pt] (10.926,-0.425) -- (10.836,-0.265) -- (11.016,-0.265) -- cycle;
\draw[poscol, line width=0.8pt] (10.661,-0.680) -- (11.216,-0.680);
\fill[poscol] (10.939,-0.680) circle (0.07);
\filldraw[fill=predone, draw=black, line width=0.22pt] (10.727,-0.595) -- (10.637,-0.755) -- (10.817,-0.755) -- cycle;
\filldraw[fill=predtwo, draw=black, line width=0.22pt] (10.569,-0.765) -- (10.479,-0.605) -- (10.659,-0.605) -- cycle;
\draw[neucol, line width=0.8pt] (10.537,-1.020) -- (11.025,-1.020);
\fill[neucol] (10.781,-1.020) circle (0.07);
\filldraw[fill=predone, draw=black, line width=0.22pt] (10.764,-0.935) -- (10.674,-1.095) -- (10.854,-1.095) -- cycle;
\filldraw[fill=predtwo, draw=black, line width=0.22pt] (10.764,-1.105) -- (10.674,-0.945) -- (10.854,-0.945) -- cycle;
\draw[black!75, line width=0.55pt] (8.190,0.153) rectangle (12.990,-1.120);
\fill[white] (13.540,0.153) rectangle (18.340,-1.120);
\draw[black!75, line width=0.55pt] (13.540,0.153) rectangle (18.340,-1.120);
\draw[black!70, line width=0.3pt] (15.940,0.153) -- (15.940,-1.120);
\draw[gray!30, line width=0.2pt] (13.807,0.153) -- (13.807,-1.120);
\node[font=\scriptsize, anchor=north] at (13.807,-1.200) {-4};
\draw[gray!30, line width=0.2pt] (14.340,0.153) -- (14.340,-1.120);
\node[font=\scriptsize, anchor=north] at (14.340,-1.200) {-3};
\draw[gray!30, line width=0.2pt] (14.873,0.153) -- (14.873,-1.120);
\node[font=\scriptsize, anchor=north] at (14.873,-1.200) {-2};
\draw[gray!30, line width=0.2pt] (15.407,0.153) -- (15.407,-1.120);
\node[font=\scriptsize, anchor=north] at (15.407,-1.200) {-1};
\draw[gray!30, line width=0.2pt] (15.940,0.153) -- (15.940,-1.120);
\node[font=\scriptsize, anchor=north] at (15.940,-1.200) {0};
\draw[gray!30, line width=0.2pt] (16.473,0.153) -- (16.473,-1.120);
\node[font=\scriptsize, anchor=north] at (16.473,-1.200) {1};
\draw[gray!30, line width=0.2pt] (17.007,0.153) -- (17.007,-1.120);
\node[font=\scriptsize, anchor=north] at (17.007,-1.200) {2};
\draw[gray!30, line width=0.2pt] (17.540,0.153) -- (17.540,-1.120);
\node[font=\scriptsize, anchor=north] at (17.540,-1.200) {3};
\draw[gray!30, line width=0.2pt] (18.073,0.153) -- (18.073,-1.120);
\node[font=\scriptsize, anchor=north] at (18.073,-1.200) {4};
\node[font=\small\bfseries, anchor=south] at (15.940,0.340) {High bias (3)};
\draw[poscol, line width=0.8pt] (16.282,-0.000) -- (16.794,-0.000);
\fill[poscol] (16.538,-0.000) circle (0.07);
\filldraw[fill=predone, draw=black, line width=0.22pt] (17.391,0.085) -- (17.301,-0.075) -- (17.481,-0.075) -- cycle;
\filldraw[fill=predtwo, draw=black, line width=0.22pt] (17.391,-0.085) -- (17.301,0.075) -- (17.481,0.075) -- cycle;
\draw[poscol, line width=0.8pt] (16.468,-0.340) -- (17.090,-0.340);
\fill[poscol] (16.779,-0.340) circle (0.07);
\filldraw[fill=predone, draw=black, line width=0.22pt] (18.033,-0.255) -- (17.943,-0.415) -- (18.123,-0.415) -- cycle;
\filldraw[fill=predtwo, draw=black, line width=0.22pt] (18.033,-0.425) -- (17.943,-0.265) -- (18.123,-0.265) -- cycle;
\draw[neucol, line width=0.8pt] (15.918,-0.680) -- (16.657,-0.680);
\fill[neucol] (16.287,-0.680) circle (0.07);
\filldraw[fill=predone, draw=black, line width=0.22pt] (17.003,-0.595) -- (16.913,-0.755) -- (17.093,-0.755) -- cycle;
\filldraw[fill=predtwo, draw=black, line width=0.22pt] (17.003,-0.765) -- (16.913,-0.605) -- (17.093,-0.605) -- cycle;
\draw[neucol, line width=0.8pt] (15.683,-1.020) -- (16.108,-1.020);
\fill[neucol] (15.896,-1.020) circle (0.07);
\filldraw[fill=predone, draw=black, line width=0.22pt] (16.544,-0.935) -- (16.454,-1.095) -- (16.634,-1.095) -- cycle;
\filldraw[fill=predtwo, draw=black, line width=0.22pt] (16.544,-1.105) -- (16.454,-0.945) -- (16.634,-0.945) -- cycle;
\draw[black!75, line width=0.55pt] (13.540,0.153) rectangle (18.340,-1.120);
\filldraw[fill=predone, draw=black, line width=0.22pt] (5.390,-1.835) -- (5.300,-1.995) -- (5.480,-1.995) -- cycle;
\node[font=\scriptsize, anchor=west] at (5.550,-1.920) {MEU};
\filldraw[fill=predtwo, draw=black, line width=0.22pt] (10.590,-2.005) -- (10.500,-1.845) -- (10.680,-1.845) -- cycle;
\node[font=\scriptsize, anchor=west] at (10.750,-1.920) {MLEU};
\end{tikzpicture}}
\vspace{1.0em}
\begin{minipage}{0.95\textwidth}
\footnotesize
Notes:\ Whiskers indicate pointwise 95\% confidence intervals with standard errors clustered by sender. Green and red whiskers indicate positive and negative estimates, respectively, whose pointwise 95\% confidence intervals exclude zero.
\end{minipage}
\caption{Predicted $\Delta^S(h,b)$ under MEU and MLEU and observed $\tilde{\Delta}^S(h,b)$.}
\label{fig:MLEU_S_h_whiskers}
\end{figure}

%% file: results_artifacts/Figure_20_Delta_S_and_Delta_R_over_time.tex
\begin{figure}[htp!]
\centering
\begin{subfigure}[t]{0.48\textwidth}
\centering
\pgfplotstableread[row sep=\\]{%
bias m0 e0m e0p m1 e1m e1p\\
0 0.1458 0.3403 0.3333 0.1286 0.3167 0.2845\\
1 0.3981 0.2667 0.2537 0.3330 0.2930 0.2700\\
3 0.9387 0.2900 0.3027 1.2556 0.3519 0.3556\\
}\learningR
\begin{tikzpicture}
\begin{axis}[
  width=\textwidth,
  height=6cm,
  ylabel={Mean $\tilde{\Delta}^R$},
  xlabel={Bias treatment},
  symbolic x coords={0,1,3},
  xtick=data,
  ymajorgrids=true,
  grid style={dashed,gray!40},
  legend to name=Learninglegend,
  legend columns=2,
  legend style={draw=none},
  legend cell align=left,
]
\addplot+[only marks, mark=*, mark options={draw=learningFirst, fill=learningFirst},
  xshift=-4pt, error bars/.cd, y dir=both, y explicit, error bar style={black}]
  table[x=bias, y=m0, y error minus=e0m, y error plus=e0p]{\learningR};
\addplot+[only marks, mark=square*, mark options={draw=learningSecond, fill=learningSecond},
  xshift=4pt, error bars/.cd, y dir=both, y explicit, error bar style={black}]
  table[x=bias, y=m1, y error minus=e1m, y error plus=e1p]{\learningR};
\addlegendentry{Rounds 1--6}
\addlegendentry{Rounds 7+}
\end{axis}
\end{tikzpicture}
\caption{$\tilde{\Delta}^R(b)$.}
\label{fig:Learning_receivers}
\end{subfigure}
\hfill
\begin{subfigure}[t]{0.48\textwidth}
\centering
\pgfplotstableread[row sep=\\]{%
bias m0 e0m e0p m1 e1m e1p\\
0 -0.0069 0.1736 0.1875 -0.0357 0.3893 0.3833\\
1 0.6185 0.3833 0.3574 0.6080 0.4700 0.4610\\
3 0.6520 0.4327 0.4433 1.1370 0.5111 0.5259\\
}\learningS
\begin{tikzpicture}
\begin{axis}[
  width=\textwidth,
  height=6cm,
  ylabel={Mean $\tilde{\Delta}^S$},
  xlabel={Bias treatment},
  symbolic x coords={0,1,3},
  xtick=data,
  ymajorgrids=true,
  grid style={dashed,gray!40},
]
\addplot+[only marks, mark=*, mark options={draw=learningFirst, fill=learningFirst},
  xshift=-4pt, error bars/.cd, y dir=both, y explicit, error bar style={black}]
  table[x=bias, y=m0, y error minus=e0m, y error plus=e0p]{\learningS};
\addplot+[only marks, mark=square*, mark options={draw=learningSecond, fill=learningSecond},
  xshift=4pt, error bars/.cd, y dir=both, y explicit, error bar style={black}]
  table[x=bias, y=m1, y error minus=e1m, y error plus=e1p]{\learningS};

\end{axis}
\end{tikzpicture}
\caption{$\tilde{\Delta}^S(b)$.}
\label{fig:Learning_senders}
\end{subfigure}
\vspace{0.7em}
\pgfplotslegendfromname{Learninglegend}
\vspace{0.4em}
\begin{minipage}{0.9\textwidth}
\footnotesize\centering
Notes: Error bars are subject-level bootstrap 95\% confidence intervals.
\end{minipage}
\caption{$\tilde{\Delta}^S(b)$ and $\tilde{\Delta}^R(b)$ over time.}
\label{fig:Learning_combined}
\end{figure}

%% file: instructions/Instructions.tex

\newenvironment{instructionsappendix}{
  \begingroup
  \setstretch{1.08}
\setlength{\parskip}{6pt}
\setlength{\parindent}{0pt}
  \fontfamily{phv}\selectfont\small\sansmath 
}{
  \endgroup
}

\newcommand{\Button}[1]{\textsf{\small[#1]}}

\begin{instructionsappendix}

\subsection*{Willkommen zum Experiment!}

Dieses Experiment besteht aus zwei Teilen: Der Slider-Aufgabe und der Urnen-Interaktion. Sie werden für jeden dieser beiden Teile bezahlt. Wir werden jeden Aufgabenteil im Detail beschreiben, bevor Sie damit beginnen.

Sie haben bereits 6,00 € für die Teilnahme am Experiment verdient. Das Experiment verwendet ein Punktesystem. Am Ende des Experiments werden Ihre Punkte in Geld umgerechnet. Sie erhalten \textbf{1 € pro 100 Punkte}. Das bedeutet, dass 1 Punkt = 1 Cent entspricht. Wir runden Ihre Auszahlung außerdem auf die nächsten 10 Cent auf. 

\subsection*{Slider-Aufgabe}

Willkommen zum ersten Teil des Experiments, der Slider-Aufgabe.

In dieser Aufgabe werden Sie als \textbf{aktive:r Teilnehmer:in} bezeichnet. Sie wurden zufällig mit einer anderen Person gepaart, die wir als \textbf{passive:n Teilnehmer:in} bezeichnen. Sie bleiben anonym und alle Ihre Entscheidungen sind vertraulich. Sie werden eine Reihe von 6 Entscheidungen darüber treffen, wie Punkte zwischen Ihnen und der oder dem passiven Teilnehmer:in aufgeteilt werden. Bitte geben Sie für jede der folgenden Fragen Ihre bevorzugte Verteilung aus den gegebenen Optionen an, indem Sie den \textbf{Slider bewegen}.

Ihre Entscheidungen führen dazu, dass sowohl Sie als auch die oder der passive Teilnehmer:in Punkte erhalten. Im untenstehenden Beispiel würden Sie 50 Punkte erhalten, während die oder der passive Teilnehmer:in 40 Punkte erhalten würde.

\begin{center}
\includegraphics[width=0.75\textwidth]{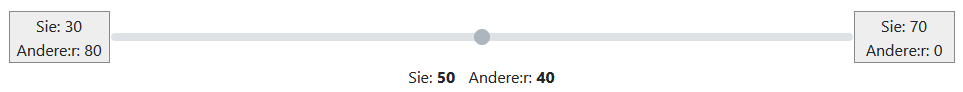}
\end{center}

Ebenso werden Sie für eine andere Person in diesem Experiment die oder der \textbf{passive Teilnehmer:in} sein und die entsprechende Auszahlung aus deren Entscheidungen erhalten. Diese Person wird jedoch nicht Ihr:e aktuelle:r Partner:in sein.

Sie und die oder der passive Teilnehmer:in werden für eine Ihrer Entscheidungen bezahlt. Diese wird zufällig und mit gleicher Wahrscheinlichkeit ausgewählt. Sie erhalten ebenfalls eine Auszahlung für eine zufällig ausgewählte Entscheidung der Person, für die Sie der oder die passive Teilnehmer:in sind.

Es gibt keine richtigen oder falschen Antworten – es geht allein um Ihre persönlichen Präferenzen. Wie Sie sehen, können Ihre Entscheidungen sowohl Ihre Auszahlung als auch die der anderen Person beeinflussen.

\subsection*{Anleitung für die Urnen-Interaktion}

\subsubsection*{Übersicht}

Willkommen zum zweiten Teil des Experiments, der Urnen-Interaktion!

Sie können jederzeit während des Experiments auf die im Folgenden erläuterte Anleitung zugreifen.

Bitte lesen Sie diese Anleitung sorgfältig durch. Am Ende der Anleitung gibt es ein paar Verständnisfragen. Sie erhalten eine zusätzliche Auszahlung von \textbf{100 Punkten}, wenn Sie die Fragen in maximal zwei Versuchen richtig beantworten.

Sie werden mit einer oder einem anderen Teilnehmer:in des Experiments interagieren. Es gibt zwei verschiedene Rollen: \\
\blue{Teilnehmer:in 1} und \redc{Teilnehmer:in 2}.

\begin{center}
\begin{minipage}{0.6\textwidth}
\begin{mdframed}[backgroundcolor=white, linewidth=0.8pt]
\textbf{Ihre Rolle}

\medskip
\textbf{Sie wurden zufällig als \blue{Teilnehmer:in 1} ausgewählt.}
\end{mdframed}
\end{minipage}
\end{center}

Ihre Rolle als Teilnehmer:in 1 bleibt für den Rest des Experiments dieselbe.

Die folgende Anleitung ist dieselbe für alle Teilnehmenden.

\subsubsection*{Urnen}

Es gibt zwei Urnen, \blue{Urne A} und \blue{Urne B}, die jeweils 10 Kugeln enthalten. Jede Kugel ist entweder \redc{rot} oder \blk{schwarz}.

\medskip
\begin{center}
\resizebox{\textwidth}{!}{
\begin{tabular}{cc}
\includegraphics[width=0.42\textwidth]{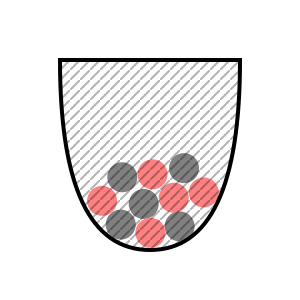} &
\includegraphics[width=0.42\textwidth]{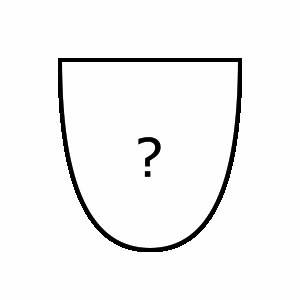} \\
\textit{\small{\blue{Urne A}: Gleiche Anzahl roter und schwarzer Kugeln}} & \textit{\small{\blue{Urne B}: Unbekannte Anzahl roter und schwarzer Kugeln}}
\end{tabular}}
\end{center}

In Urne A gibt es eine gleiche Anzahl von \redc{5 roten} und \blk{5 schwarzen} Kugeln.

In Urne B gibt es \redc{$X$ rote} und \blk{$10 - X$ schwarze} Kugeln. Keine:r der beiden Teilnehmer:innen kennt die genaue Anzahl der roten Kugeln in Urne B. Die Zahl \redc{$X$} wird in jeder Runde neu und zufällig bestimmt. Jede Zahl \redc{$X$} zwischen 0 und 10 ist dabei gleich wahrscheinlich, unabhängig von vorherigen Runden.

Es werden \blue{3} Kugeln mit Zurücklegen gezogen, \emph{jeweils} entweder aus Urne A oder Urne B. Zum Beispiel könnten eine rote und zwei schwarze Kugeln gezogen werden.

\medskip

\begin{center}
\includegraphics[width=0.75\textwidth]{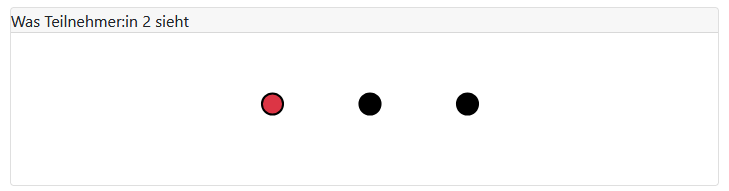}
\end{center}

Teilnehmer:in 1 weiß zusätzlich, aus welcher Urne die jeweilige Kugel gezogen wurde, Teilnehmer:in 2 nicht. Kugeln, die aus Urne A stammen, werden schraffiert dargestellt, um zu verdeutlichen, dass sie keine Informationen über die Zusammensetzung von Urne B liefern.

\begin{center}
\includegraphics[width=0.75\textwidth]{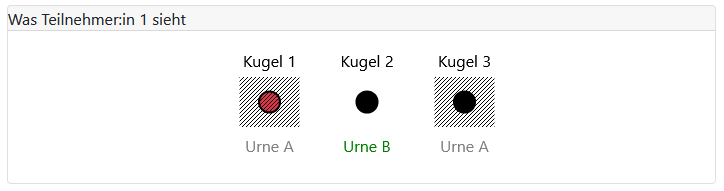}
\end{center}

Im obigen Beispiel wird Teilnehmer:in 1 darüber informiert, dass nur eine Kugel aus Urne B gezogen wurde und diese schwarz ist. Beachten Sie, dass nur die aus Urne B gezogenen Kugeln Informationen über \redc{$X$} liefern.

Im Verlauf des Experiments muss Teilnehmer:in 2 die Anzahl \redc{$X$} der roten Kugeln in Urne B schätzen. Je näher die Schätzung an der tatsächlichen Anzahl \redc{$X$} der roten Kugeln in Urne B liegt, desto höher ist die Auszahlung für Teilnehmer:in 2. Teilnehmer:in 1 hat eine Lieblingszahl: \blue{$X+3$}. Je näher die Schätzung an \blue{$X+3$} liegt, desto höher ist die Auszahlung für Teilnehmer:in 1.

\subsubsection*{Nachricht von Teilnehmer:in 1}

Teilnehmer:in 1 sendet eine Nachricht an Teilnehmer:in 2, in der sie oder er angeben kann, aus welchen Urnen die einzelnen Kugeln stammen. Diese Nachricht muss nicht wahrheitsgemäß sein. 

\begin{center}
\includegraphics[width=0.8\textwidth]{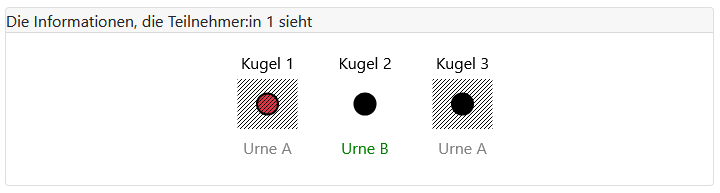}
\end{center}

\begin{center}
\includegraphics[width=0.75\textwidth]{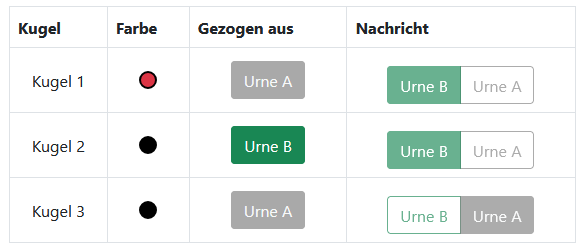}
\end{center}

\smallskip

Die Nachricht von Teilnehmer:in 1 wird grafisch dargestellt. Kugeln, die laut Teilnehmerin: 1 aus Urne A stammen, werden schraffiert dargestellt, um zu verdeutlichen, dass sie keine Informationen über die Zusammensetzung von Urne B liefern.

\begin{center}
\includegraphics[width=0.75\textwidth]{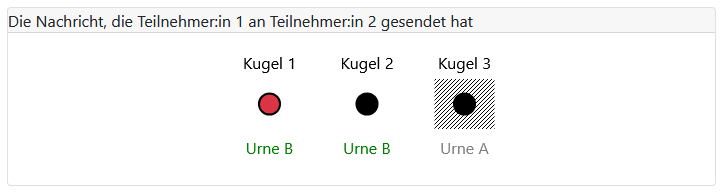}
\end{center}

Im Beispiel wurden Kugel 1 und Kugel 3 aus Urne A gezogen und Kugel 2 aus Urne B. Die Nachricht von Teilnehmer:in 1 behauptet, dass Kugel 1 und Kugel 2 aus Urne B gezogen wurden und Kugel 3 aus Urne A.

Als Rechenhilfe für Teilnehmer:in 1 erscheint das folgende Informationsfeld. Es gibt die empfohlene Schätzung für \redc{$X$} an – die Anzahl roter Kugeln, die die erwartete Auszahlung von Teilnehmer:in 2 maximiert, wenn die Kugeln tatsächlich wie in der Nachricht von Teilnehmer:in 1 angegeben gezogen wurden.

\begin{infobox}
Angenommen, die Kugeln wurden aus den von Ihnen in der rechten Spalte oberhalb dieser Nachricht gewählten Urnen gezogen. Dann maximiert die Schätzung von \redc{X=5} roten Kugeln in Urne B die erwartete Auszahlung von Teilnehmer:in 2.
\end{infobox}

\begin{center}
\begin{minipage}{0.8\textwidth}
\begin{mdframed}[backgroundcolor=white, linewidth=0.8pt]
\textbf{Wichtige Information}

Teilnehmer:in 1 muss die Informationen über die Urne nicht wahrheitsgemäß an Teilnehmer:in 2 weitergeben.
\end{mdframed}
\end{minipage}
\end{center}

\subsubsection*{Entscheidung}

Nachdem Teilnehmer:in 2 die Nachricht von Teilnehmer:in 1 erhalten hat, muss Teilnehmer:in 2 die Anzahl \redc{$X$} der roten Kugeln in Urne B schätzen.

In unserem Beispiel hat Teilnehmer:in 1 die folgende Nachricht gesendet:

\begin{center}
\includegraphics[width=0.75\textwidth]{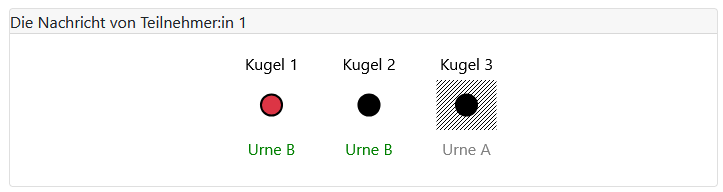}
\end{center}

In einem Informationsfeld erhält Teilnehmer:in 2 eine empfohlene Schätzung der Anzahl roter Kugeln auf Basis der Nachricht von Teilnehmer:in 1.

\begin{infobox}
Angenommen, die Kugeln wurden aus den Urnen gezogen, die Teilnehmer:in 1 in der Nachricht angegeben hat. Dann maximiert die Schätzung von \redc{X=5} roten Kugeln in Urne B Ihre erwartete Auszahlung.
\end{infobox}

Teilnehmer:in 2 muss nun die Anzahl \redc{$X$} der \redc{roten} Kugeln in Urne B schätzen. Teilnehmer:in 2 kann hierfür die folgende  Rechenhilfe benutzen, die basierend auf einer eigenen Auswahl an  relevanten Kugeln eine Empfehlung für die Wahl von  \redc{$X$} angibt.


\begin{center}
\includegraphics[width=0.75\textwidth]{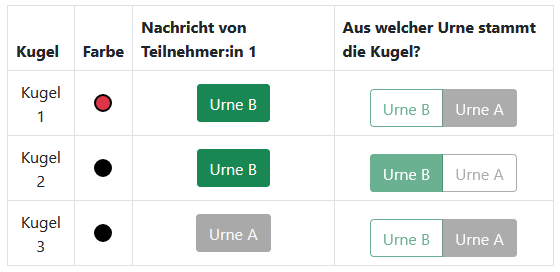}
\end{center}

\begin{infobox}
Angenommen, die Kugeln wurden aus den von Ihnen in der rechten Spalte oberhalb dieser Nachricht gewählten Urnen gezogen. Dann maximiert die Schätzung von \redc{X=2} roten Kugeln in Urne B Ihre erwartete Auszahlung.
\end{infobox}

\subsubsection*{Auszahlungen}

Teilnehmer:in 2 wird auf Grundlage ihrer oder seiner Schätzung von \redc{$X$} bezahlt. Je näher die Schätzung von Teilnehmer:in 2 an der tatsächlichen Anzahl \redc{$X$} der roten Kugeln in Urne B liegt, desto höher die Auszahlung für Teilnehmer:in 2.

Die Auszahlung für Teilnehmer:in 2 (in Punkten) wird wie folgt berechnet:

\begin{center}
\renewcommand{\arraystretch}{1.2}
\begin{tabular}{|>{\raggedright}p{5.5cm}|c|c|c|c|c|c|}
\toprule
Abstand der Schätzung zu \redc{$X$} & 0 & 1 & 2 & 3 & 4 & 5 oder mehr \\
\midrule
Auszahlung von Teilnehmer:in 2 \newline (in Punkten)
& \blk{1300} & \blk{1000} & \blk{700} & \blk{400} & \blk{100} & \blk{0} \\
\bottomrule
\end{tabular}
\end{center}

Teilnehmer:in 1 wird basierend darauf bezahlt, wie nah die Schätzung von Teilnehmer:in 2 an Ihrer \blue{Lieblingszahl X+3} ist, die größer als \redc{X} ist.

Die Auszahlung für Teilnehmer:in 1 (in Punkten) wird wie folgt berechnet:

\begin{center}
\renewcommand{\arraystretch}{1.2}
\begin{tabular}{|>{\raggedright}p{5.5cm}|c|c|c|c|c|c|}
\toprule
Abstand der Schätzung zu \blue{$X+3$} & 0 & 1 & 2 & 3 & 4 & 5 oder mehr \\
\midrule
Auszahlung von Teilnehmer:in 1 \newline (in Punkten)
& \blk{1300} & \blk{1000} & \blk{700} & \blk{400} & \blk{100} & \blk{0} \\
\bottomrule
\end{tabular}
\end{center}

Zum Beispiel: Wenn die wahre Anzahl der Kugeln in Urne B \redc{$X=6$} ist, dann ist die Lieblingszahl von Teilnehmer:in 1 \blue{$X+3=9$}. Wenn Teilnehmer:in 2 eine Schätzung von \textbf{5} gemacht hat, dann wäre die Auszahlung für Teilnehmer:in 2 \redc{1000 Punkte} (weil die Schätzung 1 Kugel von \redc{$X$} entfernt ist), und die Auszahlung von Teilnehmer:in 1 wäre \blue{100 Punkte} (weil die Schätzung 4 Kugeln von \blue{$X+3$} entfernt ist).

\medskip

\begin{center}
\scriptsize
\renewcommand{\arraystretch}{1.2}
\setlength{\tabcolsep}{4pt}

\begin{tabularx}{\textwidth}{%
  !{\color{bordergray}\vrule width 1.5pt}
  >{\raggedright\arraybackslash}p{4.5cm}
  !{\color{bordergray}\vrule width 1.5pt}
  *{5}{>{\centering\arraybackslash}X !{\color{bordergray}\vrule width 0.8pt}}
  >{\centering\arraybackslash}X !{\color{borderred}\vrule width 2pt}
  >{\centering\arraybackslash\color{borderred}}m{0.8cm} !{\color{borderred}\vrule width 2pt}
  >{\centering\arraybackslash}X !{\color{bordergray}\vrule width 0.8pt}
  >{\centering\arraybackslash}X !{\color{borderblue}\vrule width 2pt}
  >{\centering\arraybackslash\color{borderblue}}m{0.8cm} !{\color{borderblue}\vrule width 2pt}
  >{\centering\arraybackslash}X !{\color{bordergray}\vrule width 1.5pt}
}

\specialrule{2pt}{0pt}{0pt}
Schätzung & 0 & 1 & 2 & 3 & 4 & 5 & 6 & 7 & 8 & 9 & 10 \\
\specialrule{1.2pt}{0pt}{0pt}

\makecell[l]{Auszahlung von Teilnehmer:in 1\\(in Punkten)}
& 0 & 0 & 0 & 0 & 0 & 100 & 400 & 700 & 1000 & 1300 & 1000 \\
\hline
\makecell[l]{Auszahlung von Teilnehmer:in 2\\(in Punkten)}
& 0 & 0 & 100 & 400 & 700 & 1000 & 1300 & 1000 & 700 & 400 & 100 \\
\specialrule{2pt}{0pt}{0pt}

\end{tabularx}
\end{center}

\subsubsection*{Runden}

Sie werden \blue{N Runden} der Urnen-Interaktion spielen, sowie eine Übungsrunde vorab. Jedes Mal haben Sie eine oder einen neuen Partner:in, mit der oder dem Sie noch nie zuvor gespielt haben. Die Rollen von Teilnehmer:in 1 und Teilnehmer:in 2
bleiben während des gesamten Experiments gleich.

In jeder Runde wird die Anzahl der roten Kugeln in Urne B erneut zufällig gezogen. Auch die Kugeln werden erneut gezogen, möglicherweise aus anderen Urnen. Daher ist jede Runde unabhängig von den anderen.

Sie werden weder die wahre Anzahl der roten Kugeln in Urne B erfahren noch aus welchen Urnen die Kugeln gezogen wurden.

\newpage

\subsubsection*{Zusammenfassung}

Jede Runde hat folgende Phasen:

\begin{center}
\scriptsize
\renewcommand{\arraystretch}{1.2} 
\setlength{\tabcolsep}{4pt}       
\begin{tabularx}{\textwidth}{|>{\raggedright\arraybackslash}X|>{\raggedright\arraybackslash}X|}
\hline
\textbf{Teilnehmer:in 1} & \textbf{Teilnehmer:in 2} \\
\hline
\multicolumn{2}{|c|}{Die Kugeln werden neu und unabhängig von vorherigen Runden gezogen und beiden Teilnehmer:innen angezeigt.} \\
\hline
Die Urnen, aus denen die jeweiligen Kugeln gezogen wurden, werden Teilnehmer:in 1 angezeigt. Teilnehmer:in 1 sendet eine Nachricht an Teilnehmer:in 2. & \\
\hline
& Die Nachricht von Teilnehmer:in 1 wird Teilnehmer:in 2 angezeigt. \\
\hline
& Teilnehmer:in 2 schätzt die Anzahl der roten Kugeln in Urne B. \\
\hline
\end{tabularx}
\end{center}

Je näher die Schätzung von Teilnehmer:in 2 an der tatsächlichen Anzahl \redc{$X$} der roten Kugeln in Urne B liegt, desto höher ist die Auszahlung für Teilnehmer:in 2.

Je näher die Schätzung von Teilnehmer:in 2 an der Lieblingszahl \blue{$X + 3$} von Teilnehmer:in 1 liegt, desto höher ist die Auszahlung für Teilnehmer:in 1.

Sie werden basierend auf einer Runde bezahlt, die als  Zahlungsrunde bezeichnet wird. Diese wird nach Abschluss der letzten Runde vom Computer zufällig ausgewählt und Ihre Auszahlungen angezeigt.

\end{instructionsappendix}